\documentclass[twocolumn,showkeys,noprintnumbers,amssymb,aps,superscriptaddress,prr,longbibliography]{revtex4-2}
\usepackage{graphicx}
\usepackage{float}
\usepackage{dcolumn}
\usepackage{bm}
\usepackage{color}
\usepackage{xcolor}
\usepackage[T1]{fontenc}
\usepackage{amsmath}
\usepackage[breaklinks,colorlinks,bookmarks=false,citecolor=blue,linkcolor=red,urlcolor=blue]{hyperref}

\newcommand{\rsquare}{\rotatebox[origin=c]{45}{$\scriptstyle \square$}}

\begin{document}

\onecolumngrid
\vspace*{-11.5mm}

\rightline{published in \href{https://doi.org/10.1103/PhysRevB.110.214429}{Phys. Rev. B {\bfseries 110}, 214429 (2024)}}
\vspace*{2mm}
\twocolumngrid

\title{Thermodynamic properties  of the macroscopically degenerate tetramer-dimer phase of the spin-1/2 Heisenberg model on the  diamond-decorated square lattice}

\author{Katar\'ina Kar{l}'ov\'a}
\email{katarina.karlova@cyu.fr}
\affiliation{Laboratoire de Physique Th\'eorique et
Mod\'elisation, CNRS UMR 8089, CY Cergy Paris Universit\'e, Cergy-Pontoise, France}
\author{Andreas Honecker}
\affiliation{Laboratoire de Physique Th\'eorique et
Mod\'elisation, CNRS UMR 8089, CY Cergy Paris Universit\'e, Cergy-Pontoise, France}
\author{Nils Caci}
\affiliation{Laboratoire Kastler Brossel, Coll\`ege de France, CNRS, \'Ecole Normale Sup\'erieure - Universit\'e PSL, Sorbonne Universit\'e, 75005 Paris, France}
\author{Stefan Wessel}
\affiliation{Institute for Theoretical Solid State Physics, RWTH Aachen University,
Otto-Blumenthal-Str.\ 26, 52074  Aachen, Germany}
\author{Jozef Stre\v{c}ka}
\affiliation{Institute of Physics, Faculty of Science, P. J. \v{S}af\'{a}rik University, Park Angelinum 9, 04001 Ko\v{s}ice, Slovakia}
\author{Taras Verkholyak}
\affiliation{Institute for Condensed Matter Physics, National Academy of Sciences of Ukraine, Svientsitskii Street 1, 79011, L'viv, Ukraine}

\date{September 2, 2024; revised November 26, 2024}

\begin{abstract}
The spin-1/2 Heisenberg antiferromagnet on the diamond-decorated square lattice in the presence of a magnetic field displays  various quantum phases including the Lieb-Mattis ferrimagnetic, dimer-tetramer, monomer-dimer, and spin-canted phases, in addition to the trivial fully saturated state. Thermodynamic properties of this model are investigated using several complementary analytical and numerical methods such as exact diagonalization up to the systems of 40 spins, an effective monomer-dimer description, sign-problem-free quantum Monte Carlo simulations for up to 180 spins, and a decoupling approximation. Our particular attention is focused on the parameter region favoring the dimer-tetramer phase. This ground state can be represented by a classical hard-dimer model on the square lattice and retains a macroscopic degeneracy even under a magnetic field. However, the description of the low-temperature thermodynamics close to the boundary between the macroscopically degenerate dimer-tetramer and the non-degenerate monomer-dimer phases requires an extended classical monomer-dimer lattice-gas model. Anomalous thermodynamic properties emerging in the vicinity of the dimer-tetramer phase are studied in detail. Under the adiabatic demagnetization we detect an enhanced magnetocaloric effect promoting an efficient cooling to absolute zero temperature.
\end{abstract}

\maketitle

\section{Introduction}
Over the past decades significant attention has been dedicated to the investigation of low-dimensional quantum magnets situated on geometrically frustrated lattices \cite{Lacroix2011,Honecker_2004,Lhuillier2001,Diep2013}. The hallmark of classical highly frustrated magnets are
macroscopically degenerate  ground states \cite{Diep2013,Liebmann1986,ramirez2024}. 
Quantum fluctuations instead 
possess the remarkable capability to manipulate and potentially eliminate such degeneracies through a phenomenon known as the quantum order-by-disorder mechanism \cite{Villain1980,Henley1989,Moessner2021,Bergman2007,Lee2008,Buessen2018,ramirez2024}.

Quantum spin systems thus often exhibit  ordered ground states such as long-range antiferromagnetism \cite{Joli1990,Reimers1993,Bramwell1994} or valence bond solids \cite{Majumdar1969,Shastry1981,Koga2000,Ganesh2013}. However, the interplay of  frustration and quantum fluctuations can occasionally also give rise to the emergence of exotic quantum disordered states, most prominently in quantum spin liquids \cite{Balents2010,Savary2017,Zhou2017,Broholm2020}. Characterized by  long-range entanglement, quantum spin liquids indeed exhibit a plethora of intriguing phenomena, including the fractionalization of excitations and the emergence of topological order \cite{Kim2000,Yamashita2010,Jiang2012,Han2012,Banerjee2016}. Nevertheless, it is also feasible for frustrated quantum magnets to stabilize quantum disordered phases that retain a peculiar macroscopic degeneracy in the absence of quantum spin-liquid characteristics.

Frustrated quantum spin systems constitute a challenge for theory—for example, because the infamous sign problem \cite{Henelius2000,Troyer2005}
precludes straightforward and efficient quantum Monte Carlo simulations.
For this reason, tailor-made models such as the Rokhsar-Kivelson \cite{RK1988} and Kitaev \cite{Kitaev2006} models have been proposed in order to allow for obtaining an exact solution.
However, geometric frustration can sometimes also be beneficial rather than detrimental: if kinetic energy is significantly or even completely suppressed by destructive quantum interference, the quantum problem may be mapped to an effective classical dimer model \cite{Wu2006} or at least well approximated by a quantum dimer model \cite{Moessner2011}, thus rendering a controlled approximation if not an exact treatment of the low-energy physics possible.
Examples of such phenomena include but are not limited to
models with exact dimer ground states in one \cite{Majumdar1969} and two dimensions \cite{Shastry1981},
the one-third plateau in the kagome quantum antiferromagnet \cite{Moessner2001,Cabra2005},
and localized magnons appearing just below the saturation field of highly frustrated antiferromagnets
\cite{Schnack2001,Schulenburg2002,ZT2004,ZT2005,DRHS07,Schnack2020}.

\begin{figure}[t!]
	\centering
	\includegraphics[width=0.9\columnwidth]{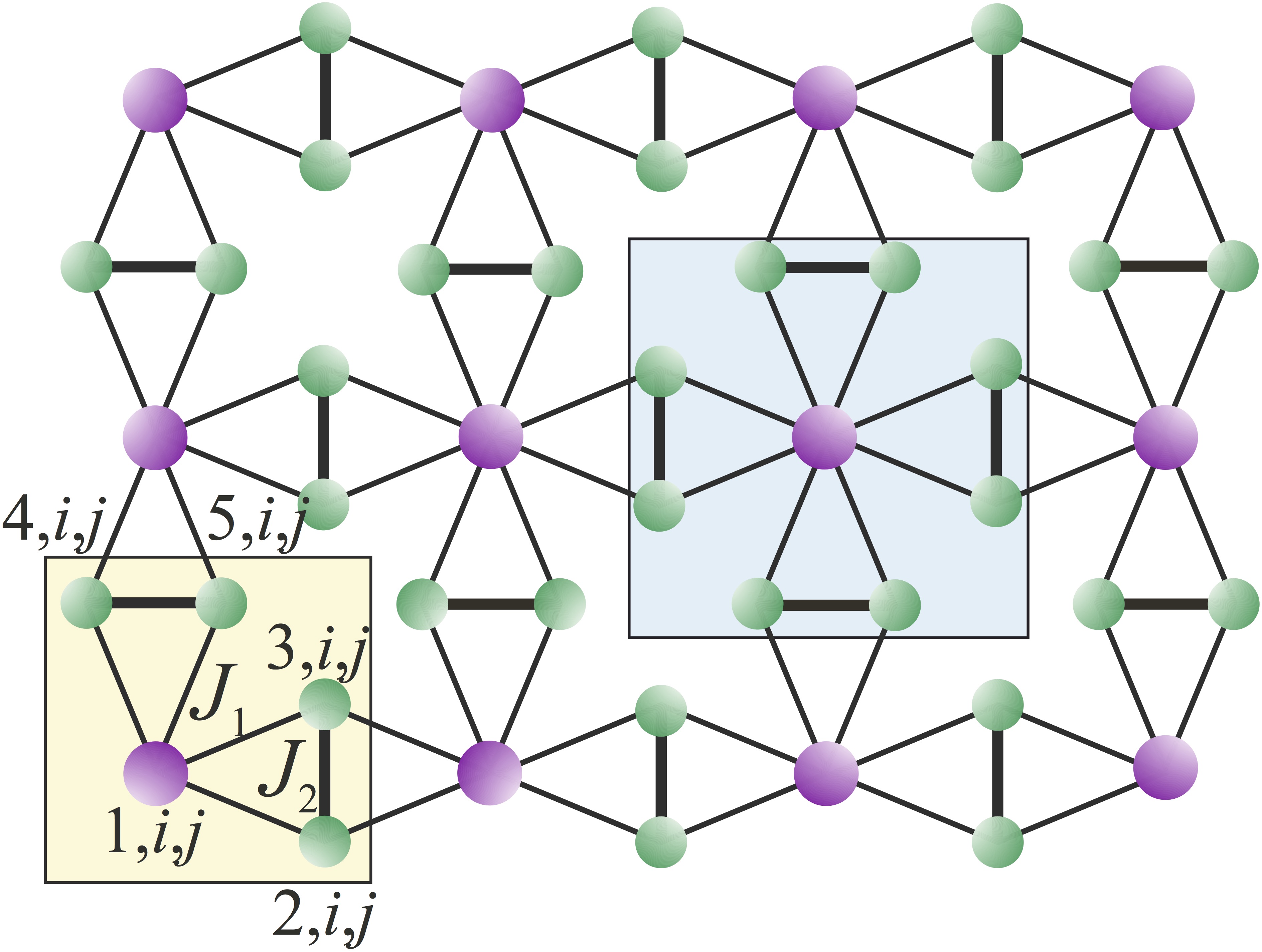}
	\caption{The spin-1/2 Heisenberg model on the diamond-decorated square lattice with one unit cell shown by the yellow square. The decoupling approximation of Sec.~\ref{spinStar} is developed from the spin-star cluster framed in a blue square.}
	\label{fig:model}
\end{figure}

In this respect, the Heisenberg model on the diamond-decorated square lattice, sketched in Fig.~\ref{fig:model}, serves as an illustrative example of the interplay between frustration, quantum fluctuations, and emergent phenomena.
Previous investigations into the ground-state characteristics of this model
in the absence of a magnetic field
demonstrate the presence of a Lieb-Mattis ferrimagnetic phase, a monomer-dimer phase, and a dimer-tetramer phase \cite{Morita2016,Hirose2016,Hirose2017,Hirose2018,Hirose2020}. The latter two phases exhibit a macroscopic ground-state degeneracy in the absence of a magnetic field and thus deserve particular attention.
The dimer-tetramer ground state retains its macroscopic degeneracy even in the presence of a finite magnetic field (in contrast to the monomer-dimer state), highlighting its exceptional character \cite{Caci2023}. This robust degeneracy can furthermore be related to the classical hard-dimer model on the square lattice, emphasizing its connection to well-known models in condensed-matter physics \cite{Lieb1967,Heilmann1970,Suzuki1971a,Suzuki1971b,Heilmann1972,Grande2011,Wu2006,Wilkins2021}.
Hence, the dimer-tetramer phase, characterized by its short-range ordering and robust macroscopic degeneracy over an extended parameter regime, offers an intriguing perspective on frustrated quantum magnetism. 
In this paper, we therefore examine the thermodynamic and magnetocaloric properties of the dimer-tetramer phase of the spin-1/2 Heisenberg model on the diamond-decorated square lattice, including a finite magnetic field. 
Indeed, we will show that this phase presents an ideal candidate for applications such as magnetic cooling during the process of adiabatic demagnetization \cite{Mike,Zhitomirsky_2004,HONECKER20061098}.

The further organization of this paper is as follows: The Hamiltonian and ground-state phase diagram of the spin-1/2 Heisenberg model on the diamond-decorated square lattice are reviewed in Sec.~\ref{model}. Our analytical and numerical approaches will be introduced in Sec.~\ref{method}. Results are then presented in Sec.~\ref{results} and a final summary is provided in Sec.~\ref{conclusion}.

\section{Heisenberg model on the diamond-decorated square lattice}
\label{model}

In the following, we consider the spin-1/2 Heisenberg model on the diamond-decorated square lattice depicted in Fig.~\ref{fig:model}, and defined through the  Hamiltonian 
\begin{eqnarray}
	{\cal H} &=& J_1 \sum \limits_{i=1}^N \Bigl[ \bm{S}_{i,1} \cdot
	\Bigl(\bm{S}_{i,2}+\bm{S}_{i,3}+\bm{S}_{i,4}+\bm{S}_{i,5}
	\nonumber \\
	&&\qquad+
	\bm{S}_{i-\hat{x},2}+\bm{S}_{i-\hat{x},3}+\bm{S}_{i-\hat{y},4}+\bm{S}_{i-\hat{y},5}\Bigr)
	\Bigr]\nonumber \\
	&&+ J_2 \sum \limits_{i=1}^N \Bigl(\bm{S}_{i,2}\cdot \bm{S}_{i,3} +
	\bm{S}_{i,4} \cdot \bm{S}_{i,5}\Bigr)\nonumber \\
	&&- h\sum \limits_{i=1}^N\sum
	\limits_{\mu=1}^{5} S_{i,\mu}^z\,.
	\label{hamos}
\end{eqnarray}
The spin-1/2 operator $\bm{S}_{i,\mu}= ({S}_{i,\mu}^x, {S}_{i,\mu}^y, {S}_{i,\mu}^z)$ is assigned to the $\mu$th spin of the $i$th unit cell. The indices $i-\hat x$  and $i-\hat y$  refer to the unit cells immediately to the left and below the $i$th unit cell, respectively. We consider a finite lattice consisting of $N$ unit cells, containing $N_s = 5N$ spins, and we apply periodic boundary conditions. Typically, we utilize square lattices with lattice sizes $L_x$, $L_y$, and the total number of unit cells $N=L_xL_y$. The two distinct exchange interactions $J_1$  and $J_2$ are depicted in Fig.~\ref{fig:model} by thin and thick lines, respectively. The magnetic field  ($h$) term  represents the standard Zeeman coupling of the spin degrees of freedom to an external magnetic field.

\begin{figure}[t!]
	\centering
	\includegraphics[width=0.9\columnwidth]{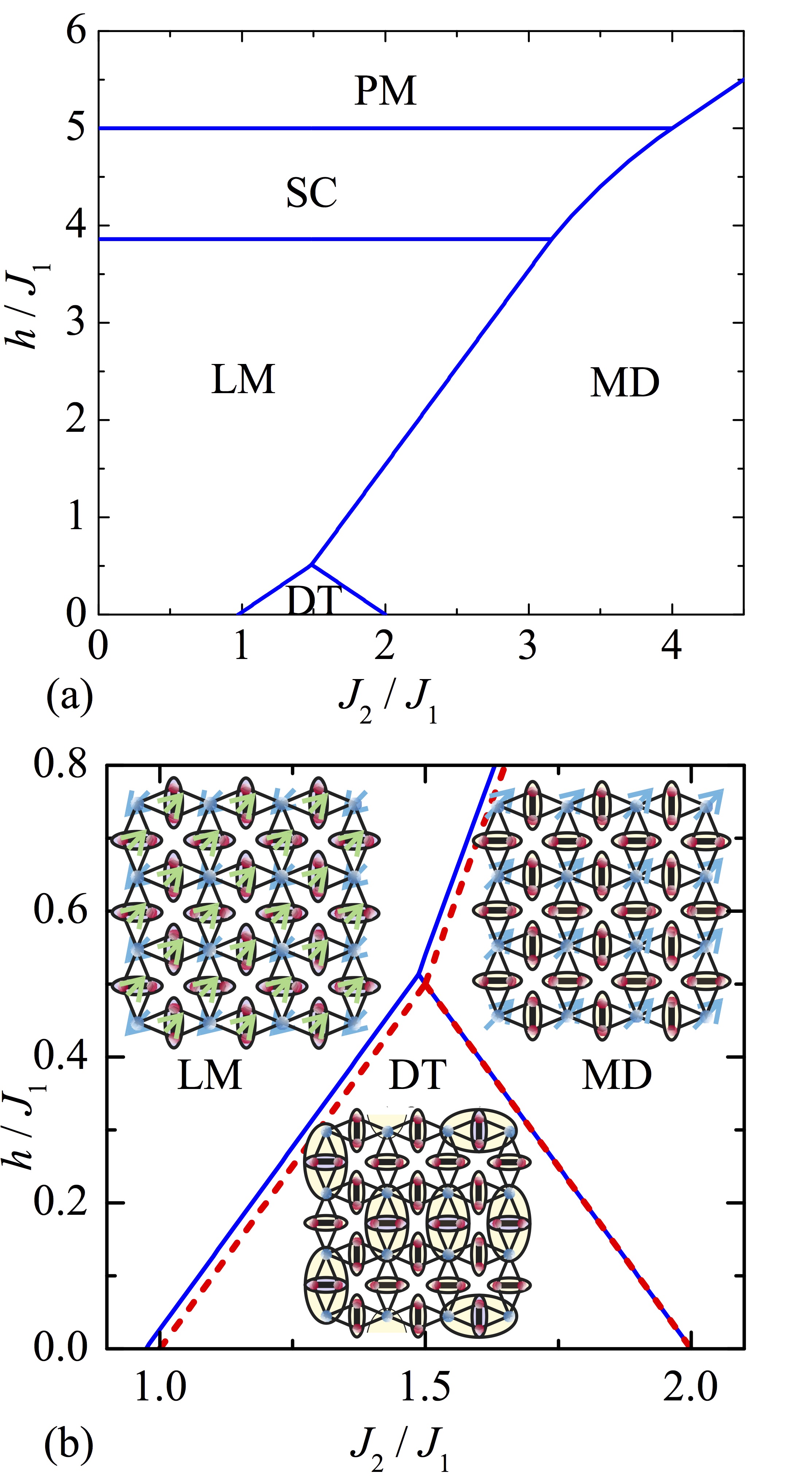}
	\caption{(a) Ground-state phase diagram of the spin-1/2 Heisenberg model on the diamond-decorated square lattice.
        The following phases were obtained in Ref.~\cite{Caci2023}:
        saturated paramagnetic (PM), spin-canted (SC), Lieb-Mattis (LM) ferrimagnetic, dimer-tetramer (DT), and monomer-dimer (MD) phases.
        Panel (b) zooms in on the DT phase that is the focus of the present paper and provides cartoons of representative ground states in the DT, MD, and LM phases. Red-broken lines show 
phase boundaries as obtained from the spin-star decoupling approximation discussed in Sec.~\ref{spinStar}.
        }
	\label{fig:gspd}
\end{figure}

Our investigations center on the thermodynamic properties of the dimer-tetramer phase. However,  we start by providing a brief overview of the ground-state phase diagram and the characteristics of various phases occurring in the low-field regime. The ground-state phase diagram of the spin-1/2 Heisenberg model on the diamond-decorated square lattice is depicted in Fig.~\ref{fig:gspd} and was extensively described previously \cite{Caci2023}. It  exhibits five phases in the parameter plane  spanned by  the interaction ratio $J_2/J_1$ and the reduced magnetic field $h/J_1$. In addition to a saturated paramagnetic (PM) state and the spin-canted (SC) phase, the phase diagram features three distinct phases: the Lieb-Mattis (LM) ferrimagnetic, the dimer-tetramer (DT), and the monomer-dimer (MD) phases. In the LM phase the Heisenberg dimers form triplets and the overall state is reminiscent of classical ferrimagnetic order.
The MD and DT phases exhibit fragmentation. In both cases, the overall wave function can be expressed as a product of individual fragments \cite{Caci2023}. In the MD phase, singlet dimers (depicted in Fig.~\ref{fig:gspd} as yellow ovals) coexist with free monomeric spins in zero field ($h=0$) and fully polarized monomeric spins in nonzero field.
More specifically,  the ground state can be presented as follows: 
\begin{eqnarray}
	|\mathrm{MD}\rangle=\left\{ 
	\begin{array}{l}
		\prod_{i=1}^N|\sigma\rangle_{i,1}\otimes\prod_{d=1}^{2N}|s\rangle_d, \, \sigma\in\{\uparrow,\downarrow\},\, h=0 \ \\
		\prod_{i=1}^N|\!\uparrow\rangle_{i,1}\otimes\prod_{d=1}^{2N}|s\rangle_d, \, h> 0,
	\end{array}\right. \
	\label{monomerDimer}
\end{eqnarray}
where $|s\rangle_d=(|\!\!\uparrow\rangle|\!\!\downarrow\rangle-|\!\!\downarrow\rangle|\!\!\uparrow\rangle)/\sqrt{2}$ represents the dimer-singlet state.
The DT phase is particularly intriguing because it comprises a macroscopic number of states that contain both singlet dimers and singlet tetramers.
The singlet tetramers contain in turn triplet dimers, as depicted in Fig.~\ref{fig:gspd} by large yellow and small purple ovals, respectively. Furthermore, each singlet tetramer is surrounded by singlet dimers within the DT ground-state manifold. 
As a result, overall these DT states are singlets and are of the following form:
\begin{eqnarray}
	|\mathrm{DT}\rangle &=& \sideset{}{'}\prod_{\{d\}}^{} \prod_{d'\neq d} |t\rangle_d |s\rangle_{d'},
	\nonumber\\
	|t\rangle_d &=& \frac{1}{\sqrt{3}}(|\!\uparrow\rangle_{i,1}|\!\downarrow\uparrow\rangle_{d}|\!\downarrow\rangle_{i',1}+|\!\downarrow\rangle_{i,1}|\!\uparrow\downarrow\rangle_{d}|\!\uparrow\rangle_{i',1}) \nonumber \\
	&&-\frac{1}{2}\left(|\!\uparrow\rangle_{i,1}|\!\uparrow\downarrow\rangle_{d}|\!\downarrow\rangle_{i',1} + |\!\uparrow\rangle_{i,1}|\!\downarrow\downarrow\rangle_{d}|\!\uparrow\rangle_{i',1}\right. \nonumber \\
	&&\quad+\left.|\!\downarrow\rangle_{i,1}|\!\uparrow\uparrow\rangle_{d}|\!\downarrow\rangle_{i',1}+ |\!\downarrow\rangle_{i,1}|\!\downarrow\uparrow\rangle_{d}|\!\uparrow\rangle_{i',1}\right)\, ,\quad
	\label{DTphase}
\end{eqnarray}
where $\sideset{}{'}\prod_{\{d\}}^{}$ denotes the product over a subset of dimers that are not nearest neighbors. 
This phase  exhibits a macroscopically high degeneracy that is related to the hard-dimer model on the square lattice \cite{Lieb1967,Heilmann1970,Heilmann1972,Grande2011,Wilkins2021,Suzuki1971a,Suzuki1971b}.
We will be especially interested in the regime  near the quantum phase transition line between the MD and DT phases. This transition line is given by the condition
\begin{eqnarray}
	h = 2J_1 - J_2, \ {\rm if} \ 1.487 J_1\lesssim J_2 \le 2J_1,
	\label{h_MD-DT}
\end{eqnarray}
and here,  the model is expected to exhibit an enhanced magnetocaloric effect, as we explain further below.  

We note that the finite-temperature properties in the vicinity of the MD-LM transition are well captured by a
Ising-Heisenberg variant of the model \cite{Strecka2023,Ghannadan2024}. By contrast, the DT phase (which is the focus of the present investigation) owes its existence to quantum fluctuations on the monomeric spins and is thus absent in the Ising-Heisenberg variant.

\section{Methods}
\label{method}
The overall 
thermodynamic properties and thermal phase transitions for various phases of this model were examined previously~\cite{Caci2023}. Here, we focus on the macroscopically degenerate DT phase, which is investigated using exact diagonalization, sign-problem-free quantum Monte Carlo simulations, an effective monomer-dimer (EMD) description, as well as a spin-star decoupling approximation. 
We also show some density-matrix-renormalization group (DMRG) ground-state
results, but no new data has been produced for the present purposes such
that we refer to Ref.~\cite{Caci2023} for technical details on these DMRG
calculations.

At the beginning of each of the following subsections, we briefly outline its purpose.
Readers primarily interested in the results and their implications may skip the technical details in the corresponding subsections,
or proceed directly to Sec.~\ref{results}, where our findings are discussed in detail.
\subsection{Effective monomer-dimer (EMD) model}
In this subsection, we derive and discuss the transfer-matrix solution of the effective monomer-dimer (EMD) model as a simplified
approach to describe the low-temperature thermodynamic properties of the spin-1/2 Heisenberg model in the DT phase. This model focuses on capturing the contributions of low-energy excitations from triplets to singlets and their influence on quantities such as entropy, specific heat, and magnetization.

In order to formulate the effective model at the boundary between the MD and DT phases, we 
 recall some notation~\cite{Caci2023}. 
The energy of singlet dimers of the MD state (\ref{monomerDimer}) can be calculated as follows:
\begin{eqnarray}
	E^{(0)}_{\mathrm{MD}} = 2N\varepsilon_{sd} = -\frac{3N}{2}J_2,
	\label{eMD0}
\end{eqnarray} 
where $\varepsilon_{sd}=-\frac{3}{4}J_2$ is the energy of one singlet dimer.
To obtain the energy of the MD state at nonzero fields, we need to add the contribution of the Zeeman term for the monomer spins to Eq.~(\ref{eMD0}):
\begin{eqnarray}
	E_{\mathrm{MD}} = E^{(0)}_{\mathrm{MD}} - Nh/2.
	\label{eMD}
\end{eqnarray} 
The DT phase of the states (\ref{DTphase}) consists of $N/2$ singlet tetramers and $3N/2$ singlet dimers.
Thus, its energy is
\begin{eqnarray}
	\label{eDT}
	E_{\rm DT} = N\left(\frac{3}{2}\varepsilon_{sd} + \frac{1}{2}\varepsilon_{st}\right)
	= - N (J_2 + J_1),
\end{eqnarray}
where $\varepsilon_{st}=-2J_1+\frac{1}{4}J_2$ is the energy of the singlet tetramer.
The condition $E_{\mathrm{MD}} = E_{\rm DT}$ yields the phase-transition line given by Eq.~(\ref{h_MD-DT}). 
The ground state is macroscopically degenerate on the phase-transition line. It contains all possible configurations of two states of decorated diamonds: (i)  singlet dimers, (ii)  singlet tetramers containing a dimer triplet. 
For all states the condition applies that the singlet tetramers cannot have common edges. 

The singlet-dimer-to-singlet-tetramer excitation is given by
the gap
\begin{eqnarray}
	\Delta_{H,V} = \varepsilon_{st} - \varepsilon_{sd} = J_2^{H,V} - 2J_1.
\end{eqnarray}
Here, we have introduced a distinction between horizontal ($H$) and vertical dimers ($V$)
with two different interaction constants $J_2^H$ and $J_2^V$
in anticipation of a transfer-matrix approach that will treat the two spatial directions differently.

\begin{figure}[t!]
	\centering
\leavevmode
\put(-20,-10){{\large (a)}}
\includegraphics[width=0.35\textwidth]{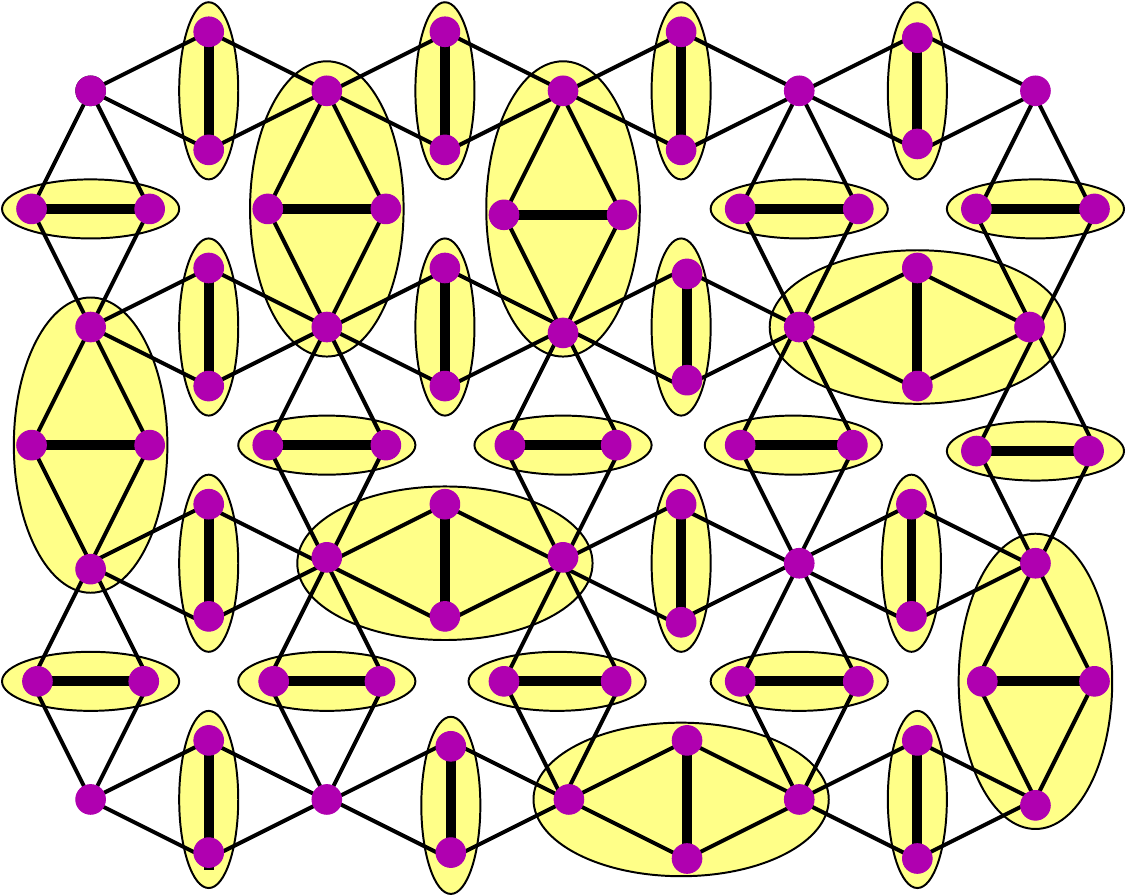}\\[12pt]
\leavevmode
\put(-20,-10){{\large (b)}}
\includegraphics[width=0.35\textwidth]{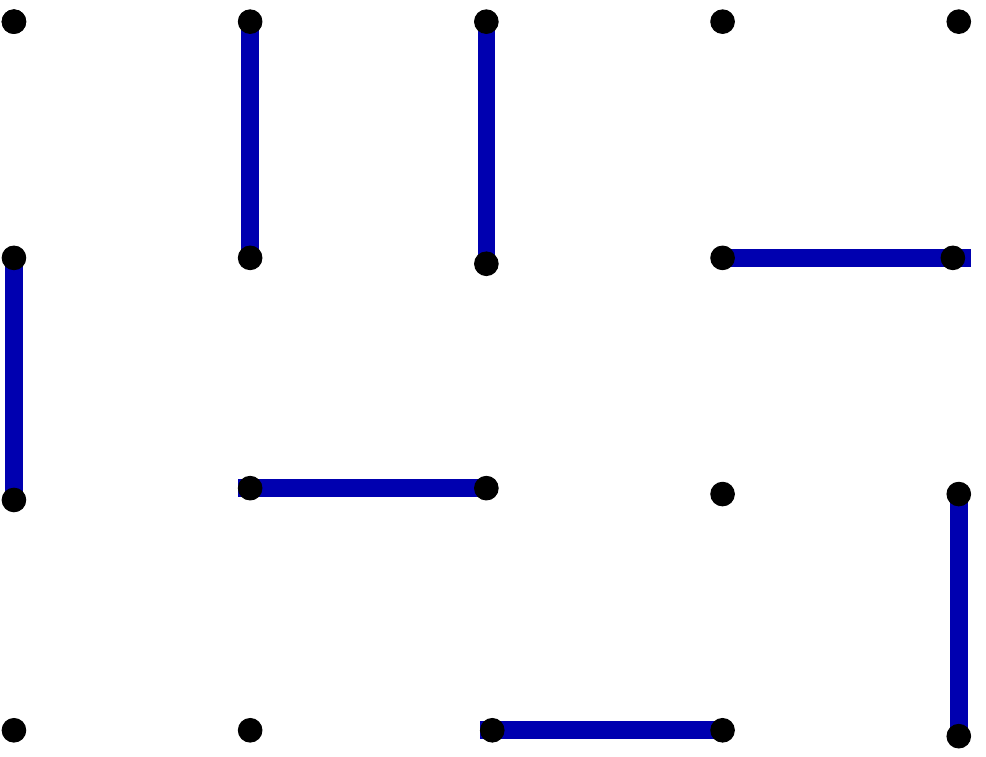}\\[12pt]
\leavevmode
\put(-20,-10){{\large (c)}}
\includegraphics[width=0.35\textwidth]{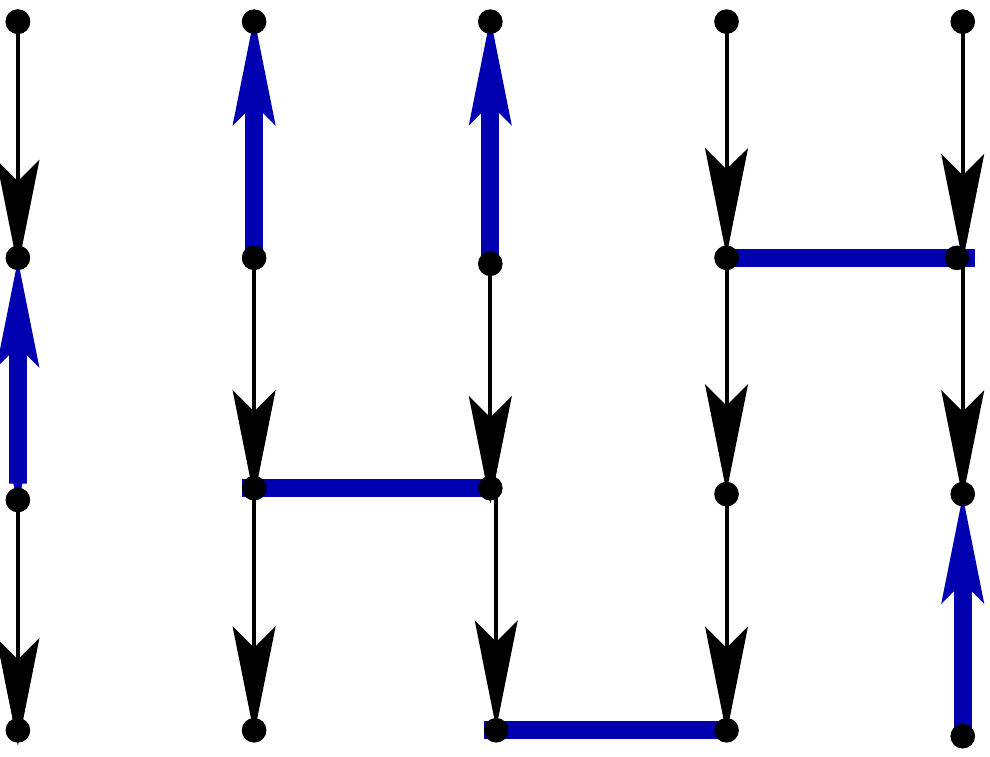}\\[12pt]
	\caption{Schematic of a particular configuration of the mixture of MD and DT phases in different representations. 
		(a) The diamond-decorated square lattice where  yellow ovals indicate a singlet state on the highlighted clusters. 
		(b) EMD representation of the same model.
		(c) The spin representation for the transfer-matrix formalism \cite{Lieb1967}.}
	\label{fig:TDphase}
\end{figure}

It should be noted that the two monomer spins in a diamond do not contribute to the Zeeman term in case they are included in the singlet tetramer. 
As a result, the effective low-temperature model can be considered as a monomer-dimer problem on the square lattice \cite{Lieb1967}, where dimers can be located on the bonds of a square lattice connecting nearest-neighbor vertices (which corresponds to the singlet tetramer state of the  diamond-decorated square lattice) and each vertex may host no more than one dimer ( i.e., different singlet tetramers cannot have common spins). This correspondence is illustrated in Figs.~\ref{fig:TDphase}(a) and \ref{fig:TDphase}(b). 
Defining the energy with respect to the dimer singlets, the partition function of such a model can be written as follows:
\begin{eqnarray}
	Z = e^{-\beta E^{(0)}_{\rm MD}}\sum_{\cal{C}} x^H y^V z^M,
\label{eq:Z_DM}
\end{eqnarray}
where $\sum_{\cal{C}}$ denotes the sum over all possible configurations of dimers on the square lattice; 
$M$, $H$, $V$ are the numbers of monomers and horizontal and vertical dimers (which obey the restriction $2H+2V+M = N$); 
$x=e^{-\beta\Delta_H}$, $y=e^{-\beta\Delta_V}$, $z=2\cosh(\beta h/2)$ are the respective activities,
$\beta = 1/T$ is the inverse temperature (where we set $k_B=1$). 
The monomer activity $z$ corresponds to the partition function of a single monomer spin in an external field.
Recall that we have introduced the anisotropic excitations $\Delta_H$ and $\Delta_V$ for the horizontal and vertical dimers in order to identify the corresponding contributions in Eq.~(\ref{eq:Z_DM}) explicitly, but fix $\Delta_H=\Delta_V$ ($J_2^H=J_2^V=J_2$) in all final calculations.

After introducing $\alpha=x/y$, $\gamma=z/y^{1/2}$, we represent the partition function as 
\begin{eqnarray}
	Z = e^{-\beta E^{(0)}_{\rm MD}}y^{N/2} \sum_{\cal{C}} \alpha^H \gamma^M, 
\end{eqnarray}
where $\alpha=e^{-\beta(\Delta_H-\Delta_V)}$,  $\gamma=2\cosh(\beta h/2)e^{\beta\Delta_V/2}$.

One can prove without solving the model that the free energy of the monomer-dimer system is analytic such that
this model cannot exhibit any finite-temperature phase transition \cite{Heilmann1970,Heilmann1972}.

In order to calculate the partition function, we resort to the transfer-matrix formalism \cite{Lieb1967} (see also Refs.~\cite{Grande2011,Morin2015,Wilkins2021}). The transfer matrix proceeds on two neighboring rows of vertical bonds. Vertical dimers are identified with spins up ($\uparrow$), and the absence of dimers on vertical bonds corresponds to spin down ($\downarrow$) [see Fig.~\ref{fig:TDphase}(c)]. The transfer matrix in operator form contains three contributions, ${\cal V}=V_3 V_2 V_1$, with
\begin{eqnarray}
\label{eq:V_i}
V_1 & = & \prod_{i=1}^{L_x} \sigma^x_i,
\nonumber\\
V_2 & = & \exp(\gamma\sum_{i=1}^{L_x} \sigma^{-}_i),
\nonumber\\
V_3 & = & \exp(\alpha\sum_{i=1}^{L_x} \sigma^{-}_i\sigma^{-}_{i+1}),
\end{eqnarray}
where $\sigma^x_i$ and $\sigma^{\pm}_i = \sigma^x_i \pm i\sigma^y_i$ are the corresponding Pauli matrices. 
Here, $V_1$ counts all possible configurations of closely packed vertical dimers, where $\sigma^x_i$ fixes the condition that the configurations with two spins up in the vertical direction, i.e., two neighboring dimers, are forbidden. 
The second term $V_2$ describes the creation of a monomer instead of a vertical dimer upon the action of the $\sigma^-_i$ operator. 
In a similar manner, $V_3$ corresponds to the creation of horizontal dimers by the annihilation of pairs of neighboring vertical dimers. 
It is easy to check that ${\cal V}={\cal V}^{\dagger}$ is Hermitian, since $V_1 V^{\dagger}_i = V_i V_1$ for $i=2,3$. 
For the algebraic calculations it is more convenient to consider ${\cal V}^2$,
\begin{eqnarray}
	{\cal V}^2 = V_3 V_2 V_1 V_3 V_2 V_1 = V_3 V_2 V^{\dagger}_3 V^{\dagger}_2,
	\label{V2matrix}
\end{eqnarray}
where $V^{\dagger}_2 {=} \exp(\gamma\sum_{i=1}^{L_x} \sigma^{+}_i)$, 
$V^{\dagger}_3 {=} \exp(\alpha\sum_{i=1}^{L_x} \sigma^{+}_i\sigma^{+}_{i+1})$.
The partition function can then be written as 
\begin{eqnarray}
    Z = e^{-\beta E^{(0)}_{\rm MD}} y^{N/2} {\rm Tr} {\cal V}^{L_y}.
\end{eqnarray}

An exact solution of this problem is available only for the pure dimer model ($\gamma=0$) \cite{Lieb1967},
where one can establish an equivalence to a generalized $XY$ model \cite{Suzuki1971a,Suzuki1971b}. 
Then, the transfer matrix is  of the form ${\cal V}=V_3 V^{\dagger}_3$, and  can be diagonalized within the Jordan-Wigner fermionization scheme \cite{Lieb1967}. In the isotropic case, $\alpha=1$, one obtains the entropy per unit cell of the closely-packed dimer model,  
$s_d\approx 0.29156$ \cite{Morita2016}. 

Here, we consider the EMD model where all the terms (\ref{eq:V_i}) in the transfer matrix are taken into account. Therefore, we proceed with a numerical calculation of the maximal eigenvalue $\Lambda_{\max}$ of the transfer matrix.
Using the Lanczos procedure \cite{Lanczos1950}, we are able to find $\Lambda_{\max}$ for systems with a number of unit cells in the horizontal direction $L_x$ up to 16. 
In the limit $L_y\to\infty$, only the maximal eigenvalue $\Lambda_{\max}$ remains essential compared to the  others, thus giving us the 
free energy per unit cell as follows:
\begin{eqnarray}
	f &=& -\frac{1}{N\beta}\ln Z = f_0 -\frac{1}{\beta L_x}\ln\Lambda_{\max},
	\nonumber\\
	f_0 
	&=& -\frac{3}{4} J_2^H - \frac{1}{4} J_2^V  - J_1.
\end{eqnarray}
All other thermodynamic quantities are finally obtained by numerical differentiation.

\subsection{Spin-star decoupling approximation}
\label{spinStar}
\begin{figure}[t!]
	\centering
	\includegraphics[width=\columnwidth]{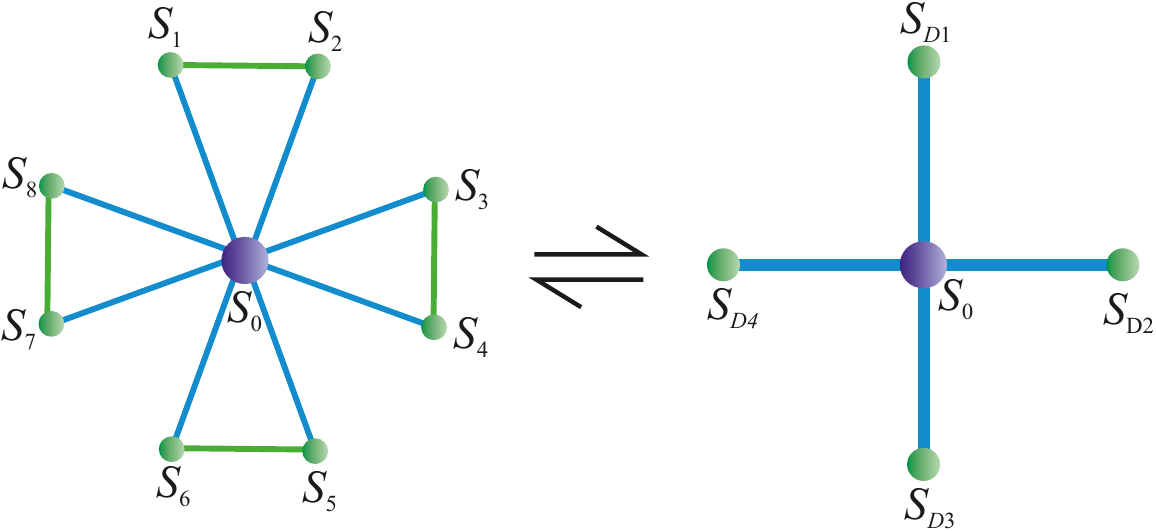}
	\caption{A spin-star fragment of the spin-1/2 Heisenberg model on the diamond-decorated square lattice (left), which can be mapped  to a simpler composite spin-star owing to local conservation of the total spin on each dimer (right).}
	\label{fig:star}
\end{figure}
This subsection introduces the spin-star decoupling approximation as a tool to describe low-energy excitations and their thermodynamic effects from the short-range perspective.
In contrast to the EMD approach of the previous subsection, here we also take triplet excitations on the dimers into account.

The spin-star decoupling approximation is developed from the idea of decomposing the total Hamiltonian of the spin-1/2 Heisenberg model on the diamond-decorated square lattice (\ref{hamos}) into a sum of non-commuting cluster Hamiltonians ${\cal H}^{0}=\sum_{j=1}^N {\cal H}_j^{0}$ ascribed to individual spin-star clusters. As a starting point of our calculation we will therefore consider the Hamiltonian ${\cal H}_j^{0}$ of a single spin-star cluster, which is composed of one monomeric spin and its four neighboring dimers. The considered spin-star cluster is schematically illustrated on the left-hand side of Fig.~\ref{fig:star} including simplified notation and is mathematically defined as follows:
\begin{eqnarray}
	{\cal H}_j^{0} &=& J_1\bm{S}_{0} \cdot \left(\bm{S}_{1}\!+\!\bm{S}_{2}\!+\!\bm{S}_{3}\!+\!\bm{S}_{4}\!+\!\bm{S}_{5}\!+\!\bm{S}_{6}\!+\!\bm{S}_{7}\!+\!\bm{S}_{8}\right)\nonumber \\
	&&+ \frac{J_2}{2}\left(\bm{S}_{1} \!\cdot\!\bm{S}_{2} \!+\!\bm{S}_{3} \!\cdot\!\bm{S}_{4}\!+\!\bm{S}_{5} \!\cdot\!\bm{S}_{6}\!+\!\bm{S}_{7} \!\cdot\!\bm{S}_{8}\right). 
\label{hamstar}
\end{eqnarray}
The inclusion of the factor of 1/2 in the second term ensures that the intradimer interaction $J_2$ will not be double-counted if one performs a summation over the spin-star Hamiltonians $H_j^{0}$ at the end of our calculation. Note furthermore that we have omitted the Zeeman term because this term can be trivially added at the end of the calculation as it only shifts the respective eigenenergies thanks to the total Hamiltonian and the Zeeman term commuting with each other. Within the spin-star cluster decoupling, the total energy of the spin-1/2 Heisenberg diamond-decorated square lattice will be obtained as the sum of eigenenergies of the spin stars given by the cluster Hamiltonian \eqref{hamstar} upon simply disregarding the non-commutative nature of the cluster Hamiltonians $[{\cal H}_i^{0},{\cal H}_j^{0}]\neq 0$ for neighbors $i$ and $j$.

 Owing to the local conservation of the total spin on each dimer, it is convenient to express the cluster Hamiltonian \eqref{hamstar} in terms of composite spin operators of four dimers denoted as $\bm{S}_{D_i} = \bm{S}_{2i-1} + \bm{S}_{2i}$ ($i=1, \ldots,4$), as well as the total spin of the dimers $\bm{S}_{D} = \sum_{i=1}^4\bm{S}_{D_i}$
 and the total spin of the spin star denoted as $\bm{S}_t=\bm{S}_0+\bm{S}_{D}$. Consequently, the eigenenergies of the spin-star cluster Hamiltonian \eqref{hamstar} can be obtained from the eigenenergies of a simpler composite spin star illustrated on the right-hand-side of Fig.~\ref{fig:star}, which depend only on the total quantum spin number ${S}_t$, the composite quantum spin numbers of the individual dimers $S_{D_i}$, and the total quantum spin number of all four dimers ${S}_D$, 
\begin{eqnarray}
E_{S_t, S_{D}, S_{D_i}}^{0} \!\! &=& \! \frac{J_1}{2} \! \left[S_t(S_t+1)-\frac{3}{4}- S_D
\left(S_D\!+1\right)\right]\nonumber\\
&&- \frac{3}{2}J_2+\frac{J_2}{4} \sum_{i=1}^4 S_{D_i} (S_{D_i} + 1).
\end{eqnarray}
We note
that each composite spin dimer  can be either in a singlet state characterized  by $S_{D_i}=0$ or a triplet state with $S_{D_i}=1$, which means that the spin star can have from zero up to four triplets.  Assuming  antiferromagnetic couplings $J_1>0$ between the central monomeric spin $S_0$ and the four neighboring spin dimers $S_{D_i}$, the energetically most favorable states of the spin star in each sector correspond to the highest
multiplicity of the dimer spins $S_D^{\rm max} = \sum_i S_{D_i}$, given by the eigenenergies $E_{S_t, S_{D}^{\rm max}}^{0}$ with
\begin{eqnarray}
\label{enehvizdy}
E_{1/2,0}^{0} &=& -\frac{3}{2}J_2,\nonumber\\
E_{1/2,1}^{0} &=& -J_1-J_2,\nonumber\\
E_{3/2,2}^{0} &=& -\frac{3}{2}J_1-\frac{J_2}{2},\nonumber\\
E_{5/2,3}^{0} &=& -2J_1,\nonumber\\
E_{7/2,4}^{0} &=& -\frac{5}{2}J_1+\frac{J_2}{2}.
\end{eqnarray}
The total number of triplets within this set of eigenstates of the spin star thus coincides with the sum of the composite quantum spin numbers $S_{D}$. 

The total energy of the spin-1/2 Heisenberg model on the diamond-decorated square lattice can be then determined by summing the energies of individual spin stars, whereby the states of the dimers can be effectively described within a quasi-particle formalism when assigning the occupation number $n_i=1$ to a dimer-triplet state and $n_i=0$ to a dimer-singlet state, respectively. By considering only the five lowest-energy eigenstates of the spin star, ranging from zero up to four triplet-dimer states as given by Eq.~(\ref{enehvizdy}), the eigenenergies can be expressed in terms of the occupation numbers of the dimer-triplet states,
\begin{eqnarray*}
\!&&\!E_{n_1,\ldots, n_4}^{0} =-\frac{3}{8}J_1-\frac{3}{2}J_2+\frac{J_2}{2}\sum_{i=1}^4n_i+\frac{J_1}{2}\prod_{i=1}^4(1-n_i)
\\
&&\!+\!\frac{J_1}{2} \! \left[\left(\sum_{i=1}^4 \! n_i\!-\!\frac{1}{2}\right)\!\!\left(\sum_{i=1}^4 \! n_i\!+\!\frac{1}{2}\right)
\!-\!\sum_{i=1}^4 \! n_i \! \left(\sum_{i=1}^4 \! n_i\!+\!1\right)\right]\!,
\end{eqnarray*}
which can be further simplified into the following compact form:
\begin{eqnarray}
E_{n_1, \ldots, n_4}^{0}\!=\!&-&\frac{1}{2}(J_1\!+\!3J_2)+\frac{1}{2}(J_2\!-\!J_1)\sum_{i=1}^4 \! n_i \nonumber \\
 &+&\frac{J_1}{2}\prod_{i=1}^4(1\!-\!n_i).
\end{eqnarray}
The last term represents the correction to the energy of free monomeric spins surrounded by four adjacent dimer singlets. 
The eigenenergy $E_{n_1, \ldots, n_4}^{0}$ corresponds to the spin multiplet with the total quantum spin number
\begin{eqnarray}
S_t= \sum_{i=1}^4 n_i+ \prod_{i=1}^4 \! (1-n_{i})-\frac{1}{2}.
\end{eqnarray}
Summing the eigenenergies of the spin stars and adding the respective Zeeman term yields the following formula for the overall energy of the spin-1/2 Heisenberg model on the diamond-decorated square lattice
$E_{T}(\{n_i\}) = E_{T}^{0}(\{n_i\})-h S_T^z$, where $E_{T}^{0}(\{n_i\})$ represents the respective zero-field energy depending on the set of all occupation numbers $\{ n_i\}$,
\begin{figure}[t!]
	\centering
	\includegraphics[width=0.5\columnwidth]{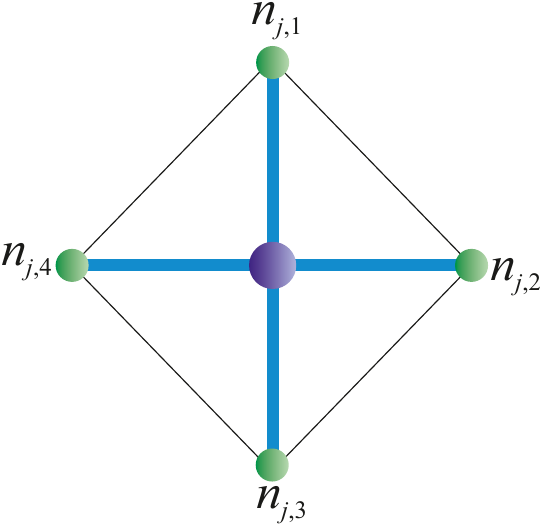}
	\caption{The diagram illustrates a central monomeric spin (depicted as a large purple sphere) surrounded by four occupation numbers $n_{j,i}$ (depicted as smaller green spheres) arranged in a square plaquette. The occupation numbers $n_{j,i}$ ($i=1-4$)  correspond to composite spin states of four dimers adjacent to the $j$th monomeric spin.}
	\label{fig:square}
\end{figure}
\begin{eqnarray}
\label{en_h_total}
E_{T}^{0}(\{n_i\})&=&-\frac{N}{2}(J_1+3J_2)+(J_2-J_1)\sum_{i=1}^{2N}n_i\nonumber\\
&&+\!\left(\frac{J_1}{2}\right)\!\sum_{j=1}^N \prod_{i\in\rsquare} \! (1-n_{j,i}).
\end{eqnarray}
The last term in Eq.~(\ref{en_h_total}) provides a correction to the energy when all four dimers surrounding the $j$th monomeric spin are in a singlet state. This situation is depicted in Fig.~\ref{fig:square}, where the central monomeric spin is surrounded by four occupation numbers assigned to  composite spins of adjacent dimers. If all four composite spins of the dimers are in the singlet state, the correction term accounts for this specific configuration, ensuring that the energy is accurately represented under this specific condition. In Eq.~(\ref{en_h_total}), the notation $i\in\rsquare$ indicates that the index $i$ runs over the four sites of a square plaquette surrounding the central monomeric spin, as depicted in Fig.~\ref{fig:square}. Finally, $S_T^z$ is the $z$ component of the total spin $S_T$ of the whole system, which can be expressed in terms of the occupation numbers of the dimer-triplet states
\begin{eqnarray}
S_T= -\frac{N}{2} +\sum_{i=1}^{2N}n_i+\sum_{j=1}^N \prod_{i\in\rsquare} \! (1-n_{j,i}).
\label{totalSpinik}
\end{eqnarray}
The first term accounts for the contribution of the monomeric spins, while the second term accounts for the contribution of the dimeric spins.  Hence, the second term involves a summation over the occupation numbers assigned to the dimer-triplet states and thus corresponds to the total number of triplets $N_{\rm trip.}$. The summation in the third term is a correction stemming from all free
monomeric spins. In fact, the third term involves a projection operator for the monomeric spin surrounded by four dimer-singlet
states and thus corresponds to the total number of free monomeric spins $N_{\rm free}$.

The free energy of the spin-1/2 Heisenberg model on the diamond-decorated square lattice within the spin-star decoupling approximation can then be calculated according to the formula
\begin{eqnarray}
	f_{\rm star} = -\frac{1}{N\beta}\ln \sum_{\{n_i\}}\sum_{S_T^z=-S_T}^{S_T} \exp\left[-\beta E_{T}^0(\{n_i\})+\beta h S_T^z\right]. \nonumber \\
	\label{freen}
\end{eqnarray}
We emphasize that the formula (\ref{totalSpinik}) for the total quantum spin number $S_T$ tacitly assumes that all free monomeric spins are fully polarized by the magnetic field. Therefore, in the zero-field limit  the formula (\ref{freen}) for the free energy does not take into account  the degeneracy from the free monomeric spins (i.e., monomeric spins surrounded by four singlet dimers), which have a significant role in determining the zero-point entropy and low-temperature thermodynamics in  the absence of a magnetic field.

To improve the zero-field estimate of the free energy within the spin-star decoupling approximation, we therefore modified the formula (\ref{freen}) for the free energy by accounting for the respective degeneracy factor of the free monomeric spins
\begin{eqnarray}
	f_{\rm star}^0 = -\frac{1}{N\beta}\ln \sum_{\{n_i\}}\sum_{S_T^z=-S_T^0}^{S_T^0}2^{N_{\rm free}} \exp\left[-\beta E_{T}^0(\{n_i\})\right], \nonumber \\
	\label{freen2}
\end{eqnarray}
where $N_{\rm free}=\sum_{j=1}^N \prod_{i\in\rsquare} \! (1-n_{j,i})$ denotes the total number of the free monomeric spins and the overall spin multiplicity $S_T^0$ in zero field can be expressed as   
\begin{eqnarray}
S_T^0=\frac{N_{\rm free}-N}{2} + \sum_{i=1}^{2N}n_i.
\end{eqnarray}

The spin-star decoupling approximation recovers the transition fields around the DT  phase in the ground-state phase diagram [see red broken lines in Fig.~\ref{fig:gspd}(b)]. By setting all $n_i=0$  in Eq.~(\ref{en_h_total}), we actually get the exact value for the ground-state energy of the MD  phase (\ref{eMD}). Requiring that there is a single triplet dimer in each spin-star configuration, we instead obtain the exact energy of the DT phase (\ref{eDT}).
Calculating the ground-state energy of the LM phase (all $n_i=1$) within the spin-star approach, $E_{\rm LM}=-\frac{N}{2}(5J_1 - J_2 + 3h)$, yields the transition field between the LM and DT phases as $h=J_2 - J_1$, which  is very close to the DMRG results shown in Fig.~\ref{fig:gspd}(b) [cf.\ the red-broken and blue-solid lines in Fig.~\ref{fig:gspd}(b)].  
Recall that on the phase boundary the energy of any configuration that contains at least one triplet in the spin-star cluster has the same energy. This  leads to an additional macroscopic degeneracy (higher than the degeneracy of the DT phase). This artifact of the decoupling approximation influences the low-temperature behavior of the entropy discussed in Sec.~\ref{results}.

The decoupling approximation based on the spin-star clusters was developed under two fundamental assumptions, (i) ignoring the non-commutative nature of the cluster Hamiltonians (\ref{hamstar}) and (ii) subsequently considering only the five energetically most favorable states of the spin star given by Eq.~(\ref{enehvizdy}). Despite these simplifications, the spin-star cluster decoupling  offers a reasonable approximation at sufficiently low temperatures whenever the system is driven towards the fragmented MD  or DT phases, whereas the collective nature of the LM ferrimagnetic phase is  
not rigorously accounted for within this treatment. Although we have attempted to improve precision of the spin-star approximation in zero magnetic field by considering degeneracies pertinent to free monomeric spins, this calculation procedure should be regarded as much more reliable and precise in
the presence of a magnetic field because of Zeeman splitting of the relevant spin multiplets. It indeed turns out that the quantitative agreement between the results derived from the spin-star approximation and a numerical treatment of the full model progressively extends to higher temperatures as the magnetic field increases,
see Sec.~\ref{results}.

\subsection{Exact diagonalization}
\label{EDecko}

\begin{figure}[t!]
	\centering
	\includegraphics[width=0.9\columnwidth]{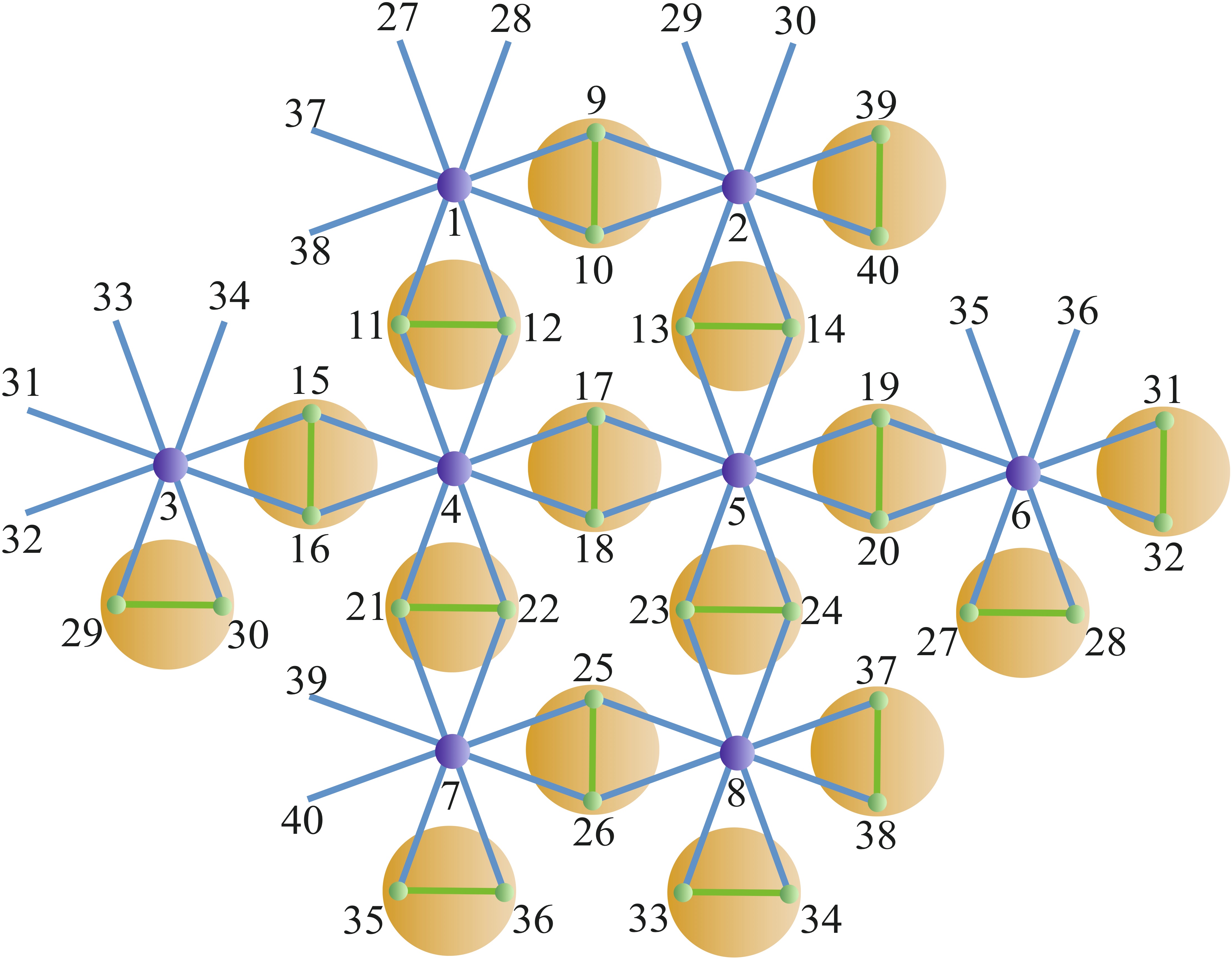}
	\caption{The $N_s = 40$ system. Periodic boundary conditions are indicated by repeated spin labels.
        Yellow spheres in the background encode the composite-spin representation on the underlying Lieb lattice.}
	\label{fig:s40}
\end{figure}
Here, we provide some details on the exact diagonalization (ED) that we used to obtain numerically precise results for small system sizes. This method allows us to calculate a broad range of thermodynamic quantities, including magnetization, specific heat, and entropy, serving as a benchmark for validating analytical and approximate models.

The full $N_s=30$ spectra were obtained previously \cite{Caci2023} and can be re-evaluated for the present purposes.
However, here we push exact diagonalization (ED) further to the $N_s=40$ system sketched in
Fig.~\ref{fig:s40}. We have followed the same strategy as in Ref.~\cite{Caci2023}, exploited conservation of the total spin
on each dimer and thus passed to an effective Lieb lattice, represented by the yellow spheres in the background of Fig.~\ref{fig:s40}.
For the present $N_s=40$ system, computer enumeration then yields 191 topologically nonequivalent arrangements of spin triplets and singlets
on the dimers with degeneracies ranging from 1 (e.g., for all dimers in the triplet state) to 1152. 
We were then able to compute the full spectra in the sectors with up
to $N_{\rm trip.}=8$ triplets. This includes in particular the sector with $N_{\rm trip.}=4$ triplets that contains the dimer-tetramer phase for the $N_s=40$ geometry.
For the sectors with nine triplets or more, we further diagonalized high-$S_z$ sectors completely and obtained low-energy levels
using the Lanczos procedure \cite{Lanczos1950} combined with the strategy of Ref.~\cite{HW09}. Specifically, in the Lieb-Mattis regime that corresponds to 16 triplets in the $N_s=40$ system,
we were only able to obtain full spectra for $S^z \ge 13$ and had to contend ourselves with low-lying levels for $S^z \le 12$,
 i.e., the ferrimagnetic state with magnetization $3/5$ and below.
Furthermore, we reconstructed the spin multiplets using $S^z$ classification and spin inversion symmetry in the $S^z=0$ sector;
the latter permits us to avoid diagonalizations in the $S^z=1$ sector, which would usually have the highest dimension.
All in all, we were able to recover an entropy of $36.082\,\ln(2)$,  i.e., 90.2\%\ of the total entropy of the $N_s=40$ system.

We note that going to the $N_s=40$ system is particularly important in the vicinity of the Lieb-Mattis phase.
Specifically, the lowest state in the sector with $N_{\rm trip.}>N_s/10$ triplets turns out to be in the sector with
spin $S=N_{\rm trip.}-N_s/10$ and approaches the Lieb-Mattis limit $S=3N_s/10$ for $N_{\rm trip.}=2N_s/5$.
At $J_2=0$, we find a ground-state energy of $ -2.46115\,J_1$ per unit cell in the $S=12$, $N_{\rm trip.}=16$ sector for $N_s =40$.
The finite-size error is about an order of magnitude smaller than the $N_s=30$ estimate of $-2.46809\,J_1$ as compared to the estimates
$-2.46083\,J_1$ \cite{Hirose2018} and $-2.46000\,J_1$ per unit cell \cite{Caci2023}, respectively. This indicates that the
$N_s=40$ system can be considered significantly closer to the thermodynamic limit than the $N_s=30$ one, even if we increased
the number of spins only by a third. This large change can be attributed to the $N_s=30$ system still having additional symmetries that arise from the periodic
boundary conditions imposed on this small system whereas such particular symmetries are absent in the $N_s=40$ geometry,
rendering the latter geometry significantly more generic.
As another indicator, we mention that the ground-state entropy per site is $s\approx0.0880$ and $0.0795$ for $N_s=30$ and $40$, respectively,
compared to $s\approx0.0583$ in the thermodynamic limit \cite{Fisher61,Kasteleyn61,Morita2016} (the corresponding ground-state degeneracy is 14 and 24, respectively).

\begin{figure}[t!]
	\centering
	\includegraphics[width=0.45\textwidth]{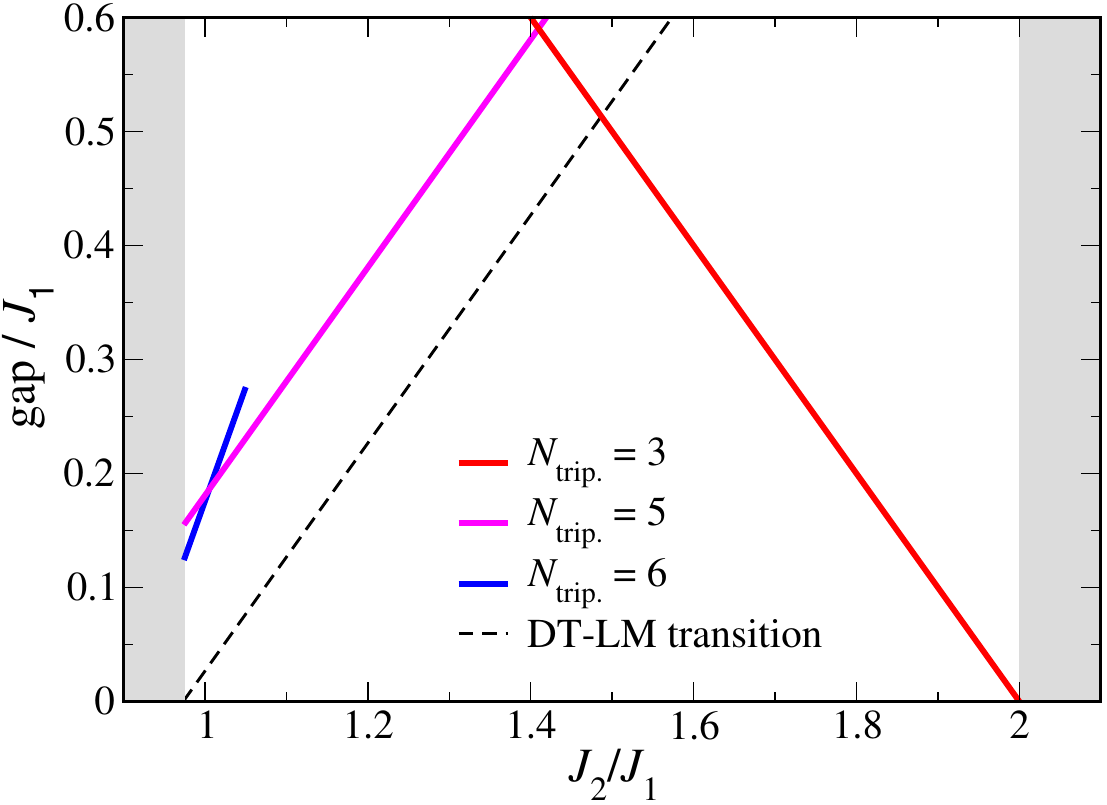}\qquad
	\caption{Lowest excitations above the  DT ground state for $N_s=40$ with  $N_{\rm trip.}=4$.
        The LM phase (on the left) and the MD phase (on the right) are shaded in gray.
        The DMRG result for the phase boundary between the DT and the LM phase from Fig.~\ref{fig:gspd} is indicated by the dashed line.}
	\label{fig:TD-gapN40}
\end{figure}

Figure~\ref{fig:TD-gapN40} shows the gaps to the lowest excitation above the DT ground state for the $N_s=40$ system ($N_{\rm trip.}=4$).
For $J_2 \gtrsim 1.5 J_1$,
the gap to the $N_{\rm trip.}=3$ excitation coincides with the DT-MD phase transition in the phase diagram Fig.~\ref{fig:gspd}, as in fact
in this case states with any number $N_{\rm trip.}\le N_s/10$ collapse.
On the other hand, the field-induced DT-LM transition in Fig.~\ref{fig:gspd} is a true first-order transition. Thus, the closing of the $N_{\rm trip.}=5$ gap in Fig.~\ref{fig:TD-gapN40} is preempted by a direct transition to a $N_{\rm trip.}=16$ configuration, reproduced from Fig.~\ref{fig:gspd} by the dashed line in
Fig.~\ref{fig:TD-gapN40}. As a consequence, the maximum of the gap is found for $J_2/J_1 \approx 1.4$
whereas the maximal extension of the DT phase in a magnetic field occurs for $J_2/J_1 \approx 1.5$.
Furthermore, one observes an $N_{\rm trip.}=6$  bound state coming down in energy when approaching the DT-LM transition at $h=0$. The nature of the DT-LM transition at $h=0$ is less clear to decide, at least based on the $N_s=40$ ED data,  i.e., one does observe higher-$N_{\rm trip.}$/higher-$S$ states coming down in energy, but it is not clear if they actually collapse in the thermodynamic limit, or if the DT-LM transition remains a conventional first-order transition at $h=0$.

\subsection{Quantum Monte Carlo}
In addition to ED, we can also use QMC in order to access thermodynamic properties of the spin-1/2 Heisenberg antiferromagnet on the diamond-decorated square lattice. In particular, this approach allows us to obtain unbiased results for system sizes that  extend beyond those accessible to ED. The QMC method that we used for our investigation has already been employed in our previous finite-field study of the spin-1/2 Heisenberg antiferromagnet on the diamond-decorated square lattice \cite{Caci2023} and therefore we summarize it only briefly.
 
More specifically, our QMC approach is based on the stochastic series expansion (SSE)  
method with directed loop updates \cite{Sandvik1991,Sandvik1999,Syljuasen2002}. In order to avoid 
the  sign problem, i.e., an exponential drop of the statistical accuracy of the QMC simulations at low temperatures and large system sizes~\cite{Henelius2000,Troyer2005,Hangleiter2020,Hen2021} because of the presence of geometric frustration, we avoid working in the conventional local spin-$S^z$ basis for the model at hand. 
 Instead, we eliminate the sign problem upon using  appropriate basis states after decomposing the Hamiltonian into separate terms based on dimers (or trimers)~\cite{Nakamura1998,Alet16,Honecker16,Weber2022}. 
For certain so-called fully frustrated models
the sign problem can indeed be completely eliminated, cf.\ Refs.~\cite{Alet16,Honecker16,NgYang17,Stapmanns2018,Fan2024} for the spin-dimer and Ref.~\cite{Weber2022} for the spin-trimer basis, respectively. 
 
In case of the diamond-decorated square lattice a finite value of the  coupling $J_2$ leads to geometric frustration. 
In order to avoid the associated sign problem, we follow Ref.~\cite{Caci2023} and  treat all $J_2$-dimer spins in the spin-dimer basis, but keep using the local $S^z$ basis for the monomer spins. Using this combined five-site cluster basis, we can simulate our model without a sign problem within the SSE framework, based on the abstract operator loop update introduced in Ref.~\cite{Weber2022}. We refer to Ref.~\cite{Weber2022} for further details on this QMC approach. 
Here, we performed QMC simulations for systems with 
up to $6\times6$ unit cells.


\section{Results}
\label{results}

\subsection{Dimer triplet density, entropy, and specific heat above the DT phase}

\begin{figure}[t!]
	\centering
	\includegraphics[width=0.8\columnwidth]{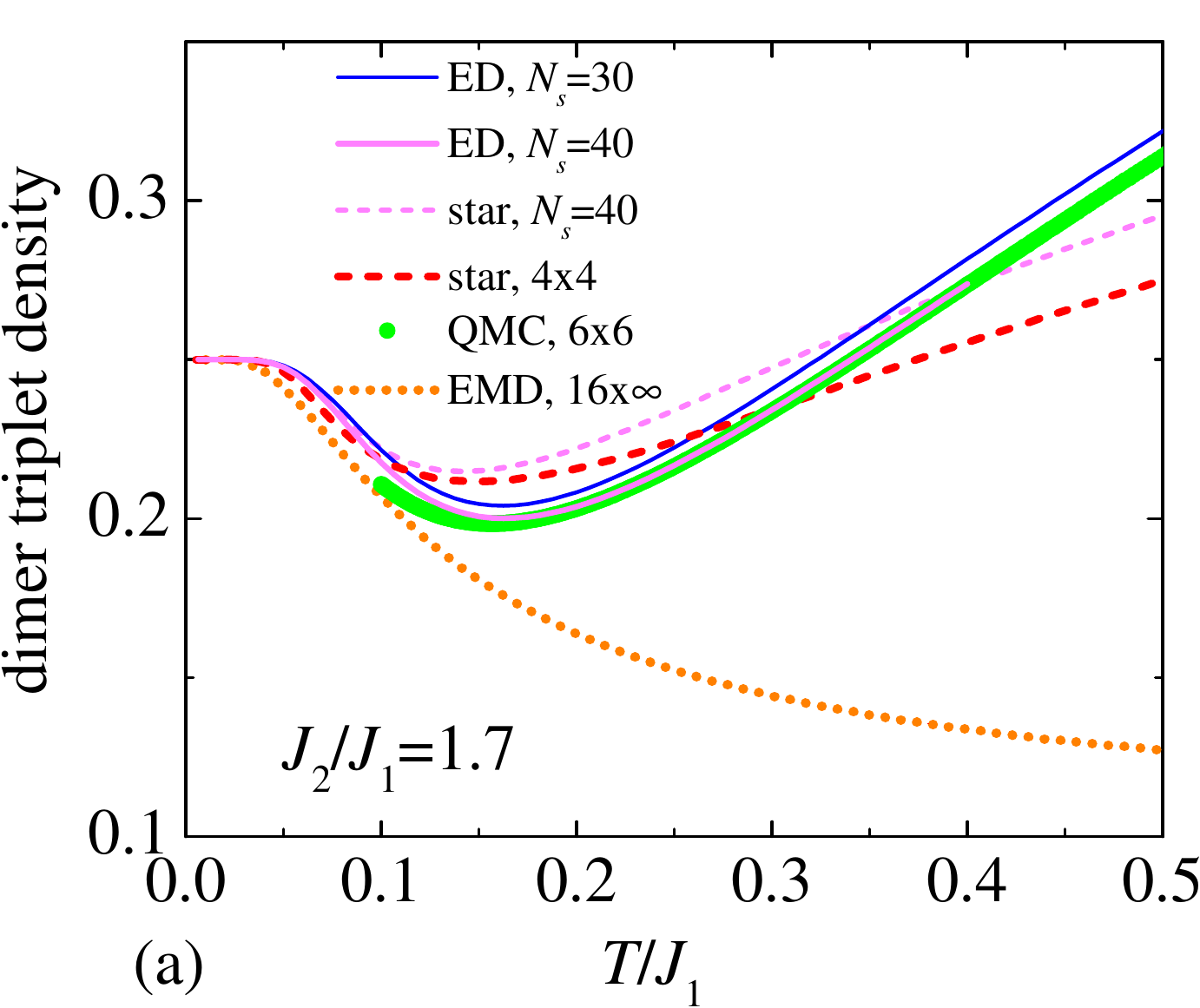}
	\includegraphics[width=0.8\columnwidth]{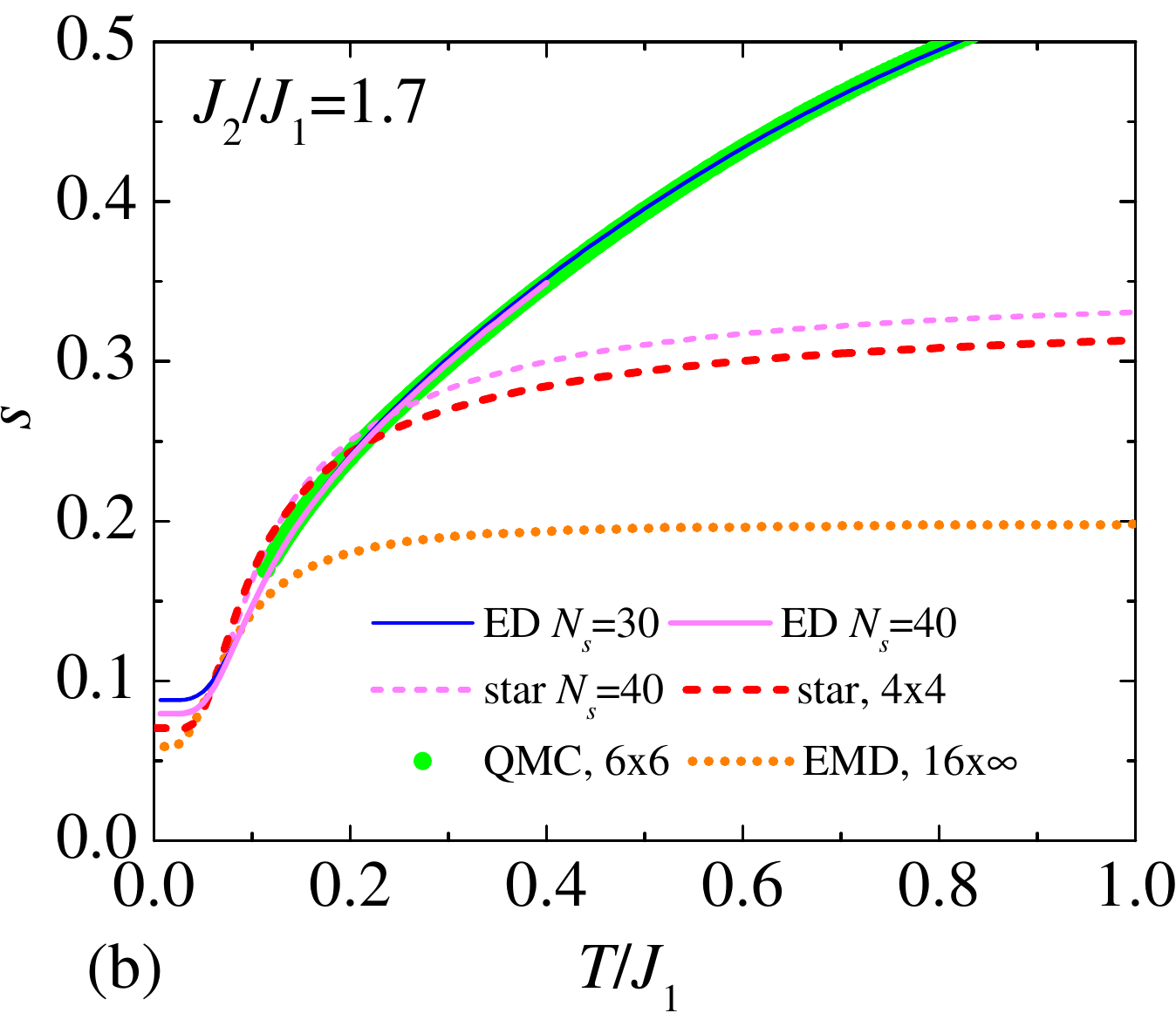}
	\includegraphics[width=0.8\columnwidth]{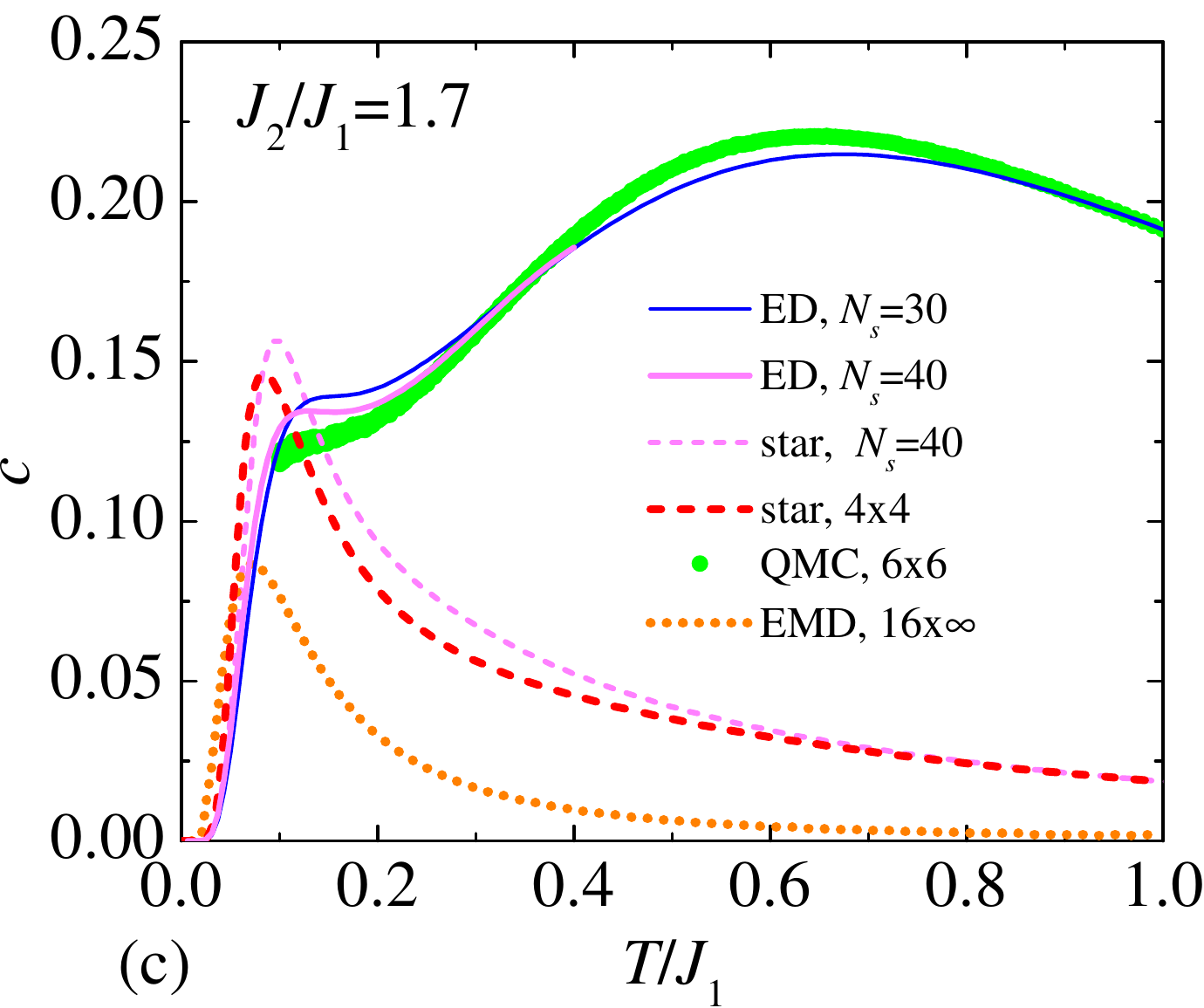}
	\caption{(a) Dimer triplet density; (b) entropy; (c) specific heat of the spin-1/2 Heisenberg model on the diamond-decorated square lattice for the interaction ratio $J_2/J_1=1.7$ as a function of temperature at zero magnetic field.}
	\label{fig:1.7}
\end{figure}
Let us start our discussion by examining the dimer triplet density, entropy, and specific heat of the spin-1/2 Heisenberg model on the diamond-decorated square lattice in zero magnetic field. We will compare results from the different methods
introduced in Sec.~\ref{method}. For the spin-star decoupling approximation we will use the zero-field formula Eq.~(\ref{freen2}) for the free energy.

Figure~\ref{fig:1.7} presents results  for the interaction ratio $J_2/J_1=1.7$. 
At zero temperature, the dimer triplet density is 0.25, reflecting the nature of the DT state, where $1/4$ of the dimers are in the dimer-triplet state within the singlet tetramers and $3/4$ of the dimers remain in the dimer-singlet state. Because of the relatively lower energy cost of creating a dimer-singlet state compared to an additional dimer-triplet state for the interaction ratio $J_2/J_1=1.7$, a decrease in the dimer-triplet density can be observed at 
low temperatures terminated at a round local minimum. With further increase of temperature, the dimer-triplet density tends towards the asymptotic value $0.75$ reached in the limit of infinite temperature owing to the threefold degeneracy of the triplet state in zero magnetic field. 

A comparison of the dimer-triplet density obtained for $N_s=30$ and $N_s=40$ spins from exact diagonalization (ED) reveals 
small
finite-size effects. ED for $N_s=40$ spins provides an interpolation between the spin-star decoupling approximation and the EMD model at low temperatures and QMC at high temperatures.
The excitation from a dimer-singlet state to a dimer-triplet state is neglected within the EMD model and the dimer-triplet density should accordingly always decrease with increasing temperature. On the other hand, the local minimum observed in the dimer-triplet density arises from the competition of the lower excitation energy for the dimer-singlet state and the higher degeneracy of the triplet state.
In fact, the lowest states not taken into account by the EMD model are at energies slightly above
$0.6J_1$ for $J_2=1.7J_1$ and these are responsible for the triplet density increasing again with increasing temperature after going through a minimum. However, the spin-star decoupling approximation takes only some of these into account [as reflected also by the lower entropy $s$ compared to ED and QMC, see Fig.~\ref{fig:1.7}(b)] and approximates their energies, thus explaining the deviation from $N_s=40$ ED and QMC, which are in good agreement for $T/J_1 \gtrsim 0.2$.

Figure~\ref{fig:1.7}(b) shows
the entropy per spin $s$
as a function of temperature at zero magnetic field for the interaction ratio $J_2/J_1=1.7$. The zero-temperature entropy of the spin-1/2 Heisenberg model on the diamond-decorated square lattice in the parameter regime of the DT phase is given by $s=\frac{1}{N_S}\ln \Omega_{N_s}$, where $\Omega_{N_s}$ represents the degeneracy of the DT phase for the system size with $N_s$ spins; for instance,
$\Omega_{30}=14$, $\Omega_{40}=24$, as already stated in the ED context in Sec.~\ref{EDecko}.
Figure~\ref{fig:1.7}(b) shows that the results obtained from the spin-star decoupling approximation for $N_s=40$ are in perfect agreement with ED data for the same system size at low enough temperature $T/J_1 \lesssim 0.1$. Similar accuracy can be expected at low temperatures for the EMD model. Notably, the residual entropy of the DT phase decreases with increasing system size, but it still remains nonzero even in the thermodynamic limit $N_s \to \infty$. From the EMD model with system size $16\times\infty$, one can for instance infer a residual entropy  $s\approx0.0587$ that quite closely converges to the known value for the hard-dimer model on the square lattice \cite{Fisher61,Kasteleyn61,Morita2016}. 

Figure~\ref{fig:1.7}(c) displays
the specific heat per spin $c$ as a function of temperature for the same value of the interaction ratio.
A notable feature is the presence of a low-temperature shoulder, superimposed on the high-temperature maximum of the specific heat, as seen clearly in ED and QMC results. As anticipated, the spin-star and EMD approaches fail to capture the behavior of the specific heat at higher temperatures as they neglect high-energy excitations above the DT phase.

On the other hand, these approaches accurately account for low-lying excitations which leads to a close agreement between the ED data for $N_s=40$ and the analytical results obtained from the spin-star decoupling approximation at low enough temperatures. A comparison of the peak height in the specific heat obtained from ED and QMC reveals relatively large finite-size effects at low temperatures. The EMD model provides further insight into the behavior of the specific heat at low temperatures, which effectively bridges the gap between the available QMC data ($T/J_1>0.1$) and extends it further towards lower temperatures. Figure~\ref{fig:1.7}(c) shows that thermal activation of the specific heat indeed
occurs at progressively lower temperatures as the system size increases such that the EMD model provides the best approximation to the thermodynamic limit at low temperatures.

\begin{figure}[t!]
	\centering
	\includegraphics[width=0.8\columnwidth]{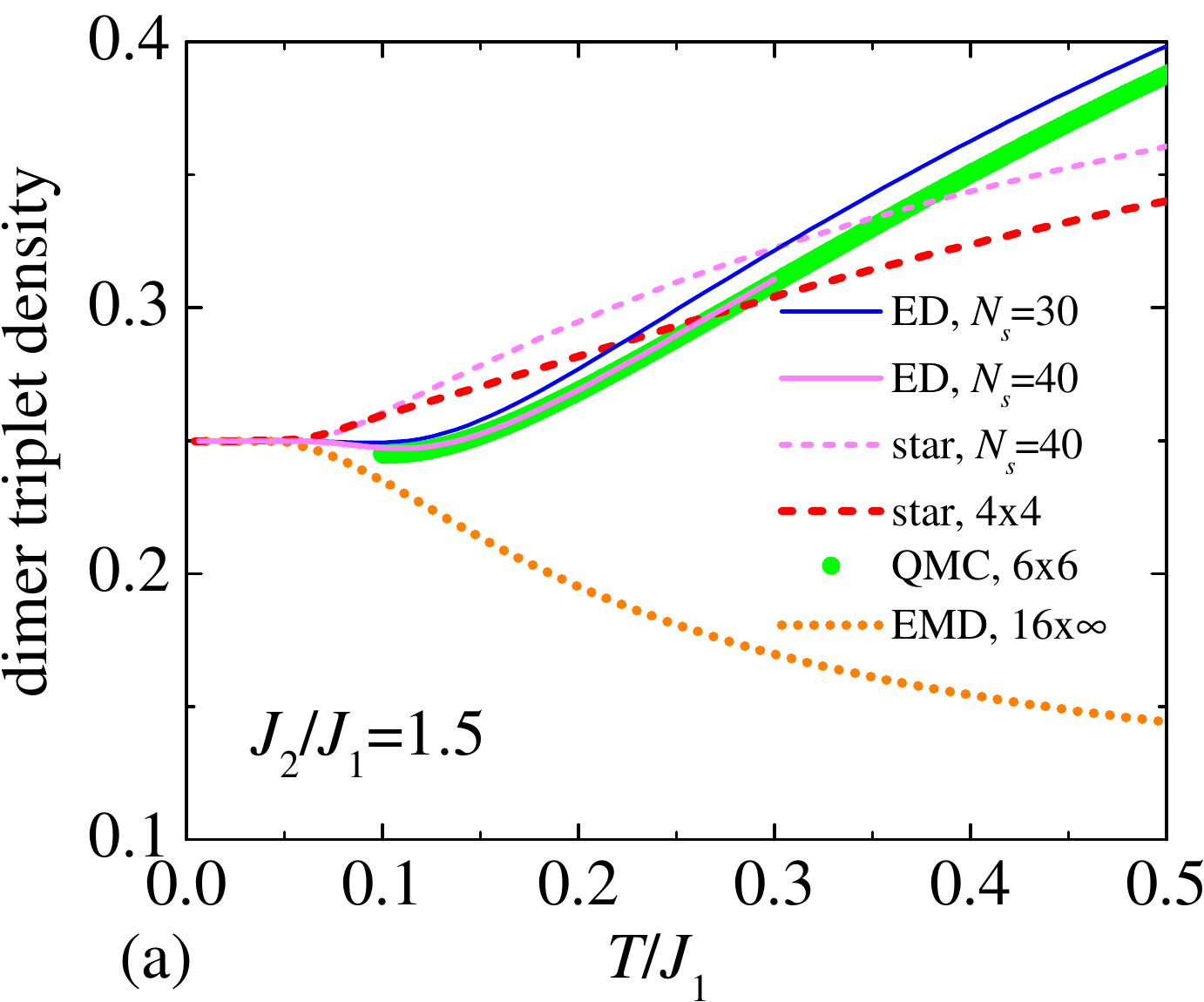}
	\includegraphics[width=0.8\columnwidth]{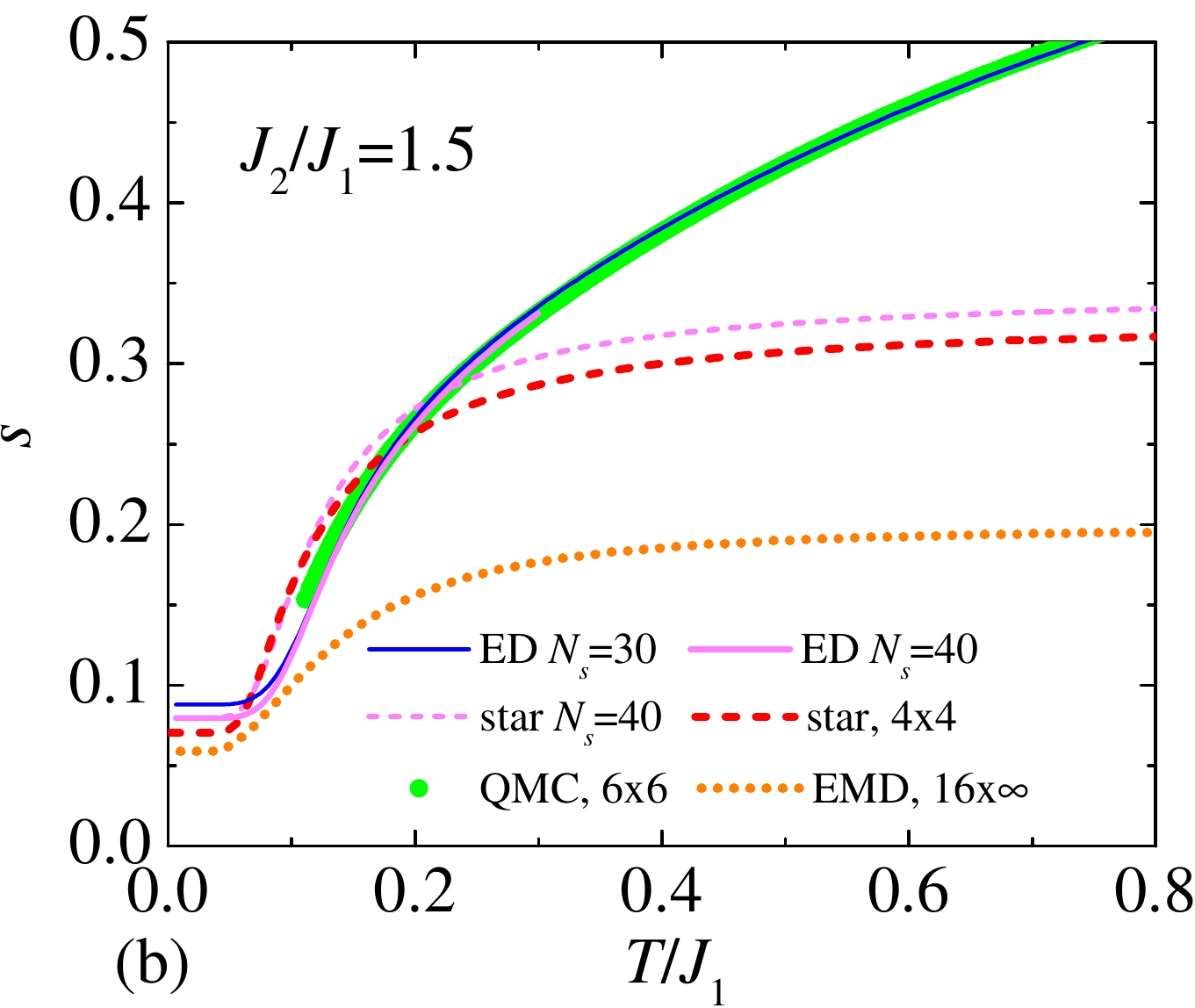}
	\includegraphics[width=0.8\columnwidth]{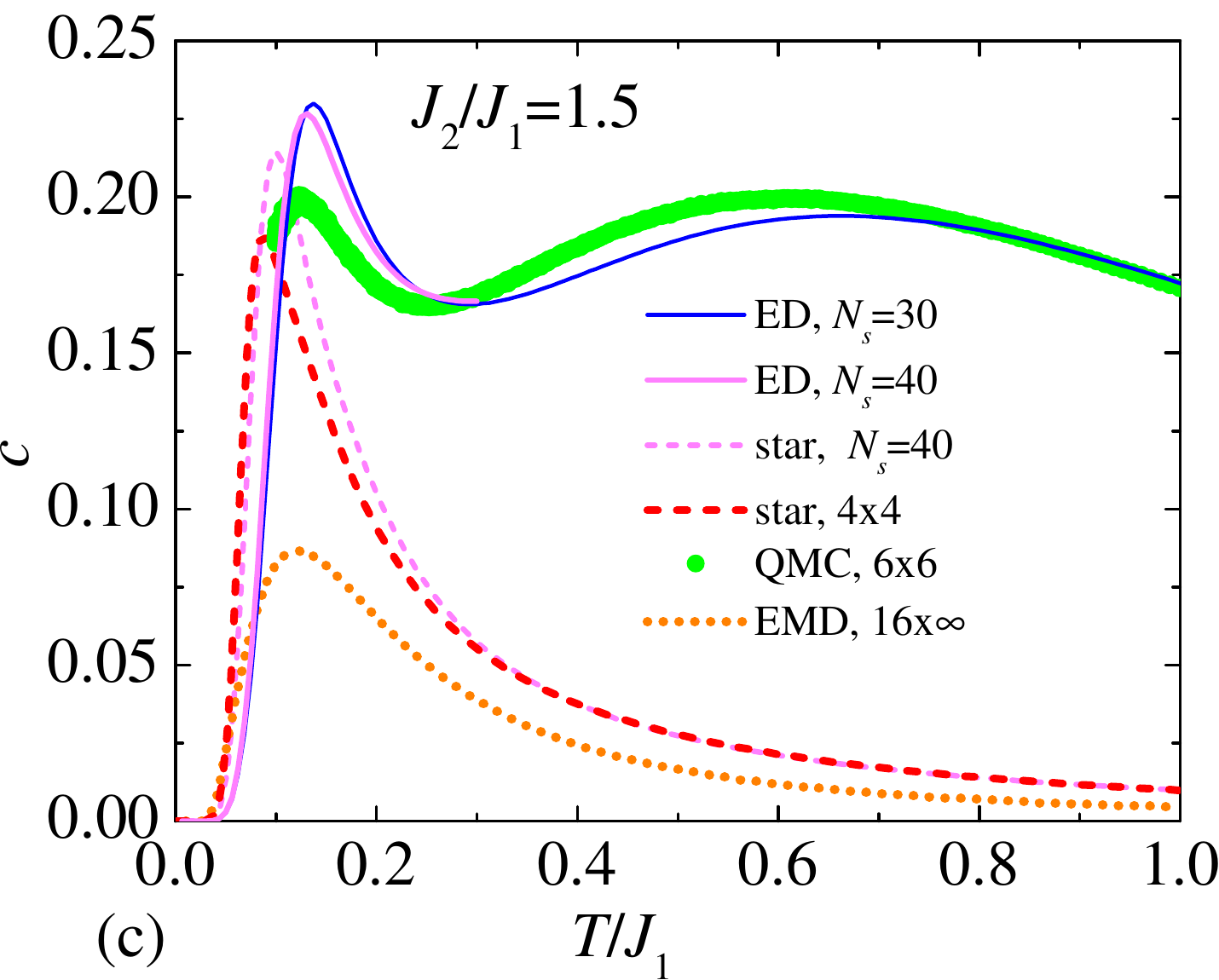}
	\caption{(a) Dimer triplet density; (b) entropy; (c) specific heat 
for the interaction ratio $J_2/J_1=1.5$ as a function of temperature at zero magnetic field.}
	\label{fig:1.5}
\end{figure}

We now turn to a discussion of the dimer-triplet density, entropy, and specific heat in zero magnetic field for the case  $J_2/J_1=1.5$  plotted in Fig.~\ref{fig:1.5}. 
Here the numerical data for the dimer-triplet density
exhibits only a shallow minimum, which is consistent with the gap for one triplet less being only slightly smaller than that for an additional triplet at  $J_2/J_1=1.5$, compare Fig.~\ref{fig:TD-gapN40}.
By contrast, the spin-star decoupling approximation fails to account for this minimum. The reason is that, within the spin-star decoupling approximation, the excitation energies from the DT phase to singlets and triplets are equal for this interaction ratio.
Consequently, the spin-star approach cannot explain the thermal population into singlet states for the interaction ratio $J_2/J_1\le1.5$. Comparing the results obtained from QMC simulations for a system size of $6 \times 6$ (total of 180 spins) with those from ED for 40 spins reveals nearly perfect agreement, indicating negligible finite-size effects. 
However, the discrepancy between ED results for 30 spins and 40 spins at intermediate temperatures ($T/J_1>0.1$) may stem from artificial symmetries imposed for the system size of 30 spins by considering periodic boundary conditions, which are absent for the system size of 40 spins.

The entropy and specific heat
for the interaction ratio $J_2/J_1=1.5$ in zero magnetic field are shown in Figs.~\ref{fig:1.5}(b) and \ref{fig:1.5}(c), respectively, and exhibit qualitatively similar behavior to that for the interaction ratio $J_2/J_1=1.7$. In the high-temperature regime, small finite-size effects are observed.
While the high-temperature peak of the specific heat shifts slightly towards higher values with increasing system size, the low-temperature peak is conversely suppressed towards lower values.
The results obtained from the spin-star decoupling approximation for 40 spins are slightly overestimated compared to ED data for $N_s=40$  at  low temperatures $T/J_1 \lesssim 0.1$. The discrepancy at higher temperatures arises because of the presence of low-energy excitations to states that share the same energy as the Lieb-Mattis ferrimagnetic state at zero magnetic field (compare the discussion in Sec.~\ref{EDecko}), which is not accurately accounted for by the spin-star decoupling approximation simply ignoring the noncommutativity of the spin-star Hamiltonians. The QMC data are consistent with the intriguing double-peak temperature dependence of the zero-field specific heat with low- and high-temperature peaks located around $T/J_1\approx 0.12$ and 0.6, respectively.

\begin{figure}[t!]
	\centering
	\includegraphics[width=0.8\columnwidth]{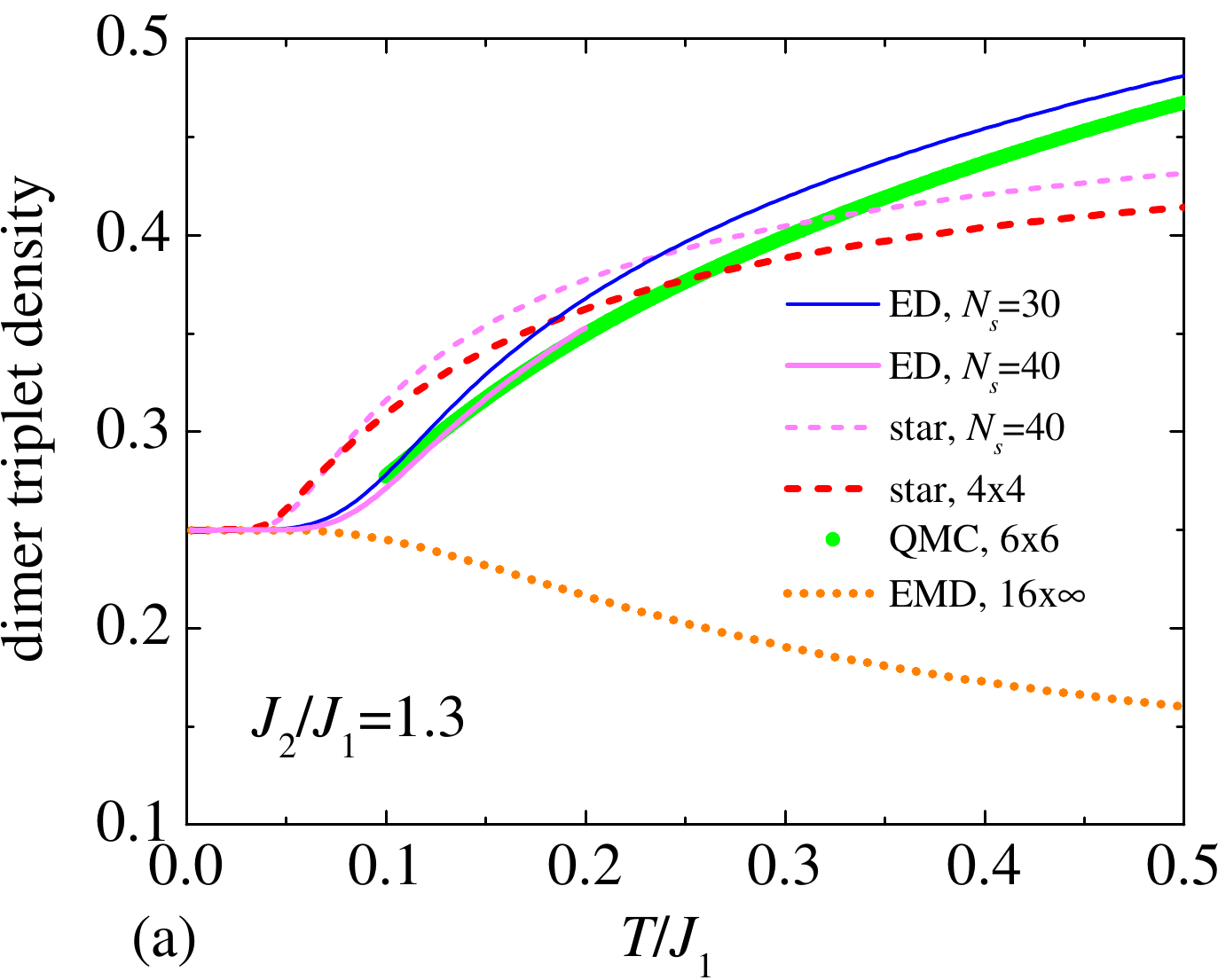}
	\includegraphics[width=0.8\columnwidth]{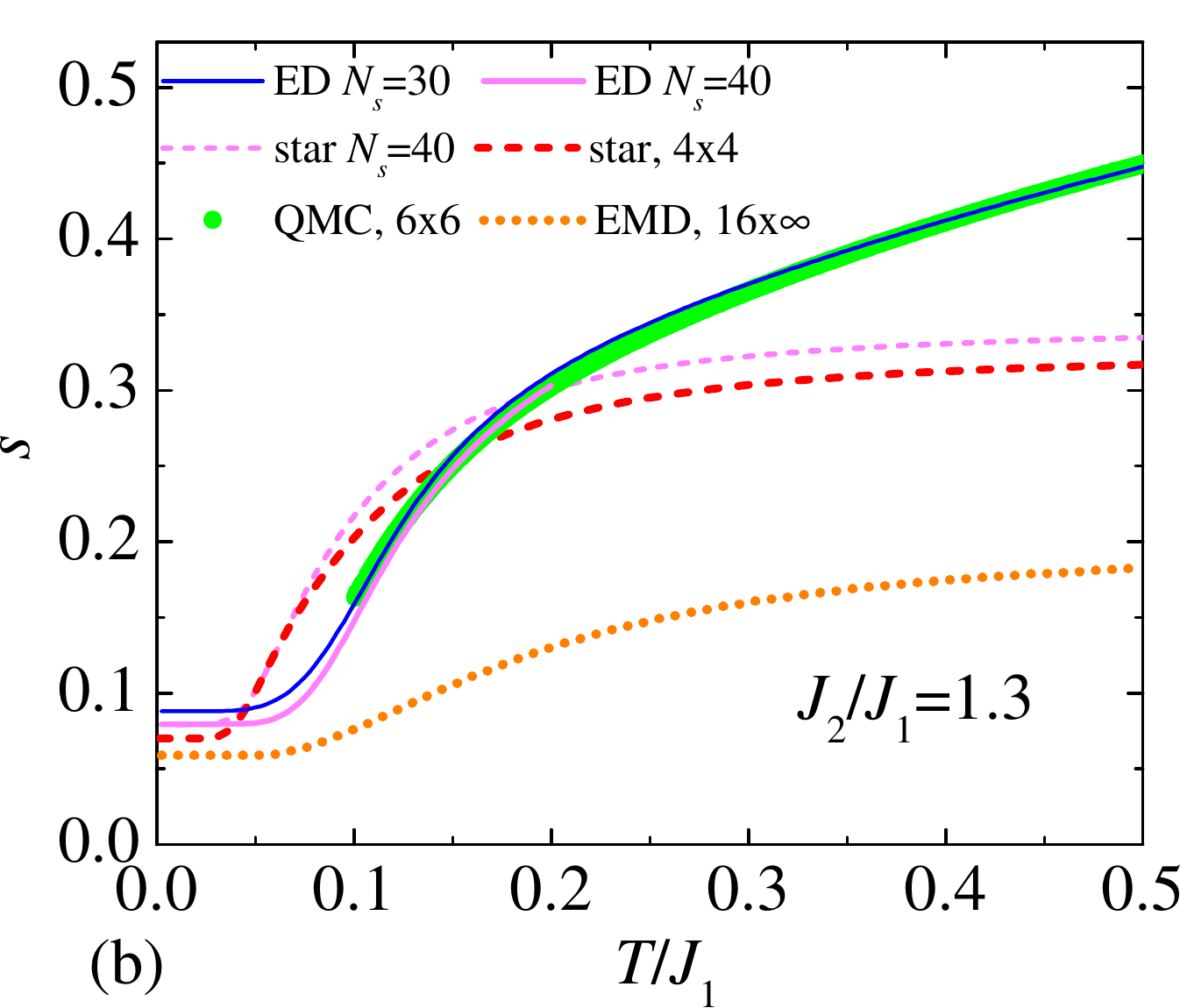}
	\includegraphics[width=0.8\columnwidth]{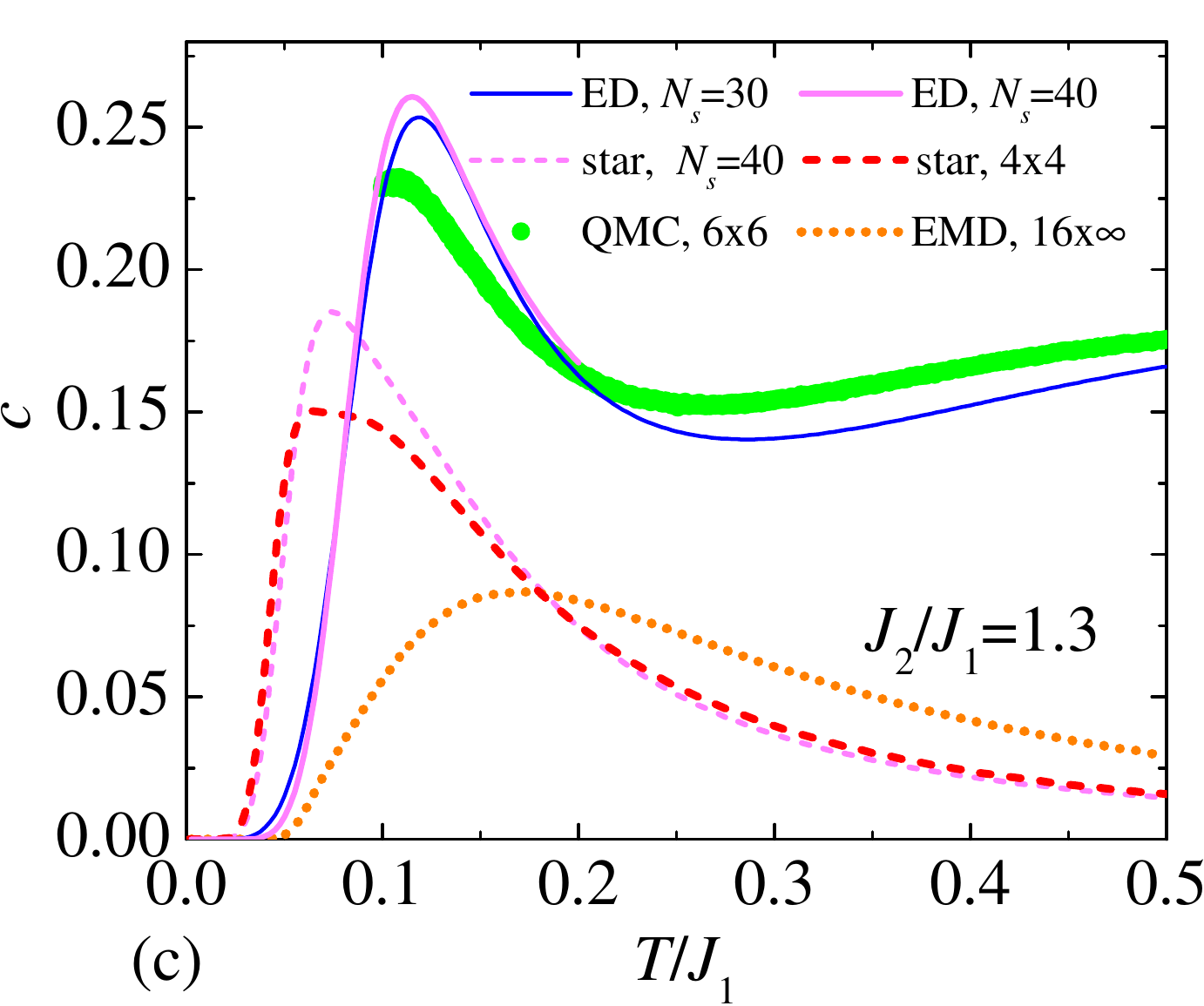}
	\caption{(a) Dimer triplet density; (b) entropy; (c) specific heat 
for the interaction ratio $J_2/J_1=1.3$ as a function of temperature at zero magnetic field.}
	\label{fig:1.3}
\end{figure}
With a yet lower value of the interaction ratio $J_2/J_1$, the analytical approaches provided by the spin-star and EMD models neglect more and more important low-energy excitations. For the interaction ratio $J_2/J_1$=1.3, both of these approaches only provide an accurate description
at very low temperatures as evidenced by the temperature dependencies of the dimer-triplet density, entropy, and specific heat depicted in Fig.~\ref{fig:1.3}. This limitation arises from the failure of these analytical approaches  to accurately capture the
degeneracy of the LM phase in zero magnetic field with possible values of the $z$ component of the total spin $S_T^z = -S_T, -S_T+1, ..., S_T$, and the low-energy excitations to these states are not properly accounted for either.
The dimer-triplet density
monotonically increases without any local minimum for the interaction ratio $J_2/J_1$=1.3, as evidenced by the ED data for $N_s=40$ spins. This behavior is caused by the
lowest excitation in this parameter regime arising from an additional dimer-triplet,
compare Fig.~\ref{fig:TD-gapN40}.
The EMD approach neglects the lowest excitations and thus only gets the $T=0$ limit right, but fails to capture the leading low-temperature asymptotics. By contrast, the star approach does account for these excitations, even if only in an approximate fashion, and thus captures at least qualitatively the low-temperature behavior of the triplet density.

While the high-temperature peak of the specific heat shown in Fig.~\ref{fig:1.3}(c) exhibits a slight variation with increasing system size, the behavior of the low-temperature peak is less straightforward. For a system size of $N_s=30$, the peak is lower than for size $N_s=40$, but the specific heat obtained from QMC for a system size of $6\times6$ unit cells ($N_s=180$) shows the lowest peak. This indicates a more complex dependency on system size.

\subsection{Magnetization, entropy, and specific heat under magnetic field}

We proceed to a discussion of the most intriguing findings
under finite magnetic fields.
Now we use the finite-field formula Eq.~(\ref{freen}) for the free energy
within the spin-star decoupling approximation.

\begin{figure}[t!]
	\centering
	\includegraphics[width=0.8\columnwidth]{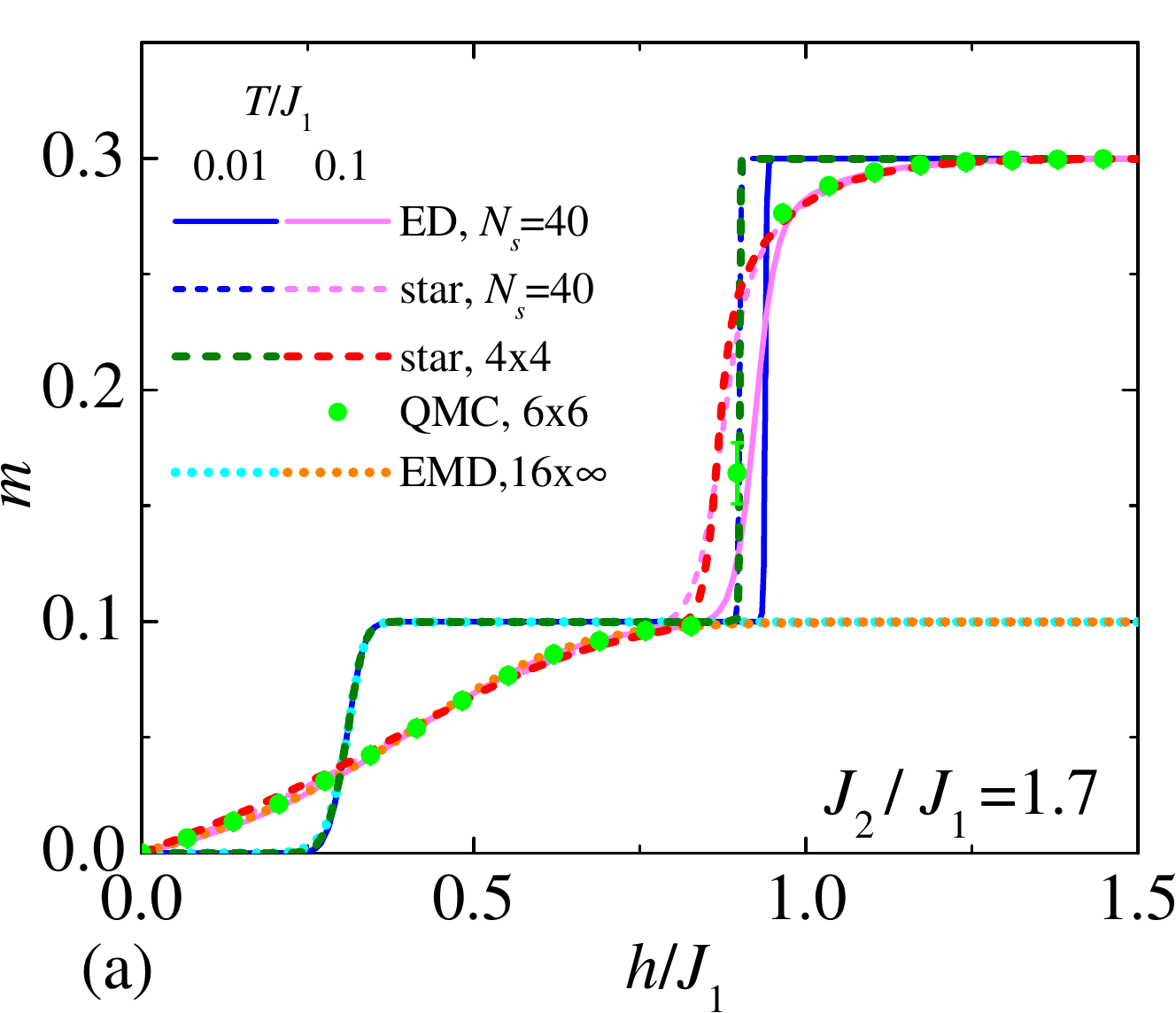}
	\includegraphics[width=0.8\columnwidth]{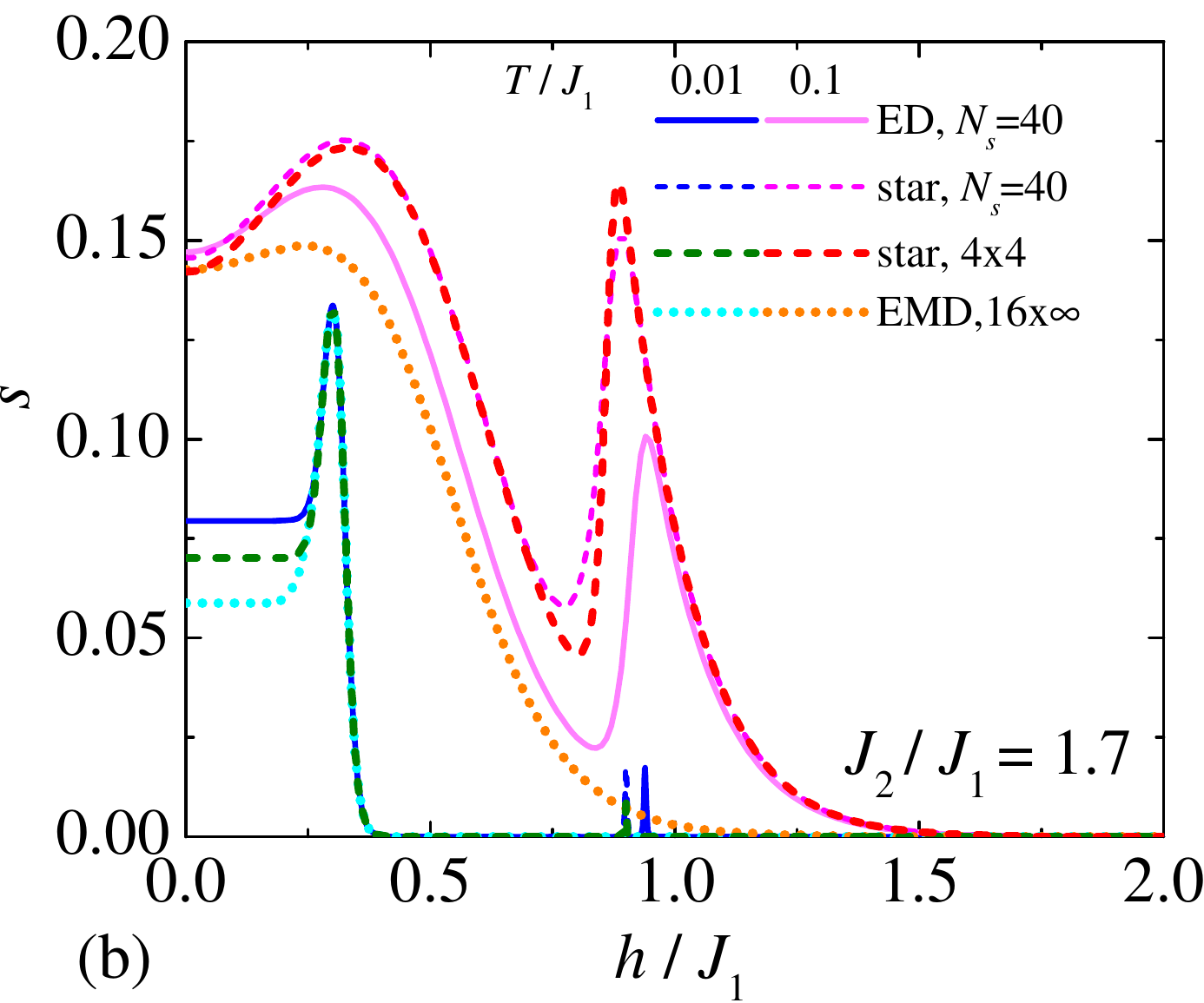}
	\includegraphics[width=0.8\columnwidth]{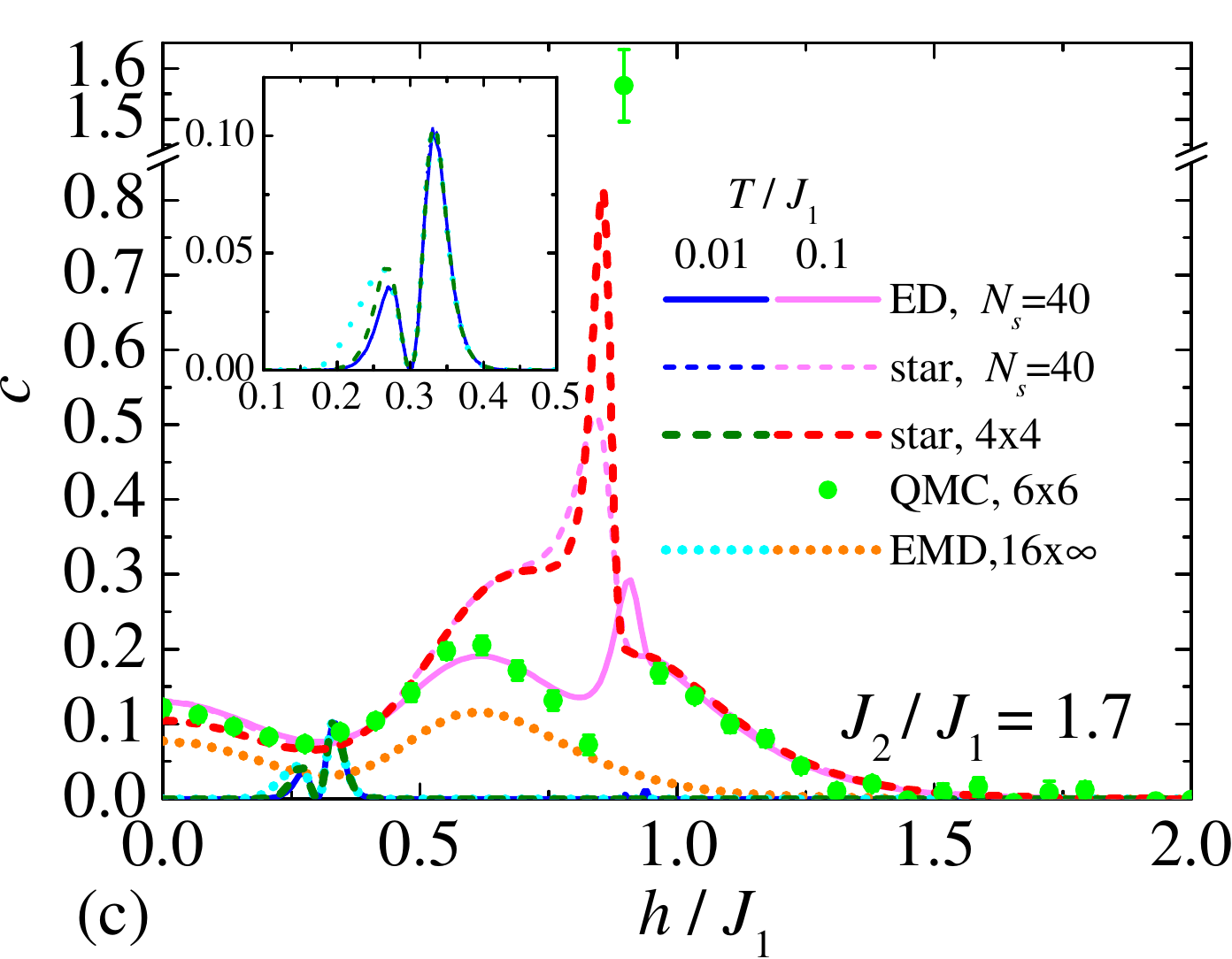}
	\caption{(a) Magnetization; (b) entropy; (c) specific heat of the spin-1/2 Heisenberg model on the diamond-decorated square lattice for the interaction ratio $J_2/J_1=1.7$ as a function of magnetic field at two different values of temperature $T/J_1=0.01$ and $T/J_1=0.1$.
The inset of panel (c) zooms into the transition region between the DT and MD phases
for $T/J_1=0.01$.
}
	\label{figh:1.7}
\end{figure}

The field dependencies of magnetization, entropy, and specific heat are illustrated for the interaction ratio $J_2/J_1=1.7$ in Fig.~\ref{figh:1.7} for two different temperatures $T/J_1=0.01$ and 0.1.
The magnetization curve at the interaction ratio $J_2/J_1=1.7$ exhibits magnetization plateaux at $m=0$ and 1/5 (full magnetization corresponding to 1/2) with the characteristics of the DT and MD phases, respectively. These can be almost perfectly described by the analytical spin-star or EMD approaches including the most important low-energy excitations above these ground states.
However, another transition from the 1/5-plateau to the 3/5-plateau corresponding to the Lieb-Mattis ferrimagnetic phase cannot be described at all by the EMD model and only qualitatively by the spin-star decoupling approximation. This finding relates to the fact that the ground-state energy of the Lieb-Mattis phase is determined within the spin-star decoupling approximation with a relative error of approximately 4.3\% (the ground-state energy of the LM phase within DMRG and spin-star decoupling methods in zero field are $E_{\rm LM}^{\rm DMRG}=-2.46J_1+2J_2$  \cite{Caci2023}
and $E_{\rm LM}^{\rm STAR} =-2.50J_1+2J_2$, respectively, compare also the related discussion in
Sec.~\ref{EDecko}).

It can be observed from Fig.~\ref{figh:1.7}(b) that the entropy starts from its nonzero residual value, which is progressively converging with increasing system size to the residual entropy of the dimer model on the square lattice per unit cell $s_d\approx 0.29156$ (per spin $\approx 0.058312$) \cite{Fisher61,Kasteleyn61,Morita2016}. Contrary to this, one detects a sizable entropy gain reaching the specific value $s=0.13256$
at the transition field between the DT and MD phases, which is almost independent of the system size as corroborated by ED, the EMD model, as well as the spin-star decoupling approximation. Similar qualitative agreement between results of all three aforementioned methods is observed in the peculiar asymmetric double-peak structure of the specific heat, which can be detected in the vicinity of the magnetic-field-driven phase transition between DT and MD phases [see the inset of Fig.~\ref{figh:1.7}(c)]. The lower peak height observed below the relevant transition field can be attributed to the macroscopic degeneracy of the DT ground state, which is not lifted by the magnetic field in contrast to the MD phase  that becomes non-degenerate at any nonzero magnetic field.

The results obtained from the spin-star decoupling approximation are in very good agreement with the ED data even for magnetic fields higher than  $h/J_1\gtrsim 1.0$, up to the end of the Lieb-Mattis ferrimagnetic phase, and even at moderately high temperatures $T/J_1=0.1$ still constitute a good approximation of the field dependence of the entropy and specific heat, as depicted in Figs.~\ref{figh:1.7}(b) and \ref{figh:1.7}(c).

\begin{figure}[t!]
	\centering
	\includegraphics[width=0.8\columnwidth]{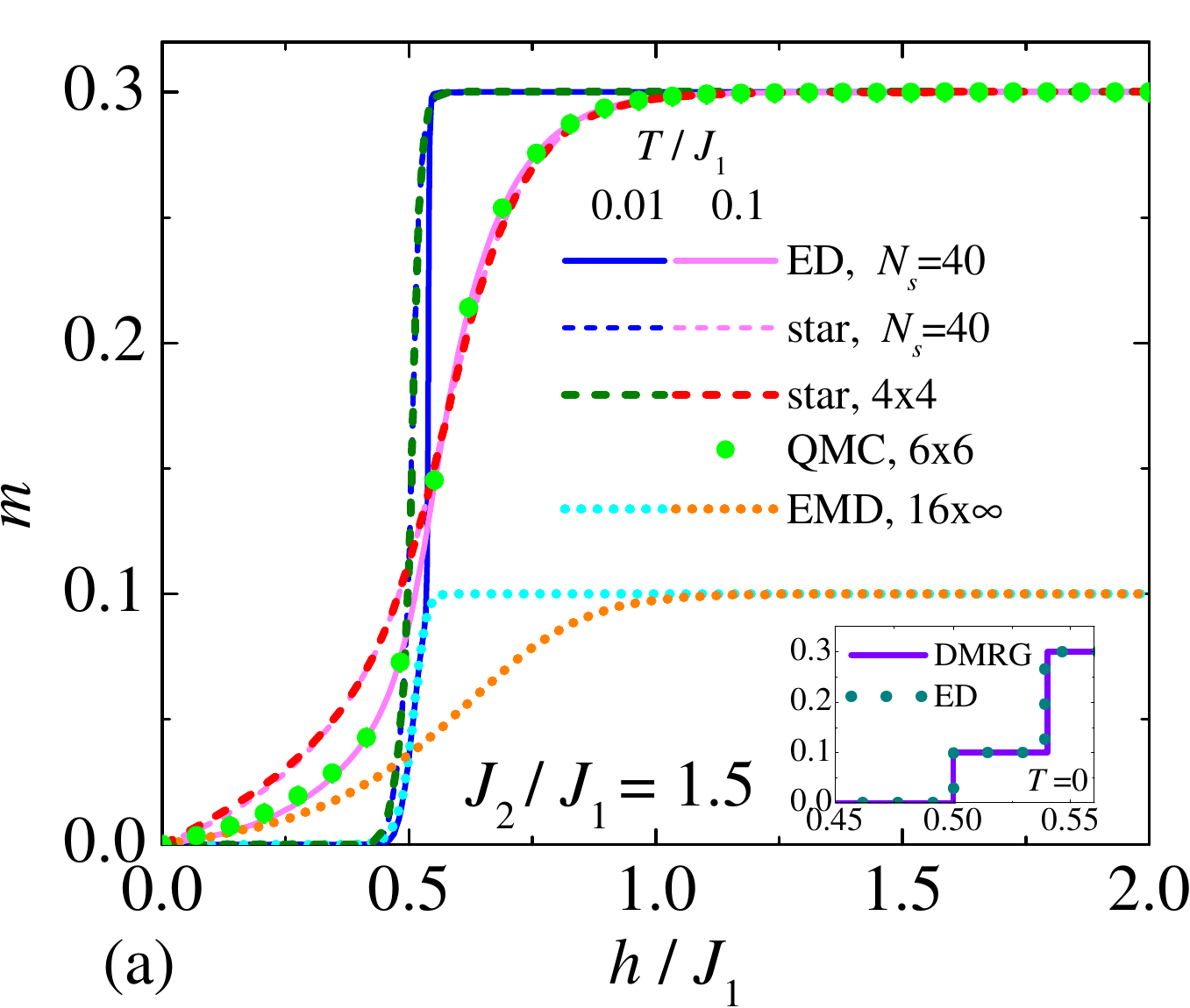}
	\includegraphics[width=0.8\columnwidth]{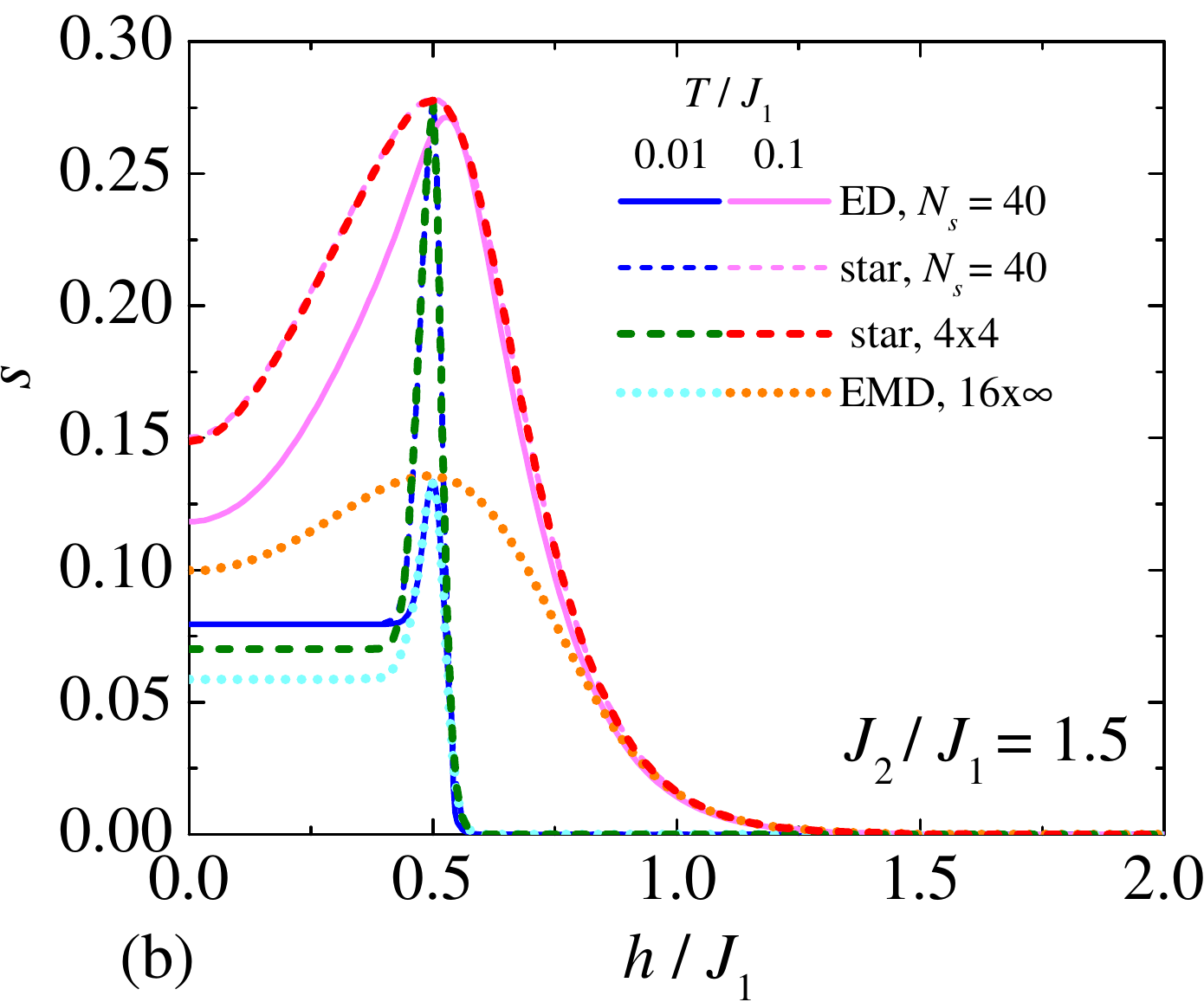}
	\includegraphics[width=0.8\columnwidth]{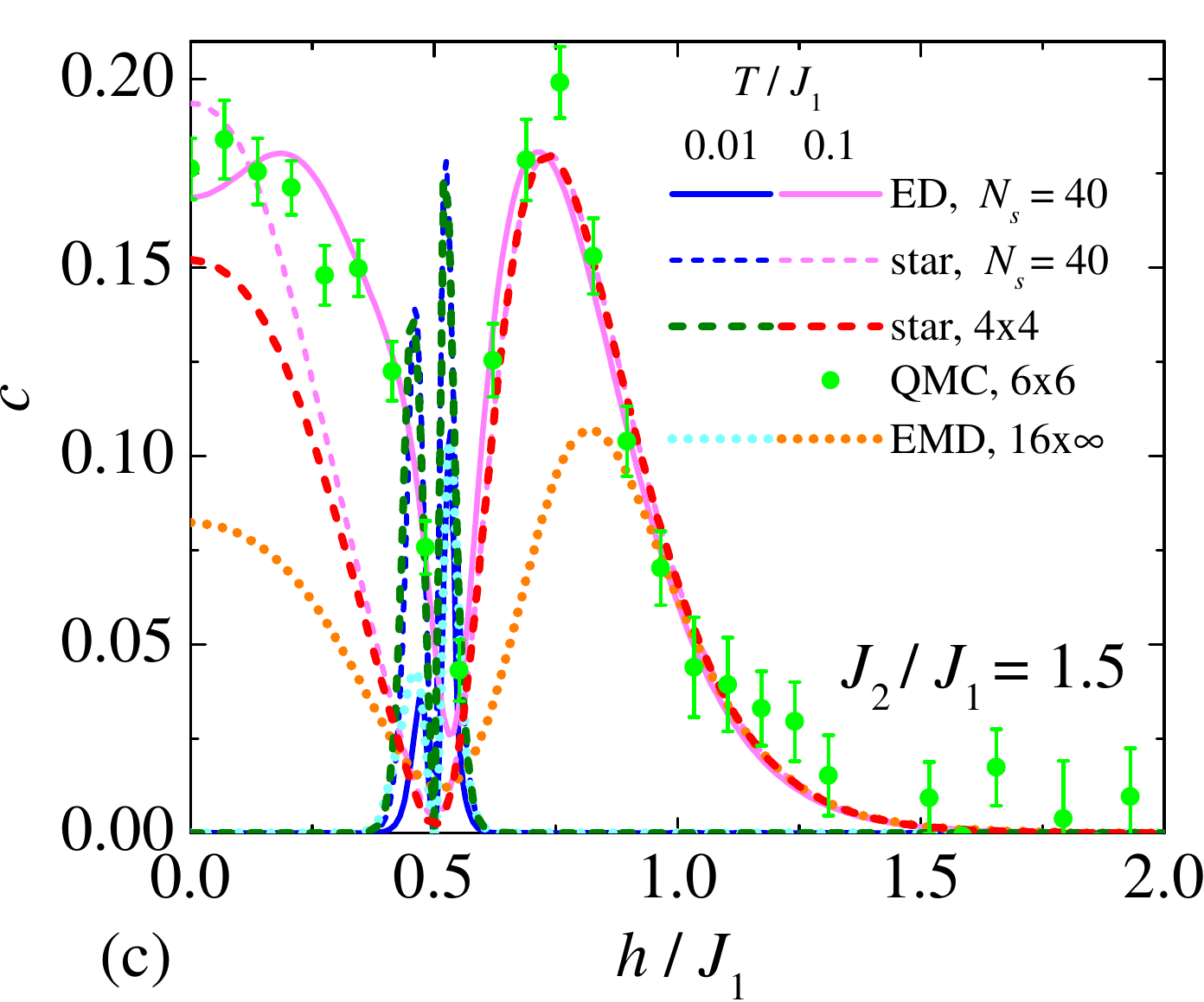}
	\caption{(a) Magnetization; (b) entropy; (c) specific heat
for the interaction ratio $J_2/J_1=1.5$ as a function of magnetic field at two different values of temperature $T/J_1=0.01$ and $T/J_1=0.1$.
The inset in (a) shows $T=0$ ED and DMRG results.}
	\label{figh:1.5}
\end{figure}

Figure~\ref{figh:1.5}(a) shows  the magnetization curves
for the interaction ratio $J_2/J_1=1.5$.
There is a distinct magnetization jump from the DT phase to the MD phase at zero temperature, as shown in the inset of Fig.~\ref{figh:1.5}(a). However, the 1/5-plateau is sufficiently small to be smeared out at finite temperatures. Consequently, there is a sharp rise in magnetization from the 0-plateau to the 3/5-plateau even at temperatures as low as $T/J_1=0.01$, which becomes smoother with increasing temperature. Owing to the presence of the MD phase within a narrow range of magnetic field strengths, the EMD model is in excellent agreement with the ED data up to a magnetic field of approximately $h/J_1\approx 0.54$ at a very low temperature $T/J_1=0.01$, although it fails to account for further increases in magnetization with increasing magnetic-field strength. On the other hand, the spin-star decoupling approximation is still capable of qualitatively reproducing the ED data even at higher magnetic fields. The comparison of ED data for $N_s=40$ spins with QMC data for 36 unit cells (i.e., $N_s=180$ spins) indicates negligible finite-size effect in this parameter region.

The results for the low-temperature entropy exhibit a relevant discrepancy between the ED and spin-star decoupling approximation only in the close vicinity of the transition field $h/J_1=0.5$ [see Fig.~\ref{figh:1.5}(b)]. The excessive macroscopic degeneracy predicted by the spin-star decoupling approximation relates to a triple phase coexistence point of the LM, DT, and MD phases, which is located exactly at the magnetic field $h/J_1=0.5$ for $J_2/J_1=1.5$.
Recall that the triple point obtained within the spin-star approach is slightly shifted towards a lower value of the magnetic field with respect to its true coordinates [see Fig.~\ref{fig:gspd}(b)]. At a higher temperature $T/J_1=0.1$, the discrepancy between ED and the spin-star decoupling approximation is significantly reduced, because all relevant low-lying excited states are thermally activated. Contrary to this, we see the opposite effect for the EMD model with much lower entropy, because high-energy excitations are missing within this description. A similar effect can also be observed in the field dependence of the specific heat [
Fig.~\ref{figh:1.5}(c)]. While the EMD model reproduces the ED data at low temperatures, the spin-star decoupling approximation correctly describes the double peak of the specific heat also at a higher temperature of $T/J_1=0.1$. At this higher
temperature, the double-peak structure of the specific heat is also confirmed by the QMC simulations
for the bigger system of $6\times6$ unit cells.

\begin{figure}[t!]
	\centering
	\includegraphics[width=0.99\columnwidth]{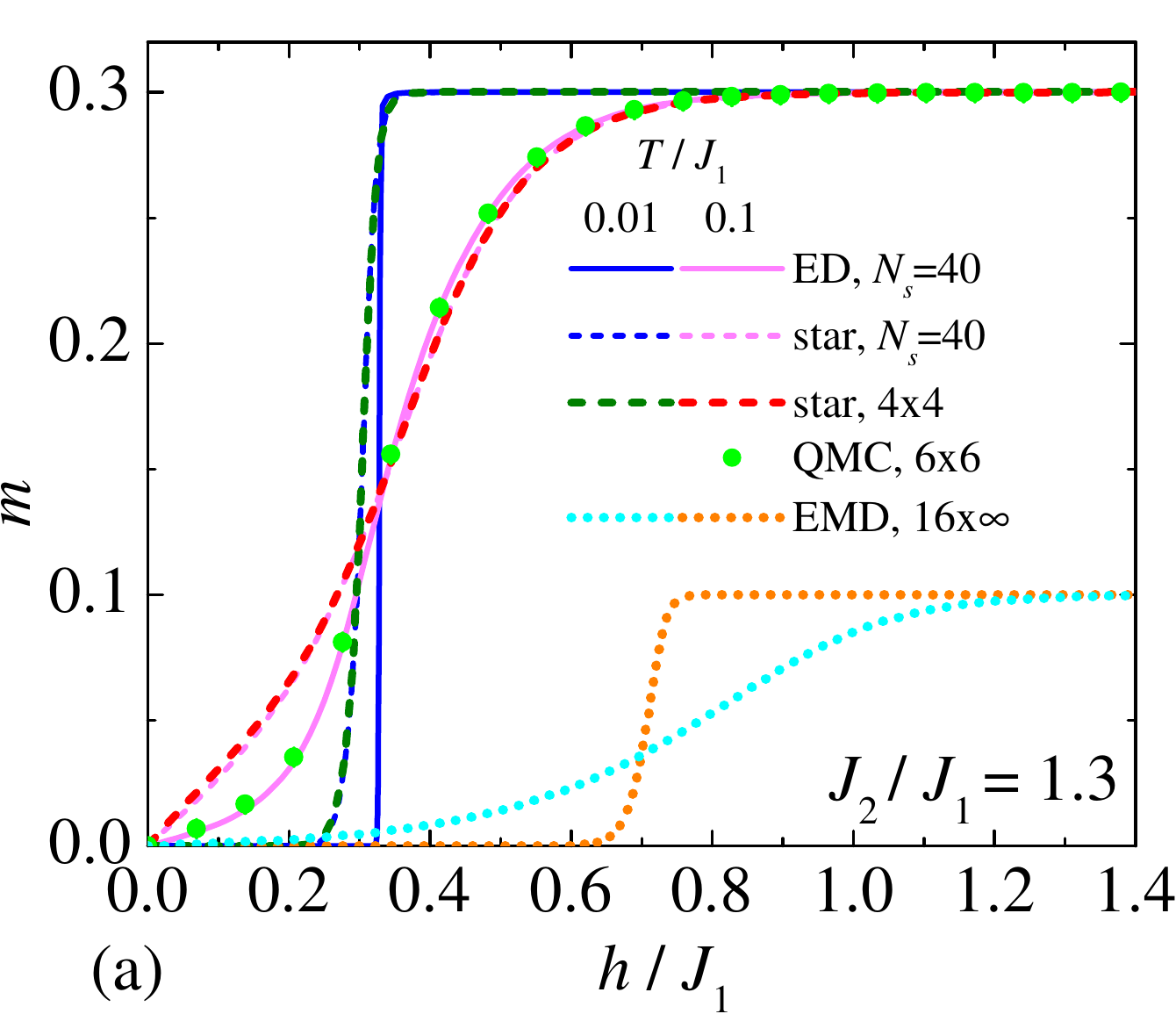}
	\includegraphics[width=0.8\columnwidth]{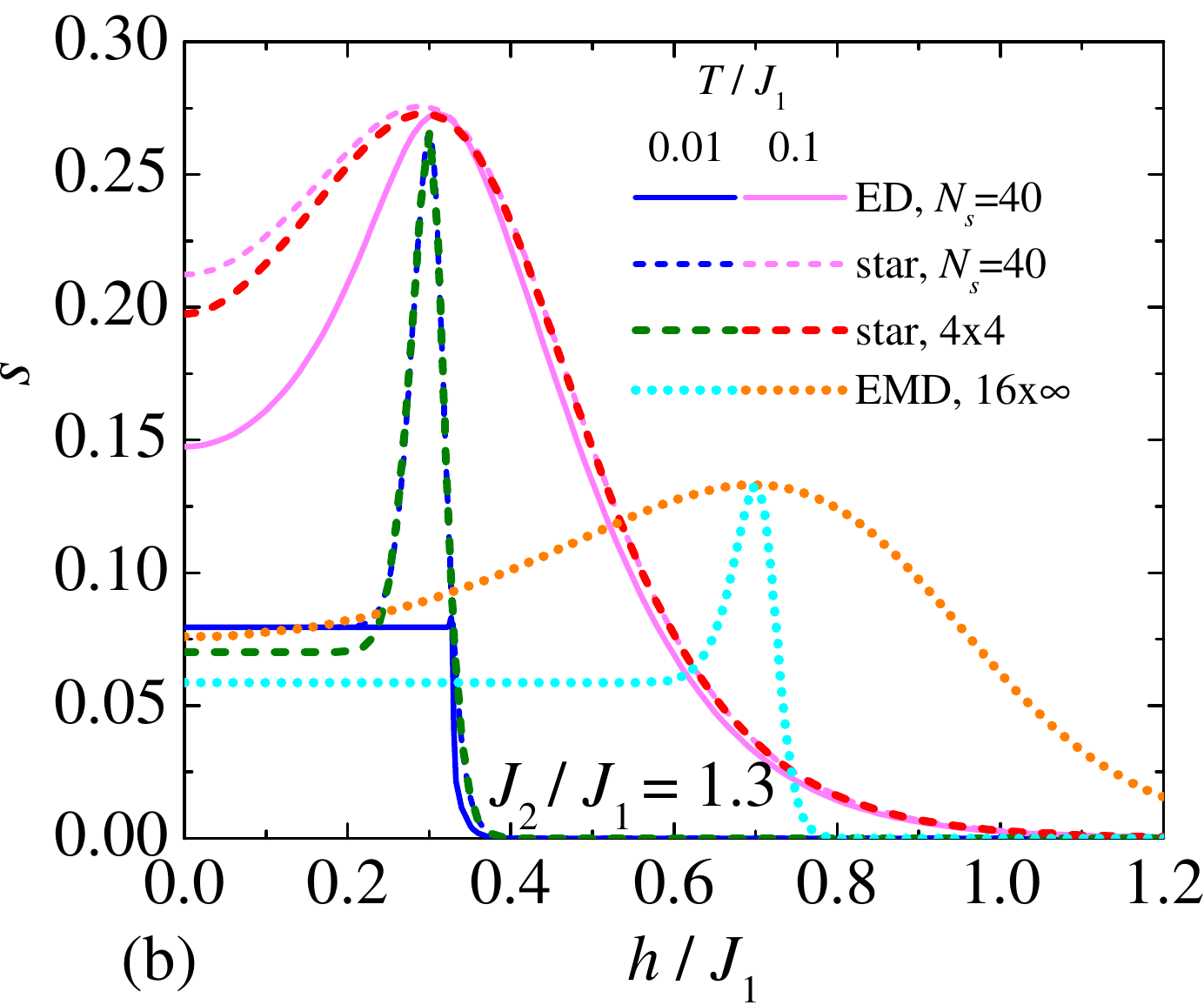}
	\includegraphics[width=0.8\columnwidth]{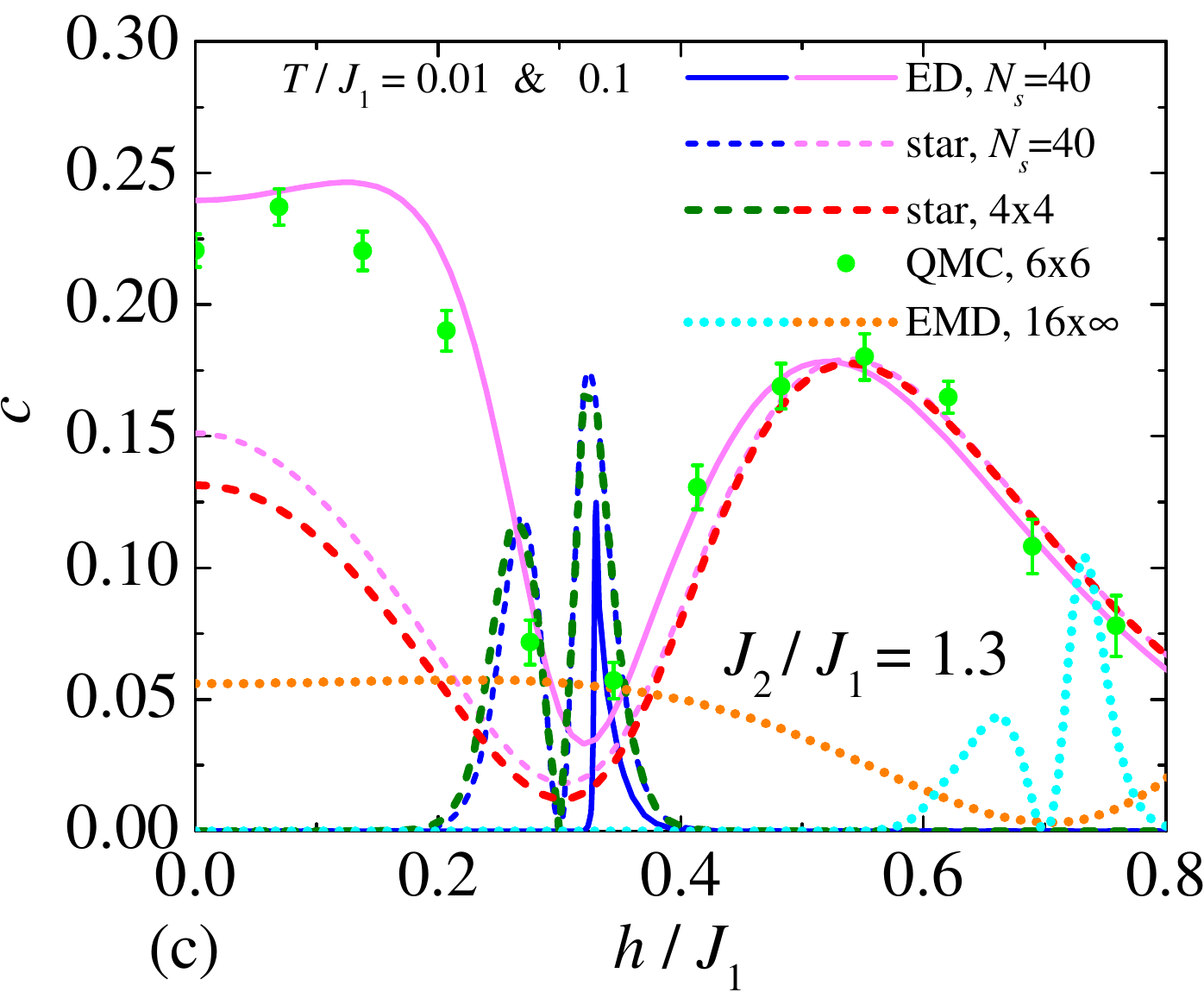}
	\caption{(a) Magnetization; (b) entropy; (c) specific heat
for the interaction ratio $J_2/J_1=1.3$ as a function of magnetic field at two different values of temperature $T/J_1=0.01$ and $T/J_1=0.1$.}
	\label{figh:1.3}
\end{figure}

Last but not least, magnetization, entropy, and specific heat as a function of temperature are depicted in Fig.~\ref{figh:1.3} for the interaction ratio $J_2/J_1=1.3$. As discussed in Sec.~\ref{EDecko}, for the interaction ratio $J_2/J_1\lesssim 1.4$, the lowest excitation above the DT phase corresponds to the creation of an additional triplet [see Fig.~\ref{fig:TD-gapN40}]. For this reason, the EMD model is unable to describe the finite-temperature properties
for the interaction ratio $J_2/J_1=1.3$.
On the other hand, our second analytical approach derived from the spin-star decoupling approximation provides a quite reliable description of all quantities above the transition field $h/J_1=0.3$.
The most important artifacts of the spin-star decoupling approximation are just below the transition field
and at $T/J_1=0.01$ 
where it yields a spike in the entropy $s$ [Fig.~\ref{figh:1.3}(b)] and a second maximum
in the specific heat $c$ [Fig.~\ref{figh:1.3}(c)] while such pronounced maxima are absent in the ED results.
The reason for this difference is that the field-induced DT-LM transition is a true first-order transition in the full model
such that excitations are scattered over a certain energy range below the transition
while the spin-star decoupling approximation treats these as a degenerate manifold that comes down at the transition field.
Conversely, a temperature $T/J_1=0.1$ is sufficiently high to thermally excite these low-lying excitations
in the full model such that at this higher temperature the $N_s=40$ ED results also exhibit a pronounced maximum
below the transition field, both in the entropy [Fig.~\ref{figh:1.3}(b)] and the specific heat [Fig.~\ref{figh:1.3}(c)].
The latter behavior is confirmed by our QMC simulations on the bigger system of $6\times 6$ unit cells.

\subsection{Enhanced magnetocaloric effect due to the macroscopically degenerate DT phase}

\begin{figure}[t!]
	\centering
	\includegraphics[width=0.8\columnwidth]{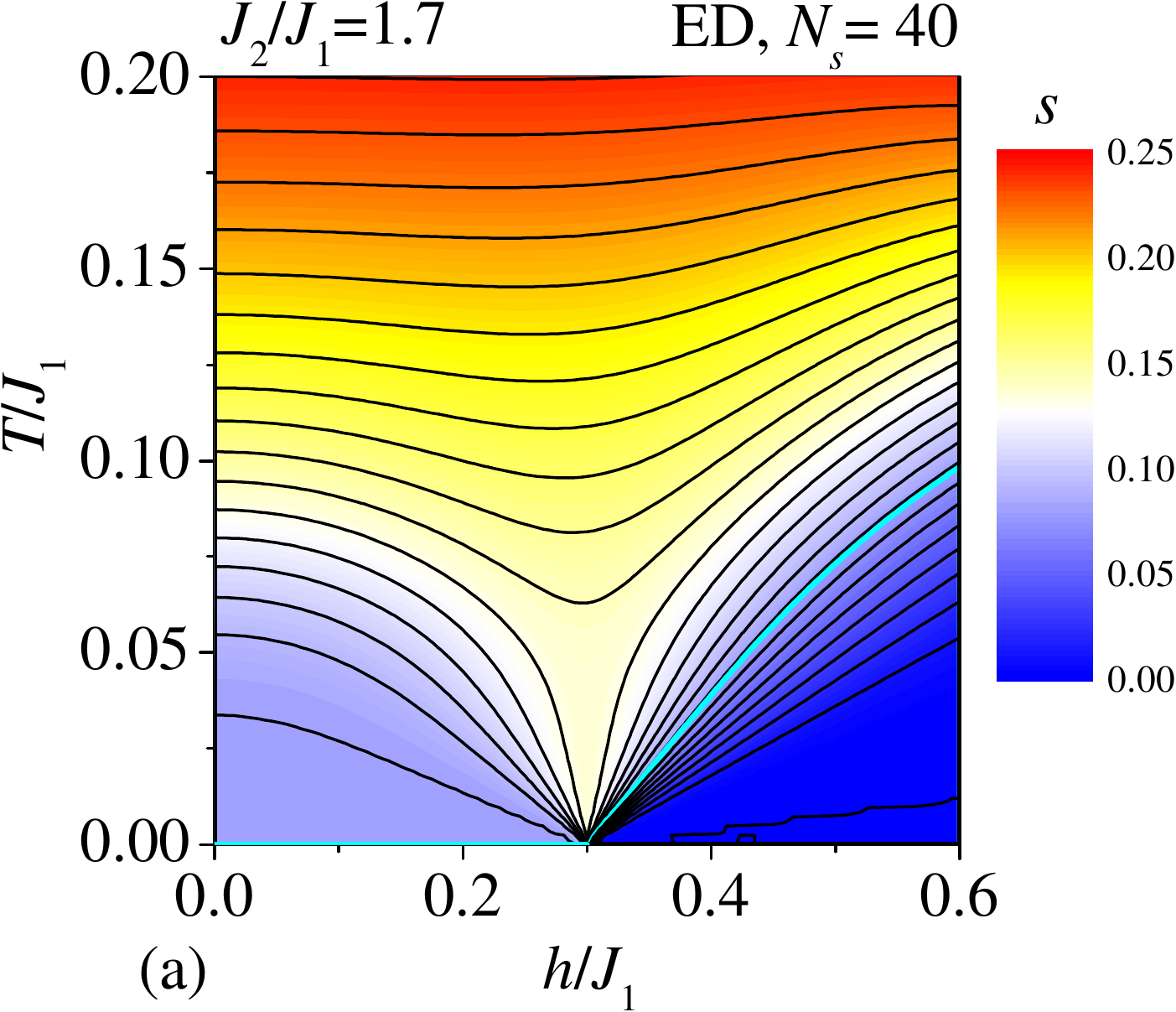}
	\includegraphics[width=0.8\columnwidth]{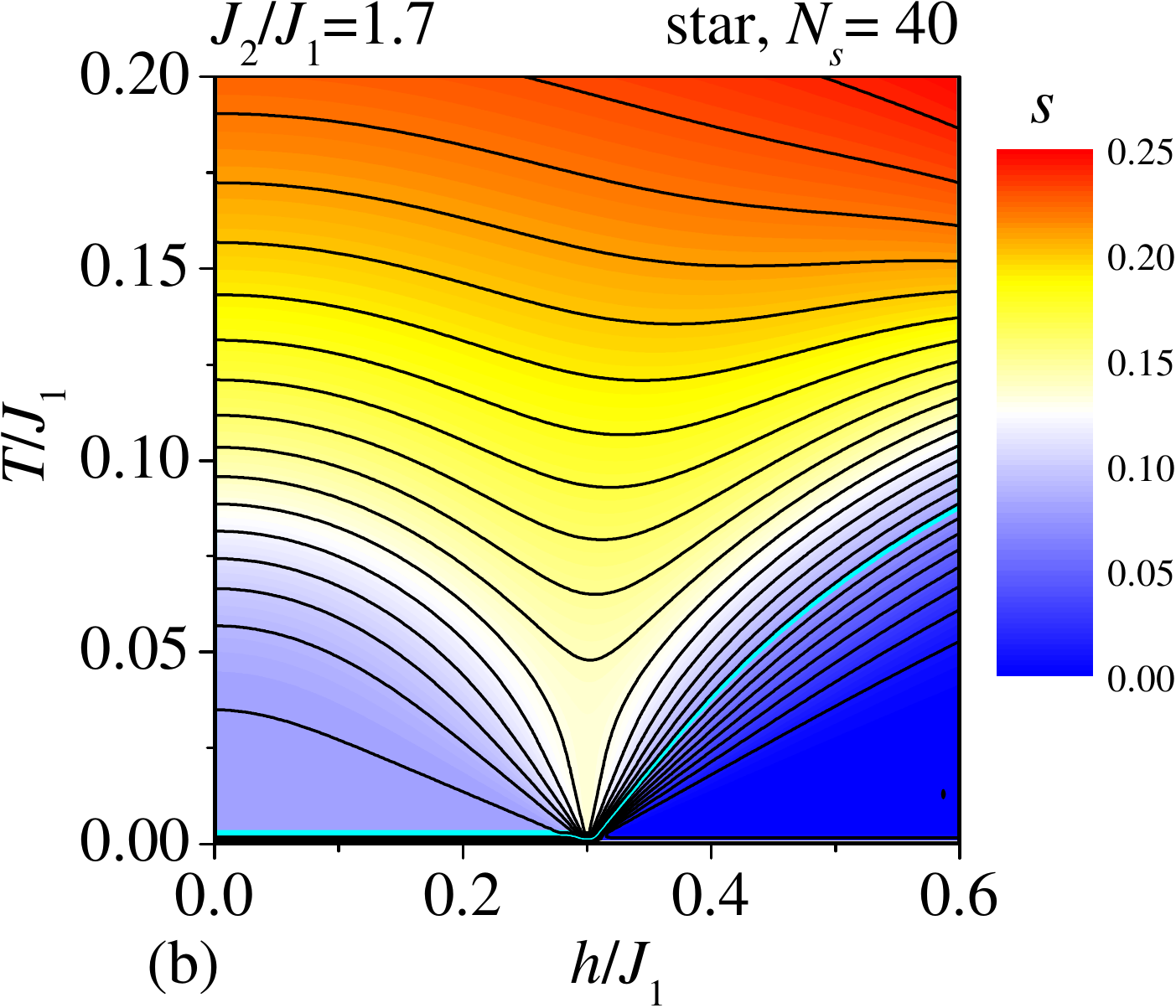}
	\includegraphics[width=0.8\columnwidth]{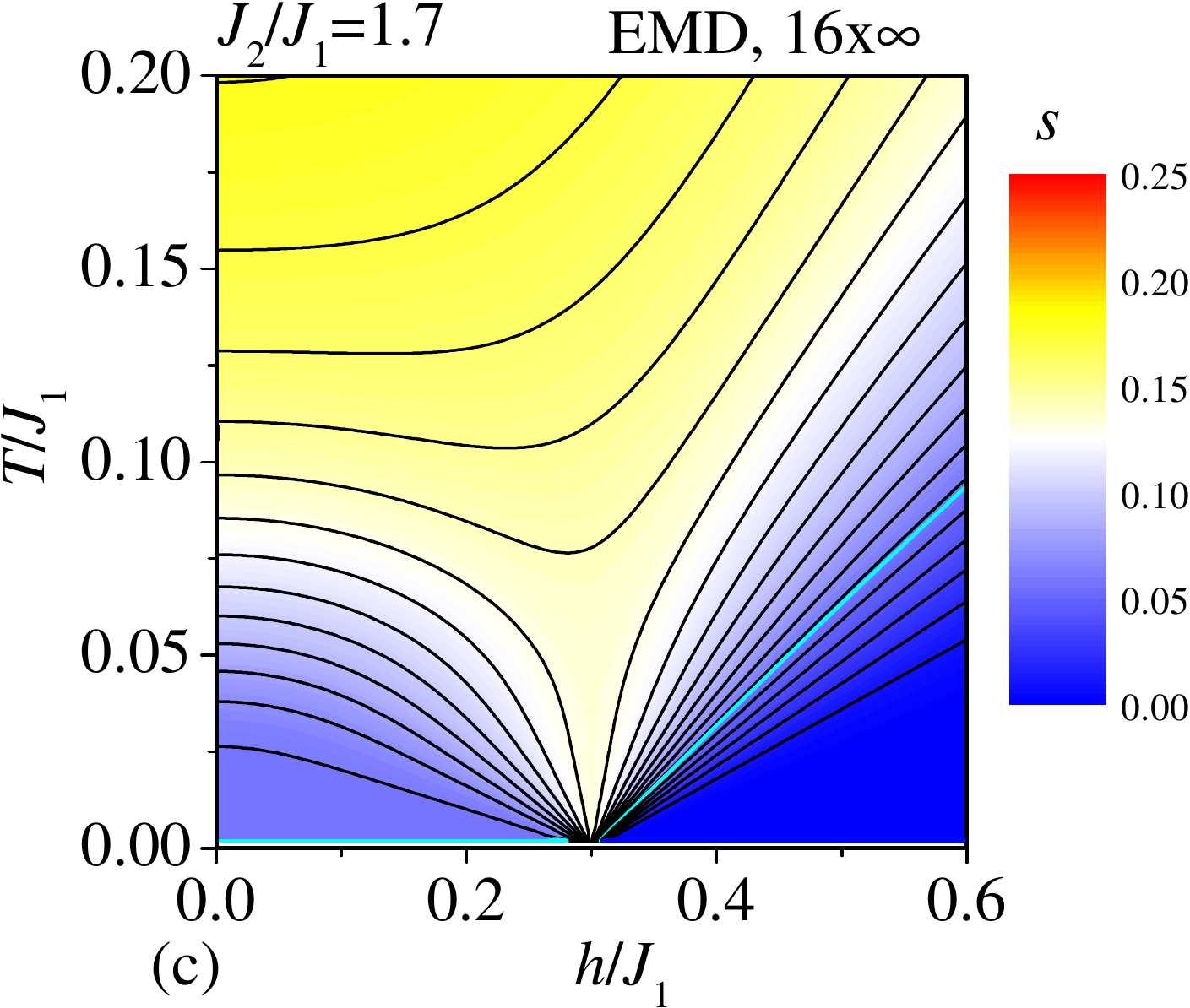}
	\caption{Density plots of the entropy in the plane $h/J_1-T/J_1$ 
for the interaction ratio $J_2/J_1=1.7$ as obtained from (a) ED for $N_s=40$ spins; (b) spin-star decoupling approximation for $N_s=40$ spins; (c) EMD model for $16\times\infty$. 
The cyan lines in panels (a) and (b) correspond to $s=0.0775$  and in panel (c) to $s=0.0575$. 
Below the cyan lines the isentropes approach zero temperature for $h/J_1\lesssim 0.3$.}
	\label{fig:mce}
\end{figure}

Finally, we discuss the magnetocaloric effect with a focus on the transition region between the macroscopically
degenerate DT phase and the MD state (that is nondegenerate in a finite field).
Figure \ref{fig:mce} shows 
density plots of the entropy in the magnetic field versus temperature plane
for the interaction ratio  $J_2/J_1=1.7$.
The superimposed constant-entropy curves correspond to the variation of the temperature $T$ during an adiabatic
(de)magnetization process.

First, we discuss the ED results for a system size of $N_s=40$ spins shown in Fig.~\ref{fig:mce}(a).
An enhanced magnetocaloric effect can clearly be seen in the vicinity of the magnetic field $h/J_1=0.3$, where the temperature sharply decreases during an adiabatic demagnetization process, whereas an inverse magnetocaloric effect associated with an increase in temperature with further reduction of the magnetic field can be mostly observed below this value of the magnetic field. 
The specific value of the magnetic field $h/J_1=0.3$ corresponds to the magnetic-field-driven phase transition from the MD phase to the DT phase for the interaction ratio $J_2/J_1=1.7$. Recall also that the residual entropy of the DT phase (per spin) for a system size of $N_s=40$ is equal to $s=\frac{\ln 24}{40} \approx 0.079$ and, hence, the system cools down to absolute zero temperature at $h/J_1=0.3$, whenever the entropy at the beginning of the adiabatic demagnetization is tuned below this value. Under this condition, the temperature stays at zero below the transition field and there is no consecutive increase in temperature upon further demagnetization,
as Fig.~\ref{fig:mce}(a) demonstrates by the absence of isentropes below the magnetic field value $h/J_1\lesssim 0.3$ (i.e., below the cyan curve). This asymmetry 
of the entropy around the transition field is caused by the special character of the DT phase, which preserves its macroscopic degeneracy 
at nonzero magnetic fields. From this point of view, the spin-1/2 Heisenberg model on the diamond-decorated square lattice offers 
promising
perspectives
for achieving ultra-low temperatures during adiabatic demagnetization.

For comparison, Fig.~\ref{fig:mce}(b) presents the density plot of the entropy, for the same set of parameters and the same system size $N_s=40$, as obtained from Eq.~(\ref{freen}) within the spin-star decoupling approximation.
Remarkably, the spin-star approach is capable of predicting the aforementioned asymmetry of the entropy, and the relevant density plot reveals qualitatively the same features of the magnetocaloric effect as obtained from ED.
Moreover, the enhanced magnetocaloric effect and cooling to absolute zero temperature should persist even in the thermodynamic limit because of the macroscopic degeneracy of the DT phase.
The EMD model shows indeed
that the asymmetric behavior of the isentropes as a function of magnetic field and temperature is preserved 
for a system size comprising of 16 elementary units in one spatial direction and infinite in the other direction, see
Fig.~\ref{fig:mce}(c). For this system size, the residual entropy of the DT phase is  $s \approx 0.0587$, which is consistent with the macroscopic degeneracy of the hard-dimer model on the square lattice
\cite{Fisher61,Kasteleyn61,Morita2016}. 
 Hence, the enhanced magnetocaloric effect accompanied with cooling to  zero temperature is detected for $h/J_1<0.3$ whenever the entropy is tuned below its zero-point value $s \lesssim 0.0587$  at the beginning of an adiabatic demagnetization process (i.e., all curves lying below the cyan curve).

\section{Conclusion}
\label{conclusion}
In this study, we have investigated the thermodynamic properties of the spin-1/2 Heisenberg model on the diamond-decorated square lattice, particularly focusing on the macroscopically degenerate DT phase. Using a combination of exact diagonalization, sign-problem-free quantum Monte Carlo simulations, an effective monomer-dimer model, and a spin-star decoupling approximation, we have provided a comprehensive analysis of this model's behavior under varying interaction ratios and external magnetic fields.
Our results confirm that the DT phase exhibits unique thermodynamic properties owing to its macroscopic degeneracy, which is linked to the classical hard-dimer model on the square lattice. The boundary between the MD and DT phases is well described by the effective monomer-dimer model, highlighting the significance of localized excitations in determining the low-temperature behavior of this system.

A particularly interesting
result of our study is the enhanced magnetocaloric effect observed upon approaching the DT phase. This effect is particularly pronounced near the transition 
field between the MD and the DT phase, where the temperature decreases sharply during adiabatic demagnetization. This phenomenon arises owing to the macroscopic degeneracy of the DT phase, which allows the system to cool to absolute zero temperature as long as the entropy content is set below the residual entropy value $s \approx 0.0583$ derived from the effective hard-dimer model on the square lattice in the thermodynamic limit. This renders the DT phase a promising candidate for achieving ultralow temperatures
by magnetic refrigeration.
The third law of thermodynamics \cite{Nernst1906} requires that this macroscopic degeneracy be lifted
in an experimental realization, for instance through a lattice distortion that breaks the conservation of the total
spin of a dimer.
Our conclusions will nevertheless remain relevant
as long as the resultant splitting of the ground-state manifold remains at an energy scale that is below the temperature scale of interest.
The challenge is more to find a suitable compound to begin with. In this respect, it may be interesting
to consider the diamond-decorated {\em honeycomb} lattice since in this case there are known compounds that exhibit
the corresponding crystal structure, see for example Refs.~\cite{Zhang2000,Travnicek2001,Hong2004}.

While the present study does not directly address specific chemical realizations, our findings offer insights into the potential applications of frustrated spin systems, particularly in magnetic cooling technologies. These insights could be relevant for materials exhibiting similar thermodynamic properties, such as those found in the aforementioned compounds with a diamond-decorated honeycomb lattice.

Our findings 
demonstrate the usefulness of effective classical-statistical physics models for the low-energy and low-temperature
behavior of a highly frustrated quantum spin system and
shed light on the rich phase diagram and thermodynamic phenomena of the spin-1/2 Heisenberg model on the diamond-decorated square lattice.
Furthermore, they suggest potential applications of such highly frustrated quantum spin systems in magnetic cooling technologies.

\begin{acknowledgments}
This project has received funding from the European Union's Horizon 2020 research and innovation programme under the Marie Sk{\l}odowska-Curie Grant Agreement No.\ 945380. We acknowledge support under the \v{S}tef\'anik+ program for Slovak-France bilateral collaboration under contract SK-FR-22-0011/49880PG. 
Computing time for the QMC simulations was provided by the IT Center at RWTH Aachen University,
part of the ED computations were carried out on the ``osaka'' cluster at the Centre de Calcul (CDC) of CY Cergy Paris Université.
T.V.\ is supported through the EURIZON project (No.~3025 ``Frustrated quantum spin models to explain the properties of magnets over wide temperature range''), which is funded by the European Union under grant agreement No.~871072. J.S.\ acknowledges financial support provided by Slovak Research and Development Agency under the Contract No.~APVV-20-0150. N.C.\ acknowledges support by the ANR through Grant
  LODIS (ANR-21-CE30-0033).
\end{acknowledgments}

\bibliography{manuscript}

\begin{thebibliography}{81}%
\makeatletter
\providecommand \@ifxundefined [1]{%
 \@ifx{#1\undefined}
}%
\providecommand \@ifnum [1]{%
 \ifnum #1\expandafter \@firstoftwo
 \else \expandafter \@secondoftwo
 \fi
}%
\providecommand \@ifx [1]{%
 \ifx #1\expandafter \@firstoftwo
 \else \expandafter \@secondoftwo
 \fi
}%
\providecommand \natexlab [1]{#1}%
\providecommand \enquote  [1]{``#1''}%
\providecommand \bibnamefont  [1]{#1}%
\providecommand \bibfnamefont [1]{#1}%
\providecommand \citenamefont [1]{#1}%
\providecommand \href@noop [0]{\@secondoftwo}%
\providecommand \href [0]{\begingroup \@sanitize@url \@href}%
\providecommand \@href[1]{\@@startlink{#1}\@@href}%
\providecommand \@@href[1]{\endgroup#1\@@endlink}%
\providecommand \@sanitize@url [0]{\catcode `\\12\catcode `\$12\catcode
  `\&12\catcode `\#12\catcode `\^12\catcode `\_12\catcode `\%12\relax}%
\providecommand \@@startlink[1]{}%
\providecommand \@@endlink[0]{}%
\providecommand \url  [0]{\begingroup\@sanitize@url \@url }%
\providecommand \@url [1]{\endgroup\@href {#1}{\urlprefix }}%
\providecommand \urlprefix  [0]{URL }%
\providecommand \Eprint [0]{\href }%
\providecommand \doibase [0]{https://doi.org/}%
\providecommand \selectlanguage [0]{\@gobble}%
\providecommand \bibinfo  [0]{\@secondoftwo}%
\providecommand \bibfield  [0]{\@secondoftwo}%
\providecommand \translation [1]{[#1]}%
\providecommand \BibitemOpen [0]{}%
\providecommand \bibitemStop [0]{}%
\providecommand \bibitemNoStop [0]{.\EOS\space}%
\providecommand \EOS [0]{\spacefactor3000\relax}%
\providecommand \BibitemShut  [1]{\csname bibitem#1\endcsname}%
\let\auto@bib@innerbib\@empty
\bibitem [{\citenamefont {Lacroix}\ \emph {et~al.}(2011)\citenamefont
  {Lacroix}, \citenamefont {Mendels},\ and\ \citenamefont
  {Mila}}]{Lacroix2011}%
  \BibitemOpen
  \bibinfo {editor} {\bibfnamefont {C.}~\bibnamefont {Lacroix}}, \bibinfo
  {editor} {\bibfnamefont {P.}~\bibnamefont {Mendels}},\ and\ \bibinfo {editor}
  {\bibfnamefont {F.}~\bibnamefont {Mila}},\ eds.,\ \href
  {https://doi.org/10.1007/978-3-642-10589-0} {\emph {\bibinfo {title}
  {Introduction to Frustrated Magnetism}}},\ \bibinfo {series} {Springer Series
  in Solid-State Sciences}, Vol.\ \bibinfo {volume} {164}\ (\bibinfo
  {publisher} {Springer Berlin, Heidelberg},\ \bibinfo {address} {Berlin,
  Heidelberg},\ \bibinfo {year} {2011})\BibitemShut {NoStop}%
\bibitem [{\citenamefont {Honecker}\ \emph {et~al.}(2004)\citenamefont
  {Honecker}, \citenamefont {Schulenburg},\ and\ \citenamefont
  {Richter}}]{Honecker_2004}%
  \BibitemOpen
  \bibfield  {author} {\bibinfo {author} {\bibfnamefont {A.}~\bibnamefont
  {Honecker}}, \bibinfo {author} {\bibfnamefont {J.}~\bibnamefont
  {Schulenburg}},\ and\ \bibinfo {author} {\bibfnamefont {J.}~\bibnamefont
  {Richter}},\ }\bibfield  {title} {\bibinfo {title} {Magnetization plateaus in
  frustrated antiferromagnetic quantum spin models},\ }\href
  {https://doi.org/10.1088/0953-8984/16/11/025} {\bibfield  {journal} {\bibinfo
   {journal} {J. Phys.: Condens. Matter}\ }\textbf {\bibinfo {volume} {16}},\
  \bibinfo {pages} {S749} (\bibinfo {year} {2004})}\BibitemShut {NoStop}%
\bibitem [{\citenamefont {Lhuillier}\ and\ \citenamefont
  {Misguich}(2001)}]{Lhuillier2001}%
  \BibitemOpen
  \bibfield  {author} {\bibinfo {author} {\bibfnamefont {C.}~\bibnamefont
  {Lhuillier}}\ and\ \bibinfo {author} {\bibfnamefont {G.}~\bibnamefont
  {Misguich}},\ }\bibinfo {title} {Frustrated quantum magnets},\ in\ \href
  {https://doi.org/10.1007/3-540-45649-X_6} {\emph {\bibinfo {booktitle} {High
  Magnetic Fields: Applications in Condensed Matter Physics and
  Spectroscopy}}},\ \bibinfo {editor} {edited by\ \bibinfo {editor}
  {\bibfnamefont {C.}~\bibnamefont {Berthier}}, \bibinfo {editor}
  {\bibfnamefont {L.~P.}\ \bibnamefont {L{\'e}vy}},\ and\ \bibinfo {editor}
  {\bibfnamefont {G.}~\bibnamefont {Martinez}}}\ (\bibinfo  {publisher}
  {Springer Berlin Heidelberg},\ \bibinfo {address} {Berlin, Heidelberg},\
  \bibinfo {year} {2001})\ pp.\ \bibinfo {pages} {161--190}\BibitemShut
  {NoStop}%
\bibitem [{\citenamefont {Diep}(2020)}]{Diep2013}%
  \BibitemOpen
  \bibfield  {author} {\bibinfo {author} {\bibfnamefont {H.~T.}\ \bibnamefont
  {Diep}},\ }\href {https://doi.org/10.1142/11660} {\emph {\bibinfo {title}
  {Frustrated Spin Systems}}},\ \bibinfo {edition} {3rd}\ ed.\ (\bibinfo
  {publisher} {World Scientific},\ \bibinfo {year} {2020})\BibitemShut
  {NoStop}%
\bibitem [{\citenamefont {Liebmann}(1986)}]{Liebmann1986}%
  \BibitemOpen
  \bibfield  {author} {\bibinfo {author} {\bibfnamefont {R.}~\bibnamefont
  {Liebmann}},\ }\href {https://doi.org/10.1007/3-540-16473-1} {\emph {\bibinfo
  {title} {Statistical Mechanics of Periodic Frustrated {I}sing Systems}}},\
  \bibinfo {series} {Lecture Notes in Physics}, Vol.\ \bibinfo {volume} {251}\
  (\bibinfo  {publisher} {Springer Berlin, Heidelberg},\ \bibinfo {address}
  {Berlin, Heidelberg},\ \bibinfo {year} {1986})\BibitemShut {NoStop}%
\bibitem [{\citenamefont {Ramirez}\ and\ \citenamefont
  {Syzranov}(2024)}]{ramirez2024}%
  \BibitemOpen
  \bibfield  {author} {\bibinfo {author} {\bibfnamefont {A.~P.}\ \bibnamefont
  {Ramirez}}\ and\ \bibinfo {author} {\bibfnamefont {S.}~\bibnamefont
  {Syzranov}},\ }\href {https://arxiv.org/abs/2408.16054} {\bibinfo {title}
  {Order and disorder in geometrically frustrated magnets}} (\bibinfo {year}
  {2024}),\ \Eprint {https://arxiv.org/abs/2408.16054} {arXiv:2408.16054
  [cond-mat.str-el]} \BibitemShut {NoStop}%
\bibitem [{\citenamefont {Villain}\ \emph {et~al.}(1980)\citenamefont
  {Villain}, \citenamefont {Bidaux}, \citenamefont {Carton},\ and\
  \citenamefont {Conte}}]{Villain1980}%
  \BibitemOpen
  \bibfield  {author} {\bibinfo {author} {\bibfnamefont {J.}~\bibnamefont
  {Villain}}, \bibinfo {author} {\bibfnamefont {R.}~\bibnamefont {Bidaux}},
  \bibinfo {author} {\bibfnamefont {J.-P.}\ \bibnamefont {Carton}},\ and\
  \bibinfo {author} {\bibfnamefont {R.}~\bibnamefont {Conte}},\ }\bibfield
  {title} {\bibinfo {title} {Order as an effect of disorder},\ }\href
  {https://doi.org/10.1051/jphys:0198000410110126300} {\bibfield  {journal}
  {\bibinfo  {journal} {J. Phys. France}\ }\textbf {\bibinfo {volume} {41}},\
  \bibinfo {pages} {1263} (\bibinfo {year} {1980})}\BibitemShut {NoStop}%
\bibitem [{\citenamefont {Henley}(1989)}]{Henley1989}%
  \BibitemOpen
  \bibfield  {author} {\bibinfo {author} {\bibfnamefont {C.~L.}\ \bibnamefont
  {Henley}},\ }\bibfield  {title} {\bibinfo {title} {Ordering due to disorder
  in a frustrated vector antiferromagnet},\ }\href
  {https://doi.org/10.1103/PhysRevLett.62.2056} {\bibfield  {journal} {\bibinfo
   {journal} {Phys. Rev. Lett.}\ }\textbf {\bibinfo {volume} {62}},\ \bibinfo
  {pages} {2056} (\bibinfo {year} {1989})}\BibitemShut {NoStop}%
\bibitem [{\citenamefont {Moessner}(2001)}]{Moessner2021}%
  \BibitemOpen
  \bibfield  {author} {\bibinfo {author} {\bibfnamefont {R.}~\bibnamefont
  {Moessner}},\ }\bibfield  {title} {\bibinfo {title} {Magnets with strong
  geometric frustration},\ }\href {https://doi.org/10.1139/p01-123} {\bibfield
  {journal} {\bibinfo  {journal} {Can. J. Phys.}\ }\textbf {\bibinfo {volume}
  {79}},\ \bibinfo {pages} {1283} (\bibinfo {year} {2001})}\BibitemShut
  {NoStop}%
\bibitem [{\citenamefont {Bergman}\ \emph {et~al.}(2007)\citenamefont
  {Bergman}, \citenamefont {Alicea}, \citenamefont {Gull}, \citenamefont
  {Trebst},\ and\ \citenamefont {Balents}}]{Bergman2007}%
  \BibitemOpen
  \bibfield  {author} {\bibinfo {author} {\bibfnamefont {D.}~\bibnamefont
  {Bergman}}, \bibinfo {author} {\bibfnamefont {J.}~\bibnamefont {Alicea}},
  \bibinfo {author} {\bibfnamefont {E.}~\bibnamefont {Gull}}, \bibinfo {author}
  {\bibfnamefont {S.}~\bibnamefont {Trebst}},\ and\ \bibinfo {author}
  {\bibfnamefont {L.}~\bibnamefont {Balents}},\ }\bibfield  {title} {\bibinfo
  {title} {Order-by-disorder and spiral spin-liquid in frustrated
  diamond-lattice antiferromagnets},\ }\href {https://doi.org/10.1038/nphys622}
  {\bibfield  {journal} {\bibinfo  {journal} {Nat. Phys.}\ }\textbf {\bibinfo
  {volume} {3}},\ \bibinfo {pages} {487} (\bibinfo {year} {2007})}\BibitemShut
  {NoStop}%
\bibitem [{\citenamefont {Lee}\ and\ \citenamefont {Balents}(2008)}]{Lee2008}%
  \BibitemOpen
  \bibfield  {author} {\bibinfo {author} {\bibfnamefont {S.}~\bibnamefont
  {Lee}}\ and\ \bibinfo {author} {\bibfnamefont {L.}~\bibnamefont {Balents}},\
  }\bibfield  {title} {\bibinfo {title} {Theory of the ordered phase in
  {$A$}-site antiferromagnetic spinels},\ }\href
  {https://doi.org/10.1103/PhysRevB.78.144417} {\bibfield  {journal} {\bibinfo
  {journal} {Phys. Rev. B}\ }\textbf {\bibinfo {volume} {78}},\ \bibinfo
  {pages} {144417} (\bibinfo {year} {2008})}\BibitemShut {NoStop}%
\bibitem [{\citenamefont {Buessen}\ \emph {et~al.}(2018)\citenamefont
  {Buessen}, \citenamefont {Hering}, \citenamefont {Reuther},\ and\
  \citenamefont {Trebst}}]{Buessen2018}%
  \BibitemOpen
  \bibfield  {author} {\bibinfo {author} {\bibfnamefont {F.~L.}\ \bibnamefont
  {Buessen}}, \bibinfo {author} {\bibfnamefont {M.}~\bibnamefont {Hering}},
  \bibinfo {author} {\bibfnamefont {J.}~\bibnamefont {Reuther}},\ and\ \bibinfo
  {author} {\bibfnamefont {S.}~\bibnamefont {Trebst}},\ }\bibfield  {title}
  {\bibinfo {title} {Quantum spin liquids in frustrated spin-1 diamond
  antiferromagnets},\ }\href {https://doi.org/10.1103/PhysRevLett.120.057201}
  {\bibfield  {journal} {\bibinfo  {journal} {Phys. Rev. Lett.}\ }\textbf
  {\bibinfo {volume} {120}},\ \bibinfo {pages} {057201} (\bibinfo {year}
  {2018})}\BibitemShut {NoStop}%
\bibitem [{\citenamefont {Jolicoeur}\ \emph {et~al.}(1990)\citenamefont
  {Jolicoeur}, \citenamefont {Dagotto}, \citenamefont {Gagliano},\ and\
  \citenamefont {Bacci}}]{Joli1990}%
  \BibitemOpen
  \bibfield  {author} {\bibinfo {author} {\bibfnamefont {T.}~\bibnamefont
  {Jolicoeur}}, \bibinfo {author} {\bibfnamefont {E.}~\bibnamefont {Dagotto}},
  \bibinfo {author} {\bibfnamefont {E.}~\bibnamefont {Gagliano}},\ and\
  \bibinfo {author} {\bibfnamefont {S.}~\bibnamefont {Bacci}},\ }\bibfield
  {title} {\bibinfo {title} {Ground-state properties of the {$S=\frac12$}
  {H}eisenberg antiferromagnet on a triangular lattice},\ }\href
  {https://doi.org/10.1103/PhysRevB.42.4800} {\bibfield  {journal} {\bibinfo
  {journal} {Phys. Rev. B}\ }\textbf {\bibinfo {volume} {42}},\ \bibinfo
  {pages} {4800} (\bibinfo {year} {1990})}\BibitemShut {NoStop}%
\bibitem [{\citenamefont {Reimers}\ and\ \citenamefont
  {Berlinsky}(1993)}]{Reimers1993}%
  \BibitemOpen
  \bibfield  {author} {\bibinfo {author} {\bibfnamefont {J.~N.}\ \bibnamefont
  {Reimers}}\ and\ \bibinfo {author} {\bibfnamefont {A.~J.}\ \bibnamefont
  {Berlinsky}},\ }\bibfield  {title} {\bibinfo {title} {Order by disorder in
  the classical {H}eisenberg kagom\'e antiferromagnet},\ }\href
  {https://doi.org/10.1103/PhysRevB.48.9539} {\bibfield  {journal} {\bibinfo
  {journal} {Phys. Rev. B}\ }\textbf {\bibinfo {volume} {48}},\ \bibinfo
  {pages} {9539} (\bibinfo {year} {1993})}\BibitemShut {NoStop}%
\bibitem [{\citenamefont {Bramwell}\ \emph {et~al.}(1994)\citenamefont
  {Bramwell}, \citenamefont {Gingras},\ and\ \citenamefont
  {Reimers}}]{Bramwell1994}%
  \BibitemOpen
  \bibfield  {author} {\bibinfo {author} {\bibfnamefont {S.~T.}\ \bibnamefont
  {Bramwell}}, \bibinfo {author} {\bibfnamefont {M.~J.~P.}\ \bibnamefont
  {Gingras}},\ and\ \bibinfo {author} {\bibfnamefont {J.~N.}\ \bibnamefont
  {Reimers}},\ }\bibfield  {title} {\bibinfo {title} {Order by disorder in an
  anisotropic pyrochlore lattice antiferromagnet},\ }\href
  {https://doi.org/10.1063/1.355676} {\bibfield  {journal} {\bibinfo  {journal}
  {J. Appl. Phys.}\ }\textbf {\bibinfo {volume} {75}},\ \bibinfo {pages} {5523}
  (\bibinfo {year} {1994})}\BibitemShut {NoStop}%
\bibitem [{\citenamefont {Majumdar}\ and\ \citenamefont
  {Ghosh}(1969)}]{Majumdar1969}%
  \BibitemOpen
  \bibfield  {author} {\bibinfo {author} {\bibfnamefont {C.~K.}\ \bibnamefont
  {Majumdar}}\ and\ \bibinfo {author} {\bibfnamefont {D.~K.}\ \bibnamefont
  {Ghosh}},\ }\bibfield  {title} {\bibinfo {title} {On next-nearest-neighbor
  interaction in linear chain. {II}},\ }\href
  {https://doi.org/10.1063/1.1664979} {\bibfield  {journal} {\bibinfo
  {journal} {J. Math. Phys.}\ }\textbf {\bibinfo {volume} {10}},\ \bibinfo
  {pages} {1399} (\bibinfo {year} {1969})}\BibitemShut {NoStop}%
\bibitem [{\citenamefont {{Sriram Shastry}}\ and\ \citenamefont
  {Sutherland}(1981)}]{Shastry1981}%
  \BibitemOpen
  \bibfield  {author} {\bibinfo {author} {\bibfnamefont {B.}~\bibnamefont
  {{Sriram Shastry}}}\ and\ \bibinfo {author} {\bibfnamefont {B.}~\bibnamefont
  {Sutherland}},\ }\bibfield  {title} {\bibinfo {title} {Exact ground state of
  a quantum mechanical antiferromagnet},\ }\href
  {https://doi.org/https://doi.org/10.1016/0378-4363(81)90838-X} {\bibfield
  {journal} {\bibinfo  {journal} {Physica B+C}\ }\textbf {\bibinfo {volume}
  {108}},\ \bibinfo {pages} {1069} (\bibinfo {year} {1981})}\BibitemShut
  {NoStop}%
\bibitem [{\citenamefont {Koga}\ and\ \citenamefont
  {Kawakami}(2000)}]{Koga2000}%
  \BibitemOpen
  \bibfield  {author} {\bibinfo {author} {\bibfnamefont {A.}~\bibnamefont
  {Koga}}\ and\ \bibinfo {author} {\bibfnamefont {N.}~\bibnamefont
  {Kawakami}},\ }\bibfield  {title} {\bibinfo {title} {Quantum phase
  transitions in the {S}hastry-{S}utherland model for
  {SrCu}$_{2}$({BO}$_{3}$)$_{2}$},\ }\href
  {https://doi.org/10.1103/PhysRevLett.84.4461} {\bibfield  {journal} {\bibinfo
   {journal} {Phys. Rev. Lett.}\ }\textbf {\bibinfo {volume} {84}},\ \bibinfo
  {pages} {4461} (\bibinfo {year} {2000})}\BibitemShut {NoStop}%
\bibitem [{\citenamefont {Ganesh}\ \emph {et~al.}(2013)\citenamefont {Ganesh},
  \citenamefont {van~den Brink},\ and\ \citenamefont {Nishimoto}}]{Ganesh2013}%
  \BibitemOpen
  \bibfield  {author} {\bibinfo {author} {\bibfnamefont {R.}~\bibnamefont
  {Ganesh}}, \bibinfo {author} {\bibfnamefont {J.}~\bibnamefont {van~den
  Brink}},\ and\ \bibinfo {author} {\bibfnamefont {S.}~\bibnamefont
  {Nishimoto}},\ }\bibfield  {title} {\bibinfo {title} {Deconfined criticality
  in the frustrated {H}eisenberg honeycomb antiferromagnet},\ }\href
  {https://doi.org/10.1103/PhysRevLett.110.127203} {\bibfield  {journal}
  {\bibinfo  {journal} {Phys. Rev. Lett.}\ }\textbf {\bibinfo {volume} {110}},\
  \bibinfo {pages} {127203} (\bibinfo {year} {2013})}\BibitemShut {NoStop}%
\bibitem [{\citenamefont {Balents}(2010)}]{Balents2010}%
  \BibitemOpen
  \bibfield  {author} {\bibinfo {author} {\bibfnamefont {L.}~\bibnamefont
  {Balents}},\ }\bibfield  {title} {\bibinfo {title} {Spin liquids in
  frustrated magnets},\ }\href {https://doi.org/10.1038/nature08917} {\bibfield
   {journal} {\bibinfo  {journal} {Nature}\ }\textbf {\bibinfo {volume}
  {464}},\ \bibinfo {pages} {199} (\bibinfo {year} {2010})}\BibitemShut
  {NoStop}%
\bibitem [{\citenamefont {Savary}\ and\ \citenamefont
  {Balents}(2017)}]{Savary2017}%
  \BibitemOpen
  \bibfield  {author} {\bibinfo {author} {\bibfnamefont {L.}~\bibnamefont
  {Savary}}\ and\ \bibinfo {author} {\bibfnamefont {L.}~\bibnamefont
  {Balents}},\ }\bibfield  {title} {\bibinfo {title} {Quantum spin liquids: a
  review},\ }\href {https://doi.org/10.1088/0034-4885/80/1/016502} {\bibfield
  {journal} {\bibinfo  {journal} {Rep. Prog. Phys.}\ }\textbf {\bibinfo
  {volume} {80}},\ \bibinfo {pages} {016502} (\bibinfo {year}
  {2017})}\BibitemShut {NoStop}%
\bibitem [{\citenamefont {Zhou}\ \emph {et~al.}(2017)\citenamefont {Zhou},
  \citenamefont {Kanoda},\ and\ \citenamefont {Ng}}]{Zhou2017}%
  \BibitemOpen
  \bibfield  {author} {\bibinfo {author} {\bibfnamefont {Y.}~\bibnamefont
  {Zhou}}, \bibinfo {author} {\bibfnamefont {K.}~\bibnamefont {Kanoda}},\ and\
  \bibinfo {author} {\bibfnamefont {T.-K.}\ \bibnamefont {Ng}},\ }\bibfield
  {title} {\bibinfo {title} {Quantum spin liquid states},\ }\href
  {https://doi.org/10.1103/RevModPhys.89.025003} {\bibfield  {journal}
  {\bibinfo  {journal} {Rev. Mod. Phys.}\ }\textbf {\bibinfo {volume} {89}},\
  \bibinfo {pages} {025003} (\bibinfo {year} {2017})}\BibitemShut {NoStop}%
\bibitem [{\citenamefont {Broholm}\ \emph {et~al.}(2020)\citenamefont
  {Broholm}, \citenamefont {Cava}, \citenamefont {Kivelson}, \citenamefont
  {Nocera}, \citenamefont {Norman},\ and\ \citenamefont
  {Senthil}}]{Broholm2020}%
  \BibitemOpen
  \bibfield  {author} {\bibinfo {author} {\bibfnamefont {C.}~\bibnamefont
  {Broholm}}, \bibinfo {author} {\bibfnamefont {R.~J.}\ \bibnamefont {Cava}},
  \bibinfo {author} {\bibfnamefont {S.~A.}\ \bibnamefont {Kivelson}}, \bibinfo
  {author} {\bibfnamefont {D.~G.}\ \bibnamefont {Nocera}}, \bibinfo {author}
  {\bibfnamefont {M.~R.}\ \bibnamefont {Norman}},\ and\ \bibinfo {author}
  {\bibfnamefont {T.}~\bibnamefont {Senthil}},\ }\bibfield  {title} {\bibinfo
  {title} {Quantum spin liquids},\ }\href
  {https://doi.org/10.1126/science.aay0668} {\bibfield  {journal} {\bibinfo
  {journal} {Science}\ }\textbf {\bibinfo {volume} {367}},\ \bibinfo {pages}
  {eaay0668} (\bibinfo {year} {2020})}\BibitemShut {NoStop}%
\bibitem [{\citenamefont {Kim}\ \emph {et~al.}(2000)\citenamefont {Kim},
  \citenamefont {F\'ath}, \citenamefont {S\'olyom},\ and\ \citenamefont
  {Scalapino}}]{Kim2000}%
  \BibitemOpen
  \bibfield  {author} {\bibinfo {author} {\bibfnamefont {E.~H.}\ \bibnamefont
  {Kim}}, \bibinfo {author} {\bibfnamefont {G.}~\bibnamefont {F\'ath}},
  \bibinfo {author} {\bibfnamefont {J.}~\bibnamefont {S\'olyom}},\ and\
  \bibinfo {author} {\bibfnamefont {D.~J.}\ \bibnamefont {Scalapino}},\
  }\bibfield  {title} {\bibinfo {title} {Phase transitions between
  topologically distinct gapped phases in isotropic spin ladders},\ }\href
  {https://doi.org/10.1103/PhysRevB.62.14965} {\bibfield  {journal} {\bibinfo
  {journal} {Phys. Rev. B}\ }\textbf {\bibinfo {volume} {62}},\ \bibinfo
  {pages} {14965} (\bibinfo {year} {2000})}\BibitemShut {NoStop}%
\bibitem [{\citenamefont {Yamashita}\ \emph {et~al.}(2010)\citenamefont
  {Yamashita}, \citenamefont {Nakata}, \citenamefont {Senshu}, \citenamefont
  {Nagata}, \citenamefont {Yamamoto}, \citenamefont {Kato}, \citenamefont
  {Shibauchi},\ and\ \citenamefont {Matsuda}}]{Yamashita2010}%
  \BibitemOpen
  \bibfield  {author} {\bibinfo {author} {\bibfnamefont {M.}~\bibnamefont
  {Yamashita}}, \bibinfo {author} {\bibfnamefont {N.}~\bibnamefont {Nakata}},
  \bibinfo {author} {\bibfnamefont {Y.}~\bibnamefont {Senshu}}, \bibinfo
  {author} {\bibfnamefont {M.}~\bibnamefont {Nagata}}, \bibinfo {author}
  {\bibfnamefont {H.~M.}\ \bibnamefont {Yamamoto}}, \bibinfo {author}
  {\bibfnamefont {R.}~\bibnamefont {Kato}}, \bibinfo {author} {\bibfnamefont
  {T.}~\bibnamefont {Shibauchi}},\ and\ \bibinfo {author} {\bibfnamefont
  {Y.}~\bibnamefont {Matsuda}},\ }\bibfield  {title} {\bibinfo {title} {Highly
  mobile gapless excitations in a two-dimensional candidate quantum spin
  liquid},\ }\href {https://doi.org/10.1126/science.1188200} {\bibfield
  {journal} {\bibinfo  {journal} {Science}\ }\textbf {\bibinfo {volume}
  {328}},\ \bibinfo {pages} {1246} (\bibinfo {year} {2010})}\BibitemShut
  {NoStop}%
\bibitem [{\citenamefont {Jiang}\ \emph {et~al.}(2012)\citenamefont {Jiang},
  \citenamefont {Wang},\ and\ \citenamefont {Balents}}]{Jiang2012}%
  \BibitemOpen
  \bibfield  {author} {\bibinfo {author} {\bibfnamefont {H.-C.}\ \bibnamefont
  {Jiang}}, \bibinfo {author} {\bibfnamefont {Z.}~\bibnamefont {Wang}},\ and\
  \bibinfo {author} {\bibfnamefont {L.}~\bibnamefont {Balents}},\ }\bibfield
  {title} {\bibinfo {title} {Identifying topological order by entanglement
  entropy},\ }\href {https://doi.org/10.1038/nphys2465} {\bibfield  {journal}
  {\bibinfo  {journal} {Nat. Phys.}\ }\textbf {\bibinfo {volume} {8}},\
  \bibinfo {pages} {902} (\bibinfo {year} {2012})}\BibitemShut {NoStop}%
\bibitem [{\citenamefont {Han}\ \emph {et~al.}(2012)\citenamefont {Han},
  \citenamefont {Helton}, \citenamefont {Chu}, \citenamefont {Nocera},
  \citenamefont {Rodriguez-Rivera}, \citenamefont {Broholm},\ and\
  \citenamefont {Lee}}]{Han2012}%
  \BibitemOpen
  \bibfield  {author} {\bibinfo {author} {\bibfnamefont {T.-H.}\ \bibnamefont
  {Han}}, \bibinfo {author} {\bibfnamefont {J.~S.}\ \bibnamefont {Helton}},
  \bibinfo {author} {\bibfnamefont {S.}~\bibnamefont {Chu}}, \bibinfo {author}
  {\bibfnamefont {D.~G.}\ \bibnamefont {Nocera}}, \bibinfo {author}
  {\bibfnamefont {J.~A.}\ \bibnamefont {Rodriguez-Rivera}}, \bibinfo {author}
  {\bibfnamefont {C.}~\bibnamefont {Broholm}},\ and\ \bibinfo {author}
  {\bibfnamefont {Y.~S.}\ \bibnamefont {Lee}},\ }\bibfield  {title} {\bibinfo
  {title} {Fractionalized excitations in the spin-liquid state of a
  kagome-lattice antiferromagnet},\ }\href
  {https://doi.org/10.1038/nature11659} {\bibfield  {journal} {\bibinfo
  {journal} {Nature}\ }\textbf {\bibinfo {volume} {492}},\ \bibinfo {pages}
  {406} (\bibinfo {year} {2012})}\BibitemShut {NoStop}%
\bibitem [{\citenamefont {Banerjee}\ \emph {et~al.}(2016)\citenamefont
  {Banerjee}, \citenamefont {Bridges}, \citenamefont {Yan}, \citenamefont
  {Aczel}, \citenamefont {Li}, \citenamefont {Stone}, \citenamefont {Granroth},
  \citenamefont {Lumsden}, \citenamefont {Yiu}, \citenamefont {Knolle},
  \citenamefont {Bhattacharjee}, \citenamefont {Kovrizhin}, \citenamefont
  {Moessner}, \citenamefont {Tennant}, \citenamefont {Mandrus},\ and\
  \citenamefont {Nagler}}]{Banerjee2016}%
  \BibitemOpen
  \bibfield  {author} {\bibinfo {author} {\bibfnamefont {A.}~\bibnamefont
  {Banerjee}}, \bibinfo {author} {\bibfnamefont {C.~A.}\ \bibnamefont
  {Bridges}}, \bibinfo {author} {\bibfnamefont {J.}~\bibnamefont {Yan}},
  \bibinfo {author} {\bibfnamefont {A.~A.}\ \bibnamefont {Aczel}}, \bibinfo
  {author} {\bibfnamefont {L.}~\bibnamefont {Li}}, \bibinfo {author}
  {\bibfnamefont {M.~B.}\ \bibnamefont {Stone}}, \bibinfo {author}
  {\bibfnamefont {G.~E.}\ \bibnamefont {Granroth}}, \bibinfo {author}
  {\bibfnamefont {M.~D.}\ \bibnamefont {Lumsden}}, \bibinfo {author}
  {\bibfnamefont {Y.}~\bibnamefont {Yiu}}, \bibinfo {author} {\bibfnamefont
  {J.}~\bibnamefont {Knolle}}, \bibinfo {author} {\bibfnamefont
  {S.}~\bibnamefont {Bhattacharjee}}, \bibinfo {author} {\bibfnamefont {D.~L.}\
  \bibnamefont {Kovrizhin}}, \bibinfo {author} {\bibfnamefont {R.}~\bibnamefont
  {Moessner}}, \bibinfo {author} {\bibfnamefont {D.~A.}\ \bibnamefont
  {Tennant}}, \bibinfo {author} {\bibfnamefont {D.~G.}\ \bibnamefont
  {Mandrus}},\ and\ \bibinfo {author} {\bibfnamefont {S.~E.}\ \bibnamefont
  {Nagler}},\ }\bibfield  {title} {\bibinfo {title} {Proximate {K}itaev quantum
  spin liquid behaviour in a honeycomb magnet},\ }\href
  {https://doi.org/10.1038/nmat4604} {\bibfield  {journal} {\bibinfo  {journal}
  {Nat. Mater.}\ }\textbf {\bibinfo {volume} {15}},\ \bibinfo {pages} {733}
  (\bibinfo {year} {2016})}\BibitemShut {NoStop}%
\bibitem [{\citenamefont {Henelius}\ and\ \citenamefont
  {Sandvik}(2000)}]{Henelius2000}%
  \BibitemOpen
  \bibfield  {author} {\bibinfo {author} {\bibfnamefont {P.}~\bibnamefont
  {Henelius}}\ and\ \bibinfo {author} {\bibfnamefont {A.~W.}\ \bibnamefont
  {Sandvik}},\ }\bibfield  {title} {\bibinfo {title} {Sign problem in {M}onte
  {C}arlo simulations of frustrated quantum spin systems},\ }\href
  {https://doi.org/10.1103/PhysRevB.62.1102} {\bibfield  {journal} {\bibinfo
  {journal} {Phys. Rev. B}\ }\textbf {\bibinfo {volume} {62}},\ \bibinfo
  {pages} {1102} (\bibinfo {year} {2000})}\BibitemShut {NoStop}%
\bibitem [{\citenamefont {Troyer}\ and\ \citenamefont
  {Wiese}(2005)}]{Troyer2005}%
  \BibitemOpen
  \bibfield  {author} {\bibinfo {author} {\bibfnamefont {M.}~\bibnamefont
  {Troyer}}\ and\ \bibinfo {author} {\bibfnamefont {U.-J.}\ \bibnamefont
  {Wiese}},\ }\bibfield  {title} {\bibinfo {title} {Computational complexity
  and fundamental limitations to fermionic quantum {M}onte {C}arlo
  simulations},\ }\href {https://doi.org/10.1103/PhysRevLett.94.170201}
  {\bibfield  {journal} {\bibinfo  {journal} {Phys. Rev. Lett.}\ }\textbf
  {\bibinfo {volume} {94}},\ \bibinfo {pages} {170201} (\bibinfo {year}
  {2005})}\BibitemShut {NoStop}%
\bibitem [{\citenamefont {Rokhsar}\ and\ \citenamefont
  {Kivelson}(1988)}]{RK1988}%
  \BibitemOpen
  \bibfield  {author} {\bibinfo {author} {\bibfnamefont {D.~S.}\ \bibnamefont
  {Rokhsar}}\ and\ \bibinfo {author} {\bibfnamefont {S.~A.}\ \bibnamefont
  {Kivelson}},\ }\bibfield  {title} {\bibinfo {title} {Superconductivity and
  the quantum hard-core dimer gas},\ }\href
  {https://doi.org/10.1103/PhysRevLett.61.2376} {\bibfield  {journal} {\bibinfo
   {journal} {Phys. Rev. Lett.}\ }\textbf {\bibinfo {volume} {61}},\ \bibinfo
  {pages} {2376} (\bibinfo {year} {1988})}\BibitemShut {NoStop}%
\bibitem [{\citenamefont {Kitaev}(2006)}]{Kitaev2006}%
  \BibitemOpen
  \bibfield  {author} {\bibinfo {author} {\bibfnamefont {A.}~\bibnamefont
  {Kitaev}},\ }\bibfield  {title} {\bibinfo {title} {Anyons in an exactly
  solved model and beyond},\ }\href
  {https://doi.org/https://doi.org/10.1016/j.aop.2005.10.005} {\bibfield
  {journal} {\bibinfo  {journal} {Annals of Physics}\ }\textbf {\bibinfo
  {volume} {321}},\ \bibinfo {pages} {2} (\bibinfo {year} {2006})}\BibitemShut
  {NoStop}%
\bibitem [{\citenamefont {Wu}(2006)}]{Wu2006}%
  \BibitemOpen
  \bibfield  {author} {\bibinfo {author} {\bibfnamefont {F.~Y.}\ \bibnamefont
  {Wu}},\ }\bibfield  {title} {\bibinfo {title} {Dimers on two-dimensional
  lattices},\ }\href {https://doi.org/10.1142/S0217979206036478} {\bibfield
  {journal} {\bibinfo  {journal} {Int. J. Mod. Phys.}\ }\textbf {\bibinfo
  {volume} {20}},\ \bibinfo {pages} {5357} (\bibinfo {year}
  {2006})}\BibitemShut {NoStop}%
\bibitem [{\citenamefont {Moessner}\ and\ \citenamefont
  {Raman}(2011)}]{Moessner2011}%
  \BibitemOpen
  \bibfield  {author} {\bibinfo {author} {\bibfnamefont {R.}~\bibnamefont
  {Moessner}}\ and\ \bibinfo {author} {\bibfnamefont {K.~S.}\ \bibnamefont
  {Raman}},\ }\bibinfo {title} {Quantum dimer models},\ in\ \href
  {https://doi.org/10.1007/978-3-642-10589-0_17} {\emph {\bibinfo {booktitle}
  {Introduction to Frustrated Magnetism: Materials, Experiments, Theory}}},\
  \bibinfo {editor} {edited by\ \bibinfo {editor} {\bibfnamefont
  {C.}~\bibnamefont {Lacroix}}, \bibinfo {editor} {\bibfnamefont
  {P.}~\bibnamefont {Mendels}},\ and\ \bibinfo {editor} {\bibfnamefont
  {F.}~\bibnamefont {Mila}}}\ (\bibinfo  {publisher} {Springer Berlin
  Heidelberg},\ \bibinfo {address} {Berlin, Heidelberg},\ \bibinfo {year}
  {2011})\ pp.\ \bibinfo {pages} {437--479}\BibitemShut {NoStop}%
\bibitem [{\citenamefont {Moessner}\ \emph {et~al.}(2001)\citenamefont
  {Moessner}, \citenamefont {Sondhi},\ and\ \citenamefont
  {Chandra}}]{Moessner2001}%
  \BibitemOpen
  \bibfield  {author} {\bibinfo {author} {\bibfnamefont {R.}~\bibnamefont
  {Moessner}}, \bibinfo {author} {\bibfnamefont {S.~L.}\ \bibnamefont
  {Sondhi}},\ and\ \bibinfo {author} {\bibfnamefont {P.}~\bibnamefont
  {Chandra}},\ }\bibfield  {title} {\bibinfo {title} {Phase diagram of the
  hexagonal lattice quantum dimer model},\ }\href
  {https://doi.org/10.1103/PhysRevB.64.144416} {\bibfield  {journal} {\bibinfo
  {journal} {Phys. Rev. B}\ }\textbf {\bibinfo {volume} {64}},\ \bibinfo
  {pages} {144416} (\bibinfo {year} {2001})}\BibitemShut {NoStop}%
\bibitem [{\citenamefont {Cabra}\ \emph {et~al.}(2005)\citenamefont {Cabra},
  \citenamefont {Grynberg}, \citenamefont {Holdsworth}, \citenamefont
  {Honecker}, \citenamefont {Pujol}, \citenamefont {Richter}, \citenamefont
  {Schmalfu{\ss}},\ and\ \citenamefont {Schulenburg}}]{Cabra2005}%
  \BibitemOpen
  \bibfield  {author} {\bibinfo {author} {\bibfnamefont {D.~C.}\ \bibnamefont
  {Cabra}}, \bibinfo {author} {\bibfnamefont {M.~D.}\ \bibnamefont {Grynberg}},
  \bibinfo {author} {\bibfnamefont {P.~C.~W.}\ \bibnamefont {Holdsworth}},
  \bibinfo {author} {\bibfnamefont {A.}~\bibnamefont {Honecker}}, \bibinfo
  {author} {\bibfnamefont {P.}~\bibnamefont {Pujol}}, \bibinfo {author}
  {\bibfnamefont {J.}~\bibnamefont {Richter}}, \bibinfo {author} {\bibfnamefont
  {D.}~\bibnamefont {Schmalfu{\ss}}},\ and\ \bibinfo {author} {\bibfnamefont
  {J.}~\bibnamefont {Schulenburg}},\ }\bibfield  {title} {\bibinfo {title}
  {Quantum kagom\'e antiferromagnet in a magnetic field: Low-lying nonmagnetic
  excitations versus valence-bond crystal order},\ }\href
  {https://doi.org/10.1103/PhysRevB.71.144420} {\bibfield  {journal} {\bibinfo
  {journal} {Phys. Rev. B}\ }\textbf {\bibinfo {volume} {71}},\ \bibinfo
  {pages} {144420} (\bibinfo {year} {2005})}\BibitemShut {NoStop}%
\bibitem [{\citenamefont {Schnack}\ \emph {et~al.}(2001)\citenamefont
  {Schnack}, \citenamefont {Schmidt}, \citenamefont {Richter},\ and\
  \citenamefont {Schulenburg}}]{Schnack2001}%
  \BibitemOpen
  \bibfield  {author} {\bibinfo {author} {\bibfnamefont {J.}~\bibnamefont
  {Schnack}}, \bibinfo {author} {\bibfnamefont {H.-J.}\ \bibnamefont
  {Schmidt}}, \bibinfo {author} {\bibfnamefont {J.}~\bibnamefont {Richter}},\
  and\ \bibinfo {author} {\bibfnamefont {J.}~\bibnamefont {Schulenburg}},\
  }\bibfield  {title} {\bibinfo {title} {Independent magnon states on magnetic
  polytopes},\ }\href {https://doi.org/10.1007/s10051-001-8701-6} {\bibfield
  {journal} {\bibinfo  {journal} {Eur. Phys. J. B}\ }\textbf {\bibinfo {volume}
  {24}},\ \bibinfo {pages} {475} (\bibinfo {year} {2001})}\BibitemShut
  {NoStop}%
\bibitem [{\citenamefont {Schulenburg}\ \emph {et~al.}(2002)\citenamefont
  {Schulenburg}, \citenamefont {Honecker}, \citenamefont {Schnack},
  \citenamefont {Richter},\ and\ \citenamefont {Schmidt}}]{Schulenburg2002}%
  \BibitemOpen
  \bibfield  {author} {\bibinfo {author} {\bibfnamefont {J.}~\bibnamefont
  {Schulenburg}}, \bibinfo {author} {\bibfnamefont {A.}~\bibnamefont
  {Honecker}}, \bibinfo {author} {\bibfnamefont {J.}~\bibnamefont {Schnack}},
  \bibinfo {author} {\bibfnamefont {J.}~\bibnamefont {Richter}},\ and\ \bibinfo
  {author} {\bibfnamefont {H.-J.}\ \bibnamefont {Schmidt}},\ }\bibfield
  {title} {\bibinfo {title} {Macroscopic magnetization jumps due to independent
  magnons in frustrated quantum spin lattices},\ }\href
  {https://doi.org/10.1103/PhysRevLett.88.167207} {\bibfield  {journal}
  {\bibinfo  {journal} {Phys. Rev. Lett.}\ }\textbf {\bibinfo {volume} {88}},\
  \bibinfo {pages} {167207} (\bibinfo {year} {2002})}\BibitemShut {NoStop}%
\bibitem [{\citenamefont {Zhitomirsky}\ and\ \citenamefont
  {Tsunetsugu}(2004)}]{ZT2004}%
  \BibitemOpen
  \bibfield  {author} {\bibinfo {author} {\bibfnamefont {M.~E.}\ \bibnamefont
  {Zhitomirsky}}\ and\ \bibinfo {author} {\bibfnamefont {H.}~\bibnamefont
  {Tsunetsugu}},\ }\bibfield  {title} {\bibinfo {title} {Exact low-temperature
  behavior of a kagom\'e antiferromagnet at high fields},\ }\href
  {https://doi.org/10.1103/PhysRevB.70.100403} {\bibfield  {journal} {\bibinfo
  {journal} {Phys. Rev. B}\ }\textbf {\bibinfo {volume} {70}},\ \bibinfo
  {pages} {100403} (\bibinfo {year} {2004})}\BibitemShut {NoStop}%
\bibitem [{\citenamefont {Zhitomirsky}\ and\ \citenamefont
  {Tsunetsugu}(2005)}]{ZT2005}%
  \BibitemOpen
  \bibfield  {author} {\bibinfo {author} {\bibfnamefont {M.~E.}\ \bibnamefont
  {Zhitomirsky}}\ and\ \bibinfo {author} {\bibfnamefont {H.}~\bibnamefont
  {Tsunetsugu}},\ }\bibfield  {title} {\bibinfo {title} {High field properties
  of geometrically frustrated magnets},\ }\href
  {https://doi.org/10.1143/PTPS.160.361} {\bibfield  {journal} {\bibinfo
  {journal} {Prog. Theor. Phys. Suppl.}\ }\textbf {\bibinfo {volume} {160}},\
  \bibinfo {pages} {361} (\bibinfo {year} {2005})}\BibitemShut {NoStop}%
\bibitem [{\citenamefont {Derzhko}\ \emph {et~al.}(2007)\citenamefont
  {Derzhko}, \citenamefont {Richter}, \citenamefont {Honecker},\ and\
  \citenamefont {Schmidt}}]{DRHS07}%
  \BibitemOpen
  \bibfield  {author} {\bibinfo {author} {\bibfnamefont {O.}~\bibnamefont
  {Derzhko}}, \bibinfo {author} {\bibfnamefont {J.}~\bibnamefont {Richter}},
  \bibinfo {author} {\bibfnamefont {A.}~\bibnamefont {Honecker}},\ and\
  \bibinfo {author} {\bibfnamefont {H.-J.}\ \bibnamefont {Schmidt}},\
  }\bibfield  {title} {\bibinfo {title} {Universal properties of highly
  frustrated quantum magnets in strong magnetic fields},\ }\href
  {https://doi.org/10.1063/1.2780166} {\bibfield  {journal} {\bibinfo
  {journal} {Low Temp. Phys.}\ }\textbf {\bibinfo {volume} {33}},\ \bibinfo
  {pages} {745} (\bibinfo {year} {2007})}\BibitemShut {NoStop}%
\bibitem [{\citenamefont {Schnack}\ \emph {et~al.}(2020)\citenamefont
  {Schnack}, \citenamefont {Schulenburg}, \citenamefont {Honecker},\ and\
  \citenamefont {Richter}}]{Schnack2020}%
  \BibitemOpen
  \bibfield  {author} {\bibinfo {author} {\bibfnamefont {J.}~\bibnamefont
  {Schnack}}, \bibinfo {author} {\bibfnamefont {J.}~\bibnamefont
  {Schulenburg}}, \bibinfo {author} {\bibfnamefont {A.}~\bibnamefont
  {Honecker}},\ and\ \bibinfo {author} {\bibfnamefont {J.}~\bibnamefont
  {Richter}},\ }\bibfield  {title} {\bibinfo {title} {Magnon crystallization in
  the kagome lattice antiferromagnet},\ }\href
  {https://doi.org/10.1103/PhysRevLett.125.117207} {\bibfield  {journal}
  {\bibinfo  {journal} {Phys. Rev. Lett.}\ }\textbf {\bibinfo {volume} {125}},\
  \bibinfo {pages} {117207} (\bibinfo {year} {2020})}\BibitemShut {NoStop}%
\bibitem [{\citenamefont {Morita}\ and\ \citenamefont
  {Shibata}(2016)}]{Morita2016}%
  \BibitemOpen
  \bibfield  {author} {\bibinfo {author} {\bibfnamefont {K.}~\bibnamefont
  {Morita}}\ and\ \bibinfo {author} {\bibfnamefont {N.}~\bibnamefont
  {Shibata}},\ }\bibfield  {title} {\bibinfo {title} {Exact nonmagnetic ground
  state and residual entropy of {$S = 1/2$} {H}eisenberg diamond spin
  lattices},\ }\href {https://doi.org/10.7566/JPSJ.85.033705} {\bibfield
  {journal} {\bibinfo  {journal} {J. Phys. Soc. Jpn}\ }\textbf {\bibinfo
  {volume} {85}},\ \bibinfo {pages} {033705} (\bibinfo {year}
  {2016})}\BibitemShut {NoStop}%
\bibitem [{\citenamefont {Hirose}\ \emph {et~al.}(2016)\citenamefont {Hirose},
  \citenamefont {Oguchi},\ and\ \citenamefont {Fukumoto}}]{Hirose2016}%
  \BibitemOpen
  \bibfield  {author} {\bibinfo {author} {\bibfnamefont {Y.}~\bibnamefont
  {Hirose}}, \bibinfo {author} {\bibfnamefont {A.}~\bibnamefont {Oguchi}},\
  and\ \bibinfo {author} {\bibfnamefont {Y.}~\bibnamefont {Fukumoto}},\
  }\bibfield  {title} {\bibinfo {title} {Exact realization of a quantum-dimer
  model in {H}eisenberg antiferromagnets on a diamond-like decorated lattice},\
  }\href {https://api.semanticscholar.org/CorpusID:118465501} {\bibfield
  {journal} {\bibinfo  {journal} {J. Phys. Soc. Jpn}\ }\textbf {\bibinfo
  {volume} {85}},\ \bibinfo {pages} {094002} (\bibinfo {year}
  {2016})}\BibitemShut {NoStop}%
\bibitem [{\citenamefont {Hirose}\ \emph {et~al.}(2017)\citenamefont {Hirose},
  \citenamefont {Miura}, \citenamefont {Yasuda},\ and\ \citenamefont
  {Fukumoto}}]{Hirose2017}%
  \BibitemOpen
  \bibfield  {author} {\bibinfo {author} {\bibfnamefont {Y.}~\bibnamefont
  {Hirose}}, \bibinfo {author} {\bibfnamefont {S.}~\bibnamefont {Miura}},
  \bibinfo {author} {\bibfnamefont {C.}~\bibnamefont {Yasuda}},\ and\ \bibinfo
  {author} {\bibfnamefont {Y.}~\bibnamefont {Fukumoto}},\ }\bibfield  {title}
  {\bibinfo {title} {Notes on ground-state properties of mixed spin-1 and
  spin-1/2 {L}ieb-lattice {H}eisenberg antiferromagnets},\ }\href
  {https://doi.org/10.7566/JPSJ.86.083705} {\bibfield  {journal} {\bibinfo
  {journal} {J. Phys. Soc. Jpn}\ }\textbf {\bibinfo {volume} {86}},\ \bibinfo
  {pages} {083705} (\bibinfo {year} {2017})}\BibitemShut {NoStop}%
\bibitem [{\citenamefont {Hirose}\ \emph {et~al.}(2018)\citenamefont {Hirose},
  \citenamefont {Miura}, \citenamefont {Yasuda},\ and\ \citenamefont
  {Fukumoto}}]{Hirose2018}%
  \BibitemOpen
  \bibfield  {author} {\bibinfo {author} {\bibfnamefont {Y.}~\bibnamefont
  {Hirose}}, \bibinfo {author} {\bibfnamefont {S.}~\bibnamefont {Miura}},
  \bibinfo {author} {\bibfnamefont {C.}~\bibnamefont {Yasuda}},\ and\ \bibinfo
  {author} {\bibfnamefont {Y.}~\bibnamefont {Fukumoto}},\ }\bibfield  {title}
  {\bibinfo {title} {Ground-state properties of spin-1/2 {H}eisenberg
  antiferromagnets with frustration on the diamond-like-decorated square and
  triangular lattices},\ }\href {https://doi.org/10.1063/1.5042792} {\bibfield
  {journal} {\bibinfo  {journal} {AIP Advances}\ }\textbf {\bibinfo {volume}
  {8}},\ \bibinfo {pages} {101427} (\bibinfo {year} {2018})}\BibitemShut
  {NoStop}%
\bibitem [{\citenamefont {Hirose}\ \emph {et~al.}(2020)\citenamefont {Hirose},
  \citenamefont {Oguchi},\ and\ \citenamefont {Fukumoto}}]{Hirose2020}%
  \BibitemOpen
  \bibfield  {author} {\bibinfo {author} {\bibfnamefont {Y.}~\bibnamefont
  {Hirose}}, \bibinfo {author} {\bibfnamefont {A.}~\bibnamefont {Oguchi}},\
  and\ \bibinfo {author} {\bibfnamefont {Y.}~\bibnamefont {Fukumoto}},\
  }\bibfield  {title} {\bibinfo {title} {Quantum dimer model containing
  {R}okhsar-{K}ivelson point expressed by spin-$\frac{1}{2}$ {H}eisenberg
  antiferromagnets},\ }\href {https://doi.org/10.1103/PhysRevB.101.174440}
  {\bibfield  {journal} {\bibinfo  {journal} {Phys. Rev. B}\ }\textbf {\bibinfo
  {volume} {101}},\ \bibinfo {pages} {174440} (\bibinfo {year}
  {2020})}\BibitemShut {NoStop}%
\bibitem [{\citenamefont {Caci}\ \emph {et~al.}(2023)\citenamefont {Caci},
  \citenamefont {Karl'ov\'a}, \citenamefont {Verkholyak}, \citenamefont
  {Stre\ifmmode~\check{c}\else \v{c}\fi{}ka}, \citenamefont {Wessel},\ and\
  \citenamefont {Honecker}}]{Caci2023}%
  \BibitemOpen
  \bibfield  {author} {\bibinfo {author} {\bibfnamefont {N.}~\bibnamefont
  {Caci}}, \bibinfo {author} {\bibfnamefont {K.}~\bibnamefont {Karl'ov\'a}},
  \bibinfo {author} {\bibfnamefont {T.}~\bibnamefont {Verkholyak}}, \bibinfo
  {author} {\bibfnamefont {J.}~\bibnamefont {Stre\ifmmode~\check{c}\else
  \v{c}\fi{}ka}}, \bibinfo {author} {\bibfnamefont {S.}~\bibnamefont
  {Wessel}},\ and\ \bibinfo {author} {\bibfnamefont {A.}~\bibnamefont
  {Honecker}},\ }\bibfield  {title} {\bibinfo {title} {Phases of the
  spin-$\frac{1}{2}$ {H}eisenberg antiferromagnet on the diamond-decorated
  square lattice in a magnetic field},\ }\href
  {https://doi.org/10.1103/PhysRevB.107.115143} {\bibfield  {journal} {\bibinfo
   {journal} {Phys. Rev. B}\ }\textbf {\bibinfo {volume} {107}},\ \bibinfo
  {pages} {115143} (\bibinfo {year} {2023})}\BibitemShut {NoStop}%
\bibitem [{\citenamefont {Lieb}(1967)}]{Lieb1967}%
  \BibitemOpen
  \bibfield  {author} {\bibinfo {author} {\bibfnamefont {E.~H.}\ \bibnamefont
  {Lieb}},\ }\bibfield  {title} {\bibinfo {title} {Solution of the dimer
  problem by the transfer matrix method},\ }\href
  {https://doi.org/10.1063/1.1705163} {\bibfield  {journal} {\bibinfo
  {journal} {J. Math. Phys.}\ }\textbf {\bibinfo {volume} {8}},\ \bibinfo
  {pages} {2339} (\bibinfo {year} {1967})}\BibitemShut {NoStop}%
\bibitem [{\citenamefont {Heilmann}\ and\ \citenamefont
  {Lieb}(1970)}]{Heilmann1970}%
  \BibitemOpen
  \bibfield  {author} {\bibinfo {author} {\bibfnamefont {O.~J.}\ \bibnamefont
  {Heilmann}}\ and\ \bibinfo {author} {\bibfnamefont {E.~H.}\ \bibnamefont
  {Lieb}},\ }\bibfield  {title} {\bibinfo {title} {Monomers and dimers},\
  }\href {https://doi.org/10.1103/PhysRevLett.24.1412} {\bibfield  {journal}
  {\bibinfo  {journal} {Phys. Rev. Lett.}\ }\textbf {\bibinfo {volume} {24}},\
  \bibinfo {pages} {1412} (\bibinfo {year} {1970})}\BibitemShut {NoStop}%
\bibitem [{\citenamefont {Suzuki}(1971{\natexlab{a}})}]{Suzuki1971a}%
  \BibitemOpen
  \bibfield  {author} {\bibinfo {author} {\bibfnamefont {M.}~\bibnamefont
  {Suzuki}},\ }\bibfield  {title} {\bibinfo {title} {The dimer problem and the
  generalized {$XY$}-model},\ }\href
  {https://doi.org/https://doi.org/10.1016/0375-9601(71)90901-7} {\bibfield
  {journal} {\bibinfo  {journal} {Phys. Lett. A}\ }\textbf {\bibinfo {volume}
  {34}},\ \bibinfo {pages} {338} (\bibinfo {year}
  {1971}{\natexlab{a}})}\BibitemShut {NoStop}%
\bibitem [{\citenamefont {Suzuki}(1971{\natexlab{b}})}]{Suzuki1971b}%
  \BibitemOpen
  \bibfield  {author} {\bibinfo {author} {\bibfnamefont {M.}~\bibnamefont
  {Suzuki}},\ }\bibfield  {title} {\bibinfo {title} {{Relationship among
  Exactly Soluble Models of Critical Phenomena. {I} --2{D} {I}sing Model, Dimer
  Problem and the Generalized {$XY$}-Model--}},\ }\href
  {https://doi.org/10.1143/PTP.46.1337} {\bibfield  {journal} {\bibinfo
  {journal} {Prog. Theor. Phys.}\ }\textbf {\bibinfo {volume} {46}},\ \bibinfo
  {pages} {1337} (\bibinfo {year} {1971}{\natexlab{b}})}\BibitemShut {NoStop}%
\bibitem [{\citenamefont {{Heilmann}}\ and\ \citenamefont
  {{Lieb}}(1972)}]{Heilmann1972}%
  \BibitemOpen
  \bibfield  {author} {\bibinfo {author} {\bibfnamefont {O.~J.}\ \bibnamefont
  {{Heilmann}}}\ and\ \bibinfo {author} {\bibfnamefont {E.~H.}\ \bibnamefont
  {{Lieb}}},\ }\bibfield  {title} {\bibinfo {title} {{Theory of monomer-dimer
  systems}},\ }\href {https://doi.org/10.1007/BF01877590} {\bibfield  {journal}
  {\bibinfo  {journal} {Comm. Math. Phys.}\ }\textbf {\bibinfo {volume} {25}},\
  \bibinfo {pages} {190} (\bibinfo {year} {1972})}\BibitemShut {NoStop}%
\bibitem [{\citenamefont {Grande}\ \emph {et~al.}(2011)\citenamefont {Grande},
  \citenamefont {Salinas},\ and\ \citenamefont {da~Costa}}]{Grande2011}%
  \BibitemOpen
  \bibfield  {author} {\bibinfo {author} {\bibfnamefont {H.~L.~C.}\
  \bibnamefont {Grande}}, \bibinfo {author} {\bibfnamefont {S.~R.}\
  \bibnamefont {Salinas}},\ and\ \bibinfo {author} {\bibfnamefont {F.~A.}\
  \bibnamefont {da~Costa}},\ }\bibfield  {title} {\bibinfo {title} {Fermionic
  representation of two-dimensional dimer models},\ }\href
  {https://doi.org/10.1007/s13538-011-0011-8} {\bibfield  {journal} {\bibinfo
  {journal} {Braz. J. Phys.}\ }\textbf {\bibinfo {volume} {41}},\ \bibinfo
  {pages} {86} (\bibinfo {year} {2011})}\BibitemShut {NoStop}%
\bibitem [{\citenamefont {Wilkins}\ and\ \citenamefont
  {Powell}(2021)}]{Wilkins2021}%
  \BibitemOpen
  \bibfield  {author} {\bibinfo {author} {\bibfnamefont {N.}~\bibnamefont
  {Wilkins}}\ and\ \bibinfo {author} {\bibfnamefont {S.}~\bibnamefont
  {Powell}},\ }\bibfield  {title} {\bibinfo {title} {Topological sectors, dimer
  correlations, and monomers from the transfer-matrix solution of the dimer
  model},\ }\href {https://doi.org/10.1103/PhysRevE.104.014145} {\bibfield
  {journal} {\bibinfo  {journal} {Phys. Rev. E}\ }\textbf {\bibinfo {volume}
  {104}},\ \bibinfo {pages} {014145} (\bibinfo {year} {2021})}\BibitemShut
  {NoStop}%
\bibitem [{\citenamefont {Zhitomirsky}(2003)}]{Mike}%
  \BibitemOpen
  \bibfield  {author} {\bibinfo {author} {\bibfnamefont {M.~E.}\ \bibnamefont
  {Zhitomirsky}},\ }\bibfield  {title} {\bibinfo {title} {Enhanced
  magnetocaloric effect in frustrated magnets},\ }\href
  {https://doi.org/10.1103/PhysRevB.67.104421} {\bibfield  {journal} {\bibinfo
  {journal} {Phys. Rev. B}\ }\textbf {\bibinfo {volume} {67}},\ \bibinfo
  {pages} {104421} (\bibinfo {year} {2003})}\BibitemShut {NoStop}%
\bibitem [{\citenamefont {Zhitomirsky}\ and\ \citenamefont
  {Honecker}(2004)}]{Zhitomirsky_2004}%
  \BibitemOpen
  \bibfield  {author} {\bibinfo {author} {\bibfnamefont {M.~E.}\ \bibnamefont
  {Zhitomirsky}}\ and\ \bibinfo {author} {\bibfnamefont {A.}~\bibnamefont
  {Honecker}},\ }\bibfield  {title} {\bibinfo {title} {Magnetocaloric effect in
  one-dimensional antiferromagnets},\ }\href
  {https://doi.org/10.1088/1742-5468/2004/07/P07012} {\bibfield  {journal}
  {\bibinfo  {journal} {J. Stat. Mech.: Theor. Exp.}\ }\textbf {\bibinfo
  {volume} {2004}},\ \bibinfo {pages} {P07012} (\bibinfo {year}
  {2004})}\BibitemShut {NoStop}%
\bibitem [{\citenamefont {Honecker}\ and\ \citenamefont
  {Wessel}(2006)}]{HONECKER20061098}%
  \BibitemOpen
  \bibfield  {author} {\bibinfo {author} {\bibfnamefont {A.}~\bibnamefont
  {Honecker}}\ and\ \bibinfo {author} {\bibfnamefont {S.}~\bibnamefont
  {Wessel}},\ }\bibfield  {title} {\bibinfo {title} {Magnetocaloric effect in
  two-dimensional spin-1/2 antiferromagnets},\ }\href
  {https://doi.org/https://doi.org/10.1016/j.physb.2006.01.436} {\bibfield
  {journal} {\bibinfo  {journal} {Physica B}\ }\textbf {\bibinfo {volume}
  {378-380}},\ \bibinfo {pages} {1098} (\bibinfo {year} {2006})}\BibitemShut
  {NoStop}%
\bibitem [{\citenamefont {Stre\v{c}ka}\ \emph {et~al.}(2023)\citenamefont
  {Stre\v{c}ka}, \citenamefont {Karl'ov\'a}, \citenamefont {Verkholyak},
  \citenamefont {Caci}, \citenamefont {Wessel},\ and\ \citenamefont
  {Honecker}}]{Strecka2023}%
  \BibitemOpen
  \bibfield  {author} {\bibinfo {author} {\bibfnamefont {J.}~\bibnamefont
  {Stre\v{c}ka}}, \bibinfo {author} {\bibfnamefont {K.}~\bibnamefont
  {Karl'ov\'a}}, \bibinfo {author} {\bibfnamefont {T.}~\bibnamefont
  {Verkholyak}}, \bibinfo {author} {\bibfnamefont {N.}~\bibnamefont {Caci}},
  \bibinfo {author} {\bibfnamefont {S.}~\bibnamefont {Wessel}},\ and\ \bibinfo
  {author} {\bibfnamefont {A.}~\bibnamefont {Honecker}},\ }\bibfield  {title}
  {\bibinfo {title} {Thermal first-order phase transitions, {I}sing critical
  points, and reentrance in the {I}sing-{H}eisenberg model on the
  diamond-decorated square lattice in a magnetic field},\ }\href
  {https://doi.org/10.1103/PhysRevB.107.134402} {\bibfield  {journal} {\bibinfo
   {journal} {Phys. Rev. B}\ }\textbf {\bibinfo {volume} {107}},\ \bibinfo
  {pages} {134402} (\bibinfo {year} {2023})}\BibitemShut {NoStop}%
\bibitem [{\citenamefont {Ghannadan}\ and\ \citenamefont
  {Stre\v{c}ka}(2024)}]{Ghannadan2024}%
  \BibitemOpen
  \bibfield  {author} {\bibinfo {author} {\bibfnamefont {A.}~\bibnamefont
  {Ghannadan}}\ and\ \bibinfo {author} {\bibfnamefont {J.}~\bibnamefont
  {Stre\v{c}ka}},\ }\bibfield  {title} {\bibinfo {title} {A possible
  coexistence of spontaneous magnetic order and thermal entanglement in a
  spin-1/2 {I}sing-{H}eisenberg diamond-decorated square lattice},\ }\href
  {https://doi.org/https://doi.org/10.1016/j.cjph.2024.01.013} {\bibfield
  {journal} {\bibinfo  {journal} {Chin. J. Phys.}\ }\textbf {\bibinfo {volume}
  {89}},\ \bibinfo {pages} {1062} (\bibinfo {year} {2024})}\BibitemShut
  {NoStop}%
\bibitem [{\citenamefont {Morin-Duchesne}\ \emph {et~al.}(2015)\citenamefont
  {Morin-Duchesne}, \citenamefont {Rasmussen},\ and\ \citenamefont
  {Ruelle}}]{Morin2015}%
  \BibitemOpen
  \bibfield  {author} {\bibinfo {author} {\bibfnamefont {A.}~\bibnamefont
  {Morin-Duchesne}}, \bibinfo {author} {\bibfnamefont {J.}~\bibnamefont
  {Rasmussen}},\ and\ \bibinfo {author} {\bibfnamefont {P.}~\bibnamefont
  {Ruelle}},\ }\bibfield  {title} {\bibinfo {title} {Dimer representations of
  the {T}emperley-{L}ieb algebra},\ }\href
  {https://doi.org/https://doi.org/10.1016/j.nuclphysb.2014.11.016} {\bibfield
  {journal} {\bibinfo  {journal} {Nucl. Phys. B}\ }\textbf {\bibinfo {volume}
  {890}},\ \bibinfo {pages} {363} (\bibinfo {year} {2015})}\BibitemShut
  {NoStop}%
\bibitem [{\citenamefont {Lanczos}(1950)}]{Lanczos1950}%
  \BibitemOpen
  \bibfield  {author} {\bibinfo {author} {\bibfnamefont {C.}~\bibnamefont
  {Lanczos}},\ }\bibfield  {title} {\bibinfo {title} {{An iteration method for
  the solution of the eigenvalue problem of linear differential and integral
  operators}},\ }\href {https://doi.org/10.6028/jres.045.026} {\bibfield
  {journal} {\bibinfo  {journal} {J. Res. Natl. Bur. Stand.}\ }\textbf
  {\bibinfo {volume} {45}},\ \bibinfo {pages} {255} (\bibinfo {year}
  {1950})}\BibitemShut {NoStop}%
\bibitem [{\citenamefont {Honecker}\ and\ \citenamefont {Wessel}(2009)}]{HW09}%
  \BibitemOpen
  \bibfield  {author} {\bibinfo {author} {\bibfnamefont {A.}~\bibnamefont
  {Honecker}}\ and\ \bibinfo {author} {\bibfnamefont {S.}~\bibnamefont
  {Wessel}},\ }\bibfield  {title} {\bibinfo {title} {Magnetocaloric effect in
  quantum spin-$s$ chains},\ }\href {https://doi.org/10.5488/CMP.12.3.399}
  {\bibfield  {journal} {\bibinfo  {journal} {Condensed Matter Physics}\
  }\textbf {\bibinfo {volume} {12}},\ \bibinfo {pages} {399} (\bibinfo {year}
  {2009})}\BibitemShut {NoStop}%
\bibitem [{\citenamefont {Fisher}(1961)}]{Fisher61}%
  \BibitemOpen
  \bibfield  {author} {\bibinfo {author} {\bibfnamefont {M.~E.}\ \bibnamefont
  {Fisher}},\ }\bibfield  {title} {\bibinfo {title} {Statistical mechanics of
  dimers on a plane lattice},\ }\href
  {https://doi.org/10.1103/PhysRev.124.1664} {\bibfield  {journal} {\bibinfo
  {journal} {Phys. Rev.}\ }\textbf {\bibinfo {volume} {124}},\ \bibinfo {pages}
  {1664} (\bibinfo {year} {1961})}\BibitemShut {NoStop}%
\bibitem [{\citenamefont {Kasteleyn}(1961)}]{Kasteleyn61}%
  \BibitemOpen
  \bibfield  {author} {\bibinfo {author} {\bibfnamefont {P.~W.}\ \bibnamefont
  {Kasteleyn}},\ }\bibfield  {title} {\bibinfo {title} {The statistics of
  dimers on a lattice: {I}. {T}he number of dimer arrangements on a quadratic
  lattice},\ }\href
  {https://doi.org/https://doi.org/10.1016/0031-8914(61)90063-5} {\bibfield
  {journal} {\bibinfo  {journal} {Physica}\ }\textbf {\bibinfo {volume} {27}},\
  \bibinfo {pages} {1209} (\bibinfo {year} {1961})}\BibitemShut {NoStop}%
\bibitem [{\citenamefont {Sandvik}\ and\ \citenamefont
  {Kurkij\"arvi}(1991)}]{Sandvik1991}%
  \BibitemOpen
  \bibfield  {author} {\bibinfo {author} {\bibfnamefont {A.~W.}\ \bibnamefont
  {Sandvik}}\ and\ \bibinfo {author} {\bibfnamefont {J.}~\bibnamefont
  {Kurkij\"arvi}},\ }\bibfield  {title} {\bibinfo {title} {Quantum {M}onte
  {C}arlo simulation method for spin systems},\ }\href
  {https://doi.org/10.1103/PhysRevB.43.5950} {\bibfield  {journal} {\bibinfo
  {journal} {Phys. Rev. B}\ }\textbf {\bibinfo {volume} {43}},\ \bibinfo
  {pages} {5950} (\bibinfo {year} {1991})}\BibitemShut {NoStop}%
\bibitem [{\citenamefont {Sandvik}(1999)}]{Sandvik1999}%
  \BibitemOpen
  \bibfield  {author} {\bibinfo {author} {\bibfnamefont {A.~W.}\ \bibnamefont
  {Sandvik}},\ }\bibfield  {title} {\bibinfo {title} {Stochastic series
  expansion method with operator-loop update},\ }\href
  {https://doi.org/10.1103/PhysRevB.59.R14157} {\bibfield  {journal} {\bibinfo
  {journal} {Phys. Rev. B}\ }\textbf {\bibinfo {volume} {59}},\ \bibinfo
  {pages} {R14157} (\bibinfo {year} {1999})}\BibitemShut {NoStop}%
\bibitem [{\citenamefont {Sylju\aa{}sen}\ and\ \citenamefont
  {Sandvik}(2002)}]{Syljuasen2002}%
  \BibitemOpen
  \bibfield  {author} {\bibinfo {author} {\bibfnamefont {O.~F.}\ \bibnamefont
  {Sylju\aa{}sen}}\ and\ \bibinfo {author} {\bibfnamefont {A.~W.}\ \bibnamefont
  {Sandvik}},\ }\bibfield  {title} {\bibinfo {title} {Quantum {M}onte {C}arlo
  with directed loops},\ }\href {https://doi.org/10.1103/PhysRevE.66.046701}
  {\bibfield  {journal} {\bibinfo  {journal} {Phys. Rev. E}\ }\textbf {\bibinfo
  {volume} {66}},\ \bibinfo {pages} {046701} (\bibinfo {year}
  {2002})}\BibitemShut {NoStop}%
\bibitem [{\citenamefont {Hangleiter}\ \emph {et~al.}(2020)\citenamefont
  {Hangleiter}, \citenamefont {Roth}, \citenamefont {Nagaj},\ and\
  \citenamefont {Eisert}}]{Hangleiter2020}%
  \BibitemOpen
  \bibfield  {author} {\bibinfo {author} {\bibfnamefont {D.}~\bibnamefont
  {Hangleiter}}, \bibinfo {author} {\bibfnamefont {I.}~\bibnamefont {Roth}},
  \bibinfo {author} {\bibfnamefont {D.}~\bibnamefont {Nagaj}},\ and\ \bibinfo
  {author} {\bibfnamefont {J.}~\bibnamefont {Eisert}},\ }\bibfield  {title}
  {\bibinfo {title} {Easing the {M}onte {C}arlo sign problem},\ }\href
  {https://doi.org/10.1126/sciadv.abb8341} {\bibfield  {journal} {\bibinfo
  {journal} {Science Advances}\ }\textbf {\bibinfo {volume} {6}},\ \bibinfo
  {pages} {eabb8341} (\bibinfo {year} {2020})}\BibitemShut {NoStop}%
\bibitem [{\citenamefont {Hen}(2021)}]{Hen2021}%
  \BibitemOpen
  \bibfield  {author} {\bibinfo {author} {\bibfnamefont {I.}~\bibnamefont
  {Hen}},\ }\bibfield  {title} {\bibinfo {title} {Determining quantum {M}onte
  {C}arlo simulability with geometric phases},\ }\href
  {https://doi.org/10.1103/PhysRevResearch.3.023080} {\bibfield  {journal}
  {\bibinfo  {journal} {Phys. Rev. Res.}\ }\textbf {\bibinfo {volume} {3}},\
  \bibinfo {pages} {023080} (\bibinfo {year} {2021})}\BibitemShut {NoStop}%
\bibitem [{\citenamefont {Nakamura}(1998)}]{Nakamura1998}%
  \BibitemOpen
  \bibfield  {author} {\bibinfo {author} {\bibfnamefont {T.}~\bibnamefont
  {Nakamura}},\ }\bibfield  {title} {\bibinfo {title} {Vanishing of the
  negative-sign problem of quantum {M}onte {C}arlo simulations in
  one-dimensional frustrated spin systems},\ }\href
  {https://doi.org/10.1103/PhysRevB.57.R3197} {\bibfield  {journal} {\bibinfo
  {journal} {Phys. Rev. B}\ }\textbf {\bibinfo {volume} {57}},\ \bibinfo
  {pages} {R3197} (\bibinfo {year} {1998})}\BibitemShut {NoStop}%
\bibitem [{\citenamefont {Alet}\ \emph {et~al.}(2016)\citenamefont {Alet},
  \citenamefont {Damle},\ and\ \citenamefont {Pujari}}]{Alet16}%
  \BibitemOpen
  \bibfield  {author} {\bibinfo {author} {\bibfnamefont {F.}~\bibnamefont
  {Alet}}, \bibinfo {author} {\bibfnamefont {K.}~\bibnamefont {Damle}},\ and\
  \bibinfo {author} {\bibfnamefont {S.}~\bibnamefont {Pujari}},\ }\bibfield
  {title} {\bibinfo {title} {Sign-problem-free {M}onte {C}arlo simulation of
  certain frustrated quantum magnets},\ }\href
  {https://doi.org/10.1103/PhysRevLett.117.197203} {\bibfield  {journal}
  {\bibinfo  {journal} {Phys. Rev. Lett.}\ }\textbf {\bibinfo {volume} {117}},\
  \bibinfo {pages} {197203} (\bibinfo {year} {2016})}\BibitemShut {NoStop}%
\bibitem [{\citenamefont {Honecker}\ \emph {et~al.}(2016)\citenamefont
  {Honecker}, \citenamefont {Wessel}, \citenamefont {Kerkdyk}, \citenamefont
  {Pruschke}, \citenamefont {Mila},\ and\ \citenamefont
  {Normand}}]{Honecker16}%
  \BibitemOpen
  \bibfield  {author} {\bibinfo {author} {\bibfnamefont {A.}~\bibnamefont
  {Honecker}}, \bibinfo {author} {\bibfnamefont {S.}~\bibnamefont {Wessel}},
  \bibinfo {author} {\bibfnamefont {R.}~\bibnamefont {Kerkdyk}}, \bibinfo
  {author} {\bibfnamefont {T.}~\bibnamefont {Pruschke}}, \bibinfo {author}
  {\bibfnamefont {F.}~\bibnamefont {Mila}},\ and\ \bibinfo {author}
  {\bibfnamefont {B.}~\bibnamefont {Normand}},\ }\bibfield  {title} {\bibinfo
  {title} {Thermodynamic properties of highly frustrated quantum spin ladders:
  Influence of many-particle bound states},\ }\href
  {https://doi.org/10.1103/PhysRevB.93.054408} {\bibfield  {journal} {\bibinfo
  {journal} {Phys. Rev. B}\ }\textbf {\bibinfo {volume} {93}},\ \bibinfo
  {pages} {054408} (\bibinfo {year} {2016})}\BibitemShut {NoStop}%
\bibitem [{\citenamefont {Weber}\ \emph {et~al.}(2022)\citenamefont {Weber},
  \citenamefont {Honecker}, \citenamefont {Normand}, \citenamefont {Corboz},
  \citenamefont {Mila},\ and\ \citenamefont {Wessel}}]{Weber2022}%
  \BibitemOpen
  \bibfield  {author} {\bibinfo {author} {\bibfnamefont {L.}~\bibnamefont
  {Weber}}, \bibinfo {author} {\bibfnamefont {A.}~\bibnamefont {Honecker}},
  \bibinfo {author} {\bibfnamefont {B.}~\bibnamefont {Normand}}, \bibinfo
  {author} {\bibfnamefont {P.}~\bibnamefont {Corboz}}, \bibinfo {author}
  {\bibfnamefont {F.}~\bibnamefont {Mila}},\ and\ \bibinfo {author}
  {\bibfnamefont {S.}~\bibnamefont {Wessel}},\ }\bibfield  {title} {\bibinfo
  {title} {Quantum {M}onte {C}arlo simulations in the trimer basis: first-order
  transitions and thermal critical points in frustrated trilayer magnets},\
  }\href {https://doi.org/10.21468/SciPostPhys.12.2.054} {\bibfield  {journal}
  {\bibinfo  {journal} {SciPost Phys.}\ }\textbf {\bibinfo {volume} {12}},\
  \bibinfo {pages} {054} (\bibinfo {year} {2022})}\BibitemShut {NoStop}%
\bibitem [{\citenamefont {Ng}\ and\ \citenamefont {Yang}(2017)}]{NgYang17}%
  \BibitemOpen
  \bibfield  {author} {\bibinfo {author} {\bibfnamefont {K.-K.}\ \bibnamefont
  {Ng}}\ and\ \bibinfo {author} {\bibfnamefont {M.-F.}\ \bibnamefont {Yang}},\
  }\bibfield  {title} {\bibinfo {title} {Field-induced quantum phases in a
  frustrated spin-dimer model: A sign-problem-free quantum {M}onte {C}arlo
  study},\ }\href {https://doi.org/10.1103/PhysRevB.95.064431} {\bibfield
  {journal} {\bibinfo  {journal} {Phys. Rev. B}\ }\textbf {\bibinfo {volume}
  {95}},\ \bibinfo {pages} {064431} (\bibinfo {year} {2017})}\BibitemShut
  {NoStop}%
\bibitem [{\citenamefont {Stapmanns}\ \emph {et~al.}(2018)\citenamefont
  {Stapmanns}, \citenamefont {Corboz}, \citenamefont {Mila}, \citenamefont
  {Honecker}, \citenamefont {Normand},\ and\ \citenamefont
  {Wessel}}]{Stapmanns2018}%
  \BibitemOpen
  \bibfield  {author} {\bibinfo {author} {\bibfnamefont {J.}~\bibnamefont
  {Stapmanns}}, \bibinfo {author} {\bibfnamefont {P.}~\bibnamefont {Corboz}},
  \bibinfo {author} {\bibfnamefont {F.}~\bibnamefont {Mila}}, \bibinfo {author}
  {\bibfnamefont {A.}~\bibnamefont {Honecker}}, \bibinfo {author}
  {\bibfnamefont {B.}~\bibnamefont {Normand}},\ and\ \bibinfo {author}
  {\bibfnamefont {S.}~\bibnamefont {Wessel}},\ }\bibfield  {title} {\bibinfo
  {title} {Thermal critical points and quantum critical end point in the
  frustrated bilayer {H}eisenberg antiferromagnet},\ }\href
  {https://doi.org/10.1103/PhysRevLett.121.127201} {\bibfield  {journal}
  {\bibinfo  {journal} {Phys. Rev. Lett.}\ }\textbf {\bibinfo {volume} {121}},\
  \bibinfo {pages} {127201} (\bibinfo {year} {2018})}\BibitemShut {NoStop}%
\bibitem [{\citenamefont {Fan}\ \emph {et~al.}(2024)\citenamefont {Fan},
  \citenamefont {Xi}, \citenamefont {Liu}, \citenamefont {Normand},\ and\
  \citenamefont {Yu}}]{Fan2024}%
  \BibitemOpen
  \bibfield  {author} {\bibinfo {author} {\bibfnamefont {Y.}~\bibnamefont
  {Fan}}, \bibinfo {author} {\bibfnamefont {N.}~\bibnamefont {Xi}}, \bibinfo
  {author} {\bibfnamefont {C.}~\bibnamefont {Liu}}, \bibinfo {author}
  {\bibfnamefont {B.}~\bibnamefont {Normand}},\ and\ \bibinfo {author}
  {\bibfnamefont {R.}~\bibnamefont {Yu}},\ }\bibfield  {title} {\bibinfo
  {title} {Field-controlled multicritical behavior and emergent universality in
  fully frustrated quantum magnets},\ }\href
  {https://doi.org/10.1038/s41535-024-00636-4} {\bibfield  {journal} {\bibinfo
  {journal} {npj Quantum Mater.}\ }\textbf {\bibinfo {volume} {9}},\ \bibinfo
  {pages} {25} (\bibinfo {year} {2024})}\BibitemShut {NoStop}%
\bibitem [{\citenamefont {Nernst}(1906)}]{Nernst1906}%
  \BibitemOpen
  \bibfield  {author} {\bibinfo {author} {\bibfnamefont {W.}~\bibnamefont
  {Nernst}},\ }\bibfield  {title} {\bibinfo {title} {{U}eber die {B}erechnung
  chemischer {G}leichgewichte aus thermischen {M}essungen},\ }\href
  {http://eudml.org/doc/58630} {\bibfield  {journal} {\bibinfo  {journal}
  {Nachrichten von der Gesellschaft der Wissenschaften zu G\"ottingen,
  Mathematisch-Physikalische Klasse}\ }\textbf {\bibinfo {volume} {1906}},\
  \bibinfo {pages} {1} (\bibinfo {year} {1906})}\BibitemShut {NoStop}%
\bibitem [{\citenamefont {Zhang}\ \emph {et~al.}(2000)\citenamefont {Zhang},
  \citenamefont {Tong}, \citenamefont {Chen}, \citenamefont {Yu},\ and\
  \citenamefont {Kang}}]{Zhang2000}%
  \BibitemOpen
  \bibfield  {author} {\bibinfo {author} {\bibfnamefont {H.-X.}\ \bibnamefont
  {Zhang}}, \bibinfo {author} {\bibfnamefont {Y.-X.}\ \bibnamefont {Tong}},
  \bibinfo {author} {\bibfnamefont {Z.-N.}\ \bibnamefont {Chen}}, \bibinfo
  {author} {\bibfnamefont {K.-B.}\ \bibnamefont {Yu}},\ and\ \bibinfo {author}
  {\bibfnamefont {B.-S.}\ \bibnamefont {Kang}},\ }\bibfield  {title} {\bibinfo
  {title} {Cyano-bridged extended heteronuclear supramolecular architectures
  with hexacyanoferrates({II}) as building blocks},\ }\href
  {https://doi.org/10.1016/S0022-328X(99)00679-8} {\bibfield  {journal}
  {\bibinfo  {journal} {J. Organomet. Chem.}\ }\textbf {\bibinfo {volume}
  {598}},\ \bibinfo {pages} {63} (\bibinfo {year} {2000})}\BibitemShut
  {NoStop}%
\bibitem [{\citenamefont {Tr\'avn\'{\i}\v{c}ek}\ \emph
  {et~al.}(2001)\citenamefont {Tr\'avn\'{\i}\v{c}ek}, \citenamefont {Sm\'ekal},
  \citenamefont {Escuer},\ and\ \citenamefont {Marek}}]{Travnicek2001}%
  \BibitemOpen
  \bibfield  {author} {\bibinfo {author} {\bibfnamefont {Z.}~\bibnamefont
  {Tr\'avn\'{\i}\v{c}ek}}, \bibinfo {author} {\bibfnamefont {Z.}~\bibnamefont
  {Sm\'ekal}}, \bibinfo {author} {\bibfnamefont {A.}~\bibnamefont {Escuer}},\
  and\ \bibinfo {author} {\bibfnamefont {J.}~\bibnamefont {Marek}},\ }\bibfield
   {title} {\bibinfo {title} {Synthesis, structure and magnetic behaviour of a
  two-dimensional cyano-bridged complex
  {[$\{$Cu(ept)$\}_3$Fe(CN)$_6$](ClO$_4$)$2\cdot5$H$_2$O} [ept
  ={$N$}-(2-aminoethyl)-1,3-diaminopropane]},\ }\href
  {https://doi.org/10.1039/B006741P} {\bibfield  {journal} {\bibinfo  {journal}
  {New J. Chem.}\ }\textbf {\bibinfo {volume} {25}},\ \bibinfo {pages} {655}
  (\bibinfo {year} {2001})}\BibitemShut {NoStop}%
\bibitem [{\citenamefont {Hong}\ and\ \citenamefont {You}(2004)}]{Hong2004}%
  \BibitemOpen
  \bibfield  {author} {\bibinfo {author} {\bibfnamefont {C.~S.}\ \bibnamefont
  {Hong}}\ and\ \bibinfo {author} {\bibfnamefont {Y.~S.}\ \bibnamefont {You}},\
  }\bibfield  {title} {\bibinfo {title} {Cyano-bridged {Fe}({II})-{Cu}({II})
  bimetallic assemblies: honeycomb-like and pentanuclear structures},\ }\href
  {https://doi.org/https://doi.org/10.1016/j.ica.2004.04.004} {\bibfield
  {journal} {\bibinfo  {journal} {Inorg. Chim. Acta}\ }\textbf {\bibinfo
  {volume} {357}},\ \bibinfo {pages} {3271} (\bibinfo {year}
  {2004})}\BibitemShut {NoStop}%
\end{thebibliography}%

\end{document}